\documentclass[11pt, lettersize]{article}
\usepackage[T1]{fontenc}
\usepackage[utf8]{inputenc}
\usepackage[margin=1in]{geometry}
\usepackage{amsmath,amssymb,amsthm}
\usepackage{booktabs,tabularx}
\usepackage{tikz}
\newcolumntype{Y}{>{\raggedright\arraybackslash}X}
\allowdisplaybreaks
\usepackage{braket}
\usepackage{comment}
\usepackage{xcolor}
\usepackage{thmtools}
\usepackage{thm-restate}

\usepackage{tcolorbox}
\tcbuselibrary{listings}
\newcounter{protocolcount}

\usepackage{xcolor}
\usepackage[colorlinks=true, citecolor=blue, linkcolor=red, urlcolor=magenta]{hyperref}
\usepackage{cleveref}

\tcbuselibrary{skins,breakable}

\newtheorem{theorem}{Theorem}
\newtheorem{lemma}[theorem]{Lemma}
\newtheorem{proposition}[theorem]{Proposition}
\newtheorem{claim}{Claim}
\newtheorem{fact}{Fact}

\newtheorem{definition}[theorem]{Definition}
\newtheorem{remark}[theorem]{Remark}

\newcommand{\dMA}{\mathsf{dMA}}
\newcommand{\dQMA}{\mathsf{dQMA}}
\newcommand{\treeq}{\mathrm{TreeEQ}}
\newcommand{\symk}{\mathrm{SYM}_D}
\newcommand{\sym}{\mathrm{SYM}}
\newcommand{\D}{\mathrm{D}}
\renewcommand{\ket}[1]{| #1 \rangle}
\renewcommand{\bra}[1]{\langle #1 |}
\newcommand{\F}{\mathbb{F}}

\newcommand{\tr}{\mathrm{tr}}

\newcommand{\A}{\mathsf{A}}
\newcommand{\B}{\mathsf{B}}

\usepackage{tcolorbox}
\tcbuselibrary{skins,breakable}

\newcommand{\eqn}{\mathrm{EQ}_N^t}
\newcommand{\eqtwo}{\mathrm{EQ}_N^2}
\newcommand{\bigo}[1]{O\left(#1\right)}

\crefname{protocolcount}{Protocol}{Protocols}
\Crefname{protocolcount}{Protocol}{Protocols}

\title{Quantum Advantage in Local Certification of Graph Properties}

\usepackage{authblk} 
\author[1,2]{Benjamin Jauregui}
\author[3]{Masayuki Miyamoto}

\affil[1]{Universidad de Chile, Chile}
\affil[2]{IRIF \& Université Paris Cité, France}
\affil[3]{University of Tsukuba, Japan}

\date{}

\begin{document}
\maketitle

    \begin{abstract}
Local certification asks how the vertices of a network
can verify a global property using certificates from an
untrusted prover and communication with their neighbors.
Its central complexity measure is the maximum certificate
size at any vertex.
Fraigniaud, Le Gall, Nishimura, and Paz (ITCS 2021)
introduced quantum local certification through the
\emph{distributed quantum Merlin--Arthur} ($\dQMA$) model
and established an exponential quantum advantage in
certificate size for distributed equality.
In this problem, $t$ vertices each hold an
$N$-bit string, and the network must verify that all
these strings are identical.
That separation is measured in the length $N$ of these
strings, rather than the number of vertices.
Hasegawa, Kundu, and Nishimura (PODC 2024) explicitly
asked whether quantum advantages also arise for graph
properties, which depend solely on the network topology.

We answer this question affirmatively, establishing
unconditional exponential and polynomial quantum
advantages measured with respect to the number $n$ of vertices.
\begin{itemize}
    \item First, for every fixed integer $\mu\ge1$, we construct
a graph property $\treeq_\mu$. We show that this property
admits a one-round $\dQMA$ protocol with
$O(\mu\log n)$-qubit certificates and messages
(hence $O(\log n)$ for constant $\mu$), whereas every $\mu$-round
$\dMA$ protocol for this property requires
$\Omega(n/(\mu\log^2 n))$-bit certificates.
\item Second, for graph symmetry, which asks whether the network
has a nontrivial automorphism, we give a one-round $\dQMA$
protocol with $\widetilde O(n^{5/3})$-qubit certificates.
This yields a polynomial quantum advantage over both
\emph{locally checkable proofs} (LCP) and one-round $\dMA$
protocols, for which an $\Omega(n^2)$-bit certificate
lower bound is known.
\end{itemize}

With respect to graph symmetry, we extend the results in two ways. On one hand, we  show that the quadratic classical lower bound
persists for any fixed number of verification rounds, or even if the graphs have bounded radius.
On the other hand, for certifying symmetry in graphs with radius at most $D$,
our quantum cost improves to $O(nD\log n)$.

These separations rely on two advances in quantum
certification of distributed equality.
First, for $N$-bit inputs at the endpoints of an $r$-edge
path, we improve the $\dQMA$ certificate bound of
Hasegawa et al.~(PODC 2024) from $O(r^2\log N)$
to $O(r\log N)$ qubits per vertex.

Second, we give a new $\dQMA$ protocol for equality on
paths using $\widetilde O(N^{2/3})$ certificate qubits
at every vertex, independently of the path length.
This avoids the linear-size certificates used at relay
vertices in the previous construction and enables
our quantum advantage for graph symmetry.

Finally, we provide an $\Omega(\log N)$ certificate lower bound
for one-round quantum equality protocols on paths of
at least three edges, even with arbitrary entanglement
and unrestricted communication.

\end{abstract}

\newpage
\tableofcontents

\newpage
\section{Introduction}

A distributed system consists of independent computing devices that communicate through a network. Examples of such networks include data-center networks, peer-to-peer systems, wireless, and sensor networks. These networks are usually seen as a connected graph in which the devices are represented as nodes, and edges represent direct communication links between devices. Then, a task to be solved within this network requires the development of a {\it distributed algorithm} through the graph, in which each node performs its own computation and exchanges messages with its neighbors. Here, a fundamental measure of efficiency is the number of
communication rounds. The LOCAL model~\cite{linial1992locality} is a standard model focusing on this resource by allowing unrestricted local computation and message sizes, with synchronous communication restricted to network edges. It is also important to {\it verify} whether a given property holds in the network. For example, the nodes may want to verify whether the network has a certain topology, such as being a tree or a planar graph, whether a distributed computation has produced a valid output, or whether some other global condition is satisfied. In this setting, the goal is to design a protocol that allows the nodes to collectively decide whether the property holds, despite each node having access only to local information and to the information obtained through communication with other nodes. Such verification protocols are relevant for several reasons.
On the one hand, distributed algorithms are often designed under specific assumptions on the network. For example, an algorithm may be more efficient on a particular class of graphs, or may only be correct when the network is a tree, a planar graph, or satisfies some other structural property. Before running such an algorithm, it is therefore important to verify that the network indeed satisfies the assumptions under which the algorithm was designed.
On the other hand, verification may be necessary for the correct operation of the distributed system itself. Nodes may need to check that a computation has produced a valid output, that the information stored across the network is consistent, or that the system remains in a valid state. Verification can also support fault tolerance: detecting that a required condition is no longer satisfied may indicate that an error has occurred and that a recovery procedure should be triggered. In all these cases, the fundamental challenge is to verify a global condition using only local communication.

\paragraph{Local certification.} The verification of such properties is studied within the framework of \emph{local certification}, where nodes use additional information, supplied as certificates, to check global properties through local communication. In a local certification protocol, a prover assigns a certificate $c(v)$ to each node $v$ of the graph, and the nodes inspect their certificates and exchange messages with their neighbors before deciding whether to accept. The prover here is not trusted: when the property holds, there must be certificates that make all nodes accept; when it does not hold, every certificate assignment must cause at least one node to reject, with a required probability. After receiving certificates, the network is allowed to perform an \textit{efficient verification algorithm}. For example, in locally checkable proofs~\cite{goos2016locally}, the network performs a constant-round LOCAL algorithm, and in proof-labeling schemes~\cite{korman2010proof}, each node only broadcasts its certificate to its neighbors in a single round. There are many other related models (see the survey~\cite{feuilloley2021introduction}). Certificates, therefore, allow the nodes to check a global claim without having to reconstruct the entire network at every node. A central question is how much information these certificates must contain.

In particular, the size of the certificates depends on the property being checked. As arguably the simplest case, consider the property of being $k$-colorable: the vertices can be assigned colors from $\{0,\ldots,k-1\}$ so that adjacent vertices receive different colors. This property admits a simple certification protocol. The prover assigns a color to each node as its certificate. Each node exchanges its color with its neighbors and accepts if its color belongs to the allowed set and differs from every neighbor's color. If the graph is $k$-colorable, the prover can supply a valid coloring that makes all nodes accept. Conversely, if all nodes accept, the supplied colors define a valid $k$-coloring. Thus, $k$-colorability can be certified in one communication round using $O(\log k)$ bits per node. Indeed, extensive research has established efficient certification protocols, with $O(\log n)$ bits per node, for multiple graph properties such as planar, bounded-genus and interval graphs~\cite{FFMRT2021,ESPERET202268,JMRR2025}, connectivity-related properties~\cite{BFP2024,chang2026local}, as well as meta-theorems covering broad
families of logically definable properties on structurally restricted
graphs~\cite{BC2025,BFP2022,CKM2025}. In contrast, it is known that some other properties require much more information. 
Indeed, $\widetilde{\Omega}(n)$ bits are required for properties such as unit-disk graphs~\cite{DELMR2024}, bounded diameter graphs~\cite{censor2020approximate}, graphs with forbidden subgraphs~\cite{bousquet2024local,BOUSQUET2025115091,miyamoto2025distributed}.
A canonical maximally hard example is
\emph{graph symmetry}, the existence of a nontrivial automorphism.
In their seminal work, G\"o\"os and Suomela established its
$\Theta(n^2)$ certificate complexity~\cite{goos2016locally}, matching the universal $O(n^2)$-bit upper bound obtained by supplying every vertex with a complete description of the network. 

\paragraph{Quantum local certification.}
Understanding when quantum computation and communication can reduce the round complexity of distributed algorithms has received growing attention in both the CONGEST and LOCAL models~\cite{LeGallMagniez2018QuantumCONGEST,LeGallNishimuraRosmanis2019QuantumLOCAL,WuYao2022QuantumDiameter,VanApeldoornDeVos2022QuantumQueries,FraigniaudEtAl2024EvenCycles,FLS2026}.
For locally checkable problems, recent work has established quantum advantages~\cite{BalliuEtAl2025QuantumLocal,BalliuEtAl2026QuantumLCL}, while also identifying limitations for graph coloring~\cite{CoiteuxRoyEtAl2024,FraigniaudMagniezZiccardi2027} and the distributed Lov{\'a}sz local lemma~\cite{BrandtGoettlicher2026}, with further structural insights coming from connections between quantum distributed computing, non-signaling distributions, and online locality~\cite{AkbariEtAl2025OnlineQuantum}.

In this paper, we focus on a complementary line of work, which investigates quantum advantages in the certificate space required for constant-round verification~\cite{fraigniaud2021distributed, LeGallEtAl2023DistributedMASynthesis, le2023distributed, hasegawa2024power} with a basic question about how much smaller quantum certificates can be than classical ones. 
Fraigniaud, Le Gall, Nishimura, and Paz~\cite{fraigniaud2021distributed} initiated this direction by introducing a framework of quantum local certification, more precisely, distributed quantum Merlin--Arthur ($\mathsf{dQMA}$) protocols (see Definition~\ref{def:dma-dqma} for a formal definition). Within this model, they proved a quantum advantage for the verification of ``replicated data'', which we will refer to as the \emph{distributed equality problem} and denote as $\eqn$:
    Given two parameters $N,t\in\mathbb N$, and a collection of $N$-bit strings $x_1,\dots,x_t$ stored at $t$ terminal nodes $u_1,\dots,u_t$, the problem $\eqn$ corresponds to verifying the equality $x_1=\dots = x_t$.
In~\cite{fraigniaud2021distributed}, they proved that $\eqn$ can be certified in the $\dQMA$ model with $O(tr^2\log N)$ qubits, where $r:=\min_{i\in\{1,\ldots,t\}}\max_{j\in\{1,\ldots,t\}}\operatorname{dist}_G(u_i,u_j)$ is the radius of the terminal set $\{u_1,\ldots,u_t\}$, but requires $\Omega(N)$ bits in the classical $\dMA$ model. Subsequent works further developed this framework~\cite{LeGallEtAl2023DistributedMASynthesis,hasegawa2024power}. In all these works, the quantum advantage established is with respect to the size $N$ of the input strings. Therefore, whether quantum advantages extend to graph properties, which depend only on the network topology rather than additional inputs supplied to its vertices, remained a central open question~\cite{hasegawa2024power}. As we explain next, we answer this question affirmatively.

\subsection{Our results}
In this paper, we compare classical distributed Merlin--Arthur protocols, abbreviated $\dMA$, with quantum distributed Merlin--Arthur protocols, abbreviated $\dQMA$, as in \cite{fraigniaud2021distributed,hasegawa2024power,LeGallEtAl2023DistributedMASynthesis}. See Definition~\ref{def:dma-dqma} for formal definitions of the models.

\subsubsection*{Quantum advantage for graph properties} 

We obtain a quantum advantage for two different graph properties. Throughout this work, we identify each graph property
$\mathcal P$ with the class of graphs satisfying it.
Thus, we use ``$G$ satisfies $\mathcal P$'' and
``$G\in\mathcal P$'' interchangeably.

\paragraph{Exponential advantage.} Inspired by the equality problem $\eqtwo$, for every constant $\mu\ge 1$, we introduce the graph property $\mu$-\emph{tree-encoded equality}, denoted by $\treeq_\mu$.  To define $\treeq_\mu$, we encode the strings to be compared in the original $\eqtwo$ problem into the network topology itself. Informally, the construction consists of two tree components joined by a path of size $\Theta(\mu)$, each encoding a binary string through its branching structure. A graph belongs to $\treeq_\mu$ if it has the prescribed structure and the two encoded strings coincide. We defer the precise construction and formal definition to \Cref{sec:tree}. We establish the following quantum advantage for $\treeq_\mu$.

\begin{theorem}[Exponential quantum advantage for $\treeq_\mu$]\label{thm:treeq-separation}
Let $\mu\ge 1$ be a fixed positive integer. There exists a graph property $\treeq_\mu$ that admits a one-round $\mathsf{dQMA}$
protocol with perfect completeness, soundness error $1/3$,
and certificates with $O(\mu\log n)$ qubits.
The implicit constant is independent of $\mu$.
In contrast, every $\mu$-round $\mathsf{dMA}$ protocol for $\treeq_\mu$
with completeness $c$ and soundness error $s$ satisfying
$2c-1>s$ requires certificates of
$\Omega(n/(\mu\log^2 n))$ bits.
\end{theorem}

\Cref{thm:treeq-separation} establishes an exponential separation between quantum and classical certificate sizes, measured as a function of the number of vertices. For example, by taking $\mu = 1$, this theorem gives a property with distributed quantum complexity equal to $\bigo{\log n}$ qubits, while any $1$-round $\dMA$ protocol requires $\Omega(n/\log^2 n)$ bits.

\paragraph{Polynomial advantage for Graph Symmetry.} The separation in \Cref{thm:treeq-separation} is obtained by encoding instances of $\eqtwo$ into a carefully constructed family of trees. Although this demonstrates the power of quantum certification even on trees, the property is specifically defined to express an underlying equality problem. We therefore complement this construction with a quantum advantage for a more natural graph property without input: the well-studied \emph{symmetry problem}, denoted by $\sym$, which asks whether the graph admits a nontrivial automorphism, that is, an adjacency-preserving permutation of the vertices other than the identity. It is known \cite{fraigniaud2019merlin} that any one-round $\dMA$ protocol for this problem requires certificates of $\Omega(n^2)$ bits. We exhibit a polynomial quantum advantage for this problem.

\begin{restatable}[Quantum advantage for graph symmetry]
{theorem}{quantumsymmetry}\label{thm:quan-upper-sym}
The graph symmetry problem $\sym$ admits a one-round
$\dQMA$ protocol with perfect completeness,
soundness error $1/3$, and certificates with
$O(n^{5/3}(\log n)^{7/3})$
qubits per node.
\end{restatable}

It is known that any $1$-round $\dMA$ protocol for $\sym$ requires $\Omega(n^2)$-bit certificates \cite{fraigniaud2019merlin}, and therefore \Cref{thm:quan-upper-sym} gives a polynomial quantum advantage for this problem.

We complement this result with smaller quantum certificates
for graphs of bounded radius.
For a known integer $D\ge1$, let $\symk$ denote the class
of graphs of radius at most $D$ that admit a nontrivial
automorphism.
We obtain the following upper bound.

\begin{restatable}{corollary}{quantumsymmetrybounded}
\label{cor:quan-upper-symk}
For every known integer $D\ge1$, the property $\symk$
admits a one-round $\dQMA$ protocol with perfect completeness,
soundness error $1/3$, and $O(nD\log n)$ certificate qubits
per node.
\end{restatable}

We also show that the quadratic classical lower bound
persists for any fixed number of verification rounds,
both for $\sym$ and for $\symk$ when $D\ge\mu+4$.
More generally, we prove the following bound.

\begin{restatable}{theorem}{classicalsymmetry}
\label{thm:BC}
Let $\mu=\mu(n)\ge1$ satisfy $\mu=o(n)$.
Every $\mu$-round $\dMA$ protocol for $\sym$
with fixed completeness $c$ and soundness error $s$
satisfying $2c-1>s$ has worst-case certificate size
$\Omega(n^2/\mu)$ bits.
The same bound holds for $\symk$ whenever $D\ge5$
and $\mu\le D-4$.
\end{restatable}

For fixed integers $\mu\ge1$ and $D\ge\mu+4$,
\Cref{cor:quan-upper-symk} and \Cref{thm:BC} therefore establish
a quantum advantage for $\symk$:
$O(n\log n)$ quantum certificate qubits in one round
versus $\Omega(n^2)$ classical certificate bits
in $\mu$ rounds.

\paragraph{Improved bound for equality problem.}
A key ingredient in establishing quantum advantages for graph properties is our improved protocol for the distributed equality problem $\eqn$, previously studied in \cite{fraigniaud2021distributed,hasegawa2024power}. 
We consider the equality problem on a path: two endpoints of a path of $r$ edges receive an $N$-bit input string. The best known $\mathsf{dQMA}$ upper bound is $O(r^2 \log N)$~\cite{fraigniaud2021distributed,hasegawa2024power}. We improve this upper bound in two ways.

\begin{theorem}[Improved protocols for equality on a path]\label{thm:equality-uniform}
For every path $P=v_0,\dots,v_r$ of length $r\geq 1$, the problem $\mathrm{EQ}_N^2$ of two $N$-bit strings held at the endpoint nodes $v_0$ and $v_r$ admits a one-round $\dQMA$ protocol with perfect completeness, soundness error at most $1/3$ and certificates of
\[
O\left( \mathrm{min}\{r\log N,N^{2/3}(\log N)^{5/3}\}\right)
\]
qubits per node.
\end{theorem}

Theorem~\ref{thm:equality-uniform} provides two complementary bounds, each obtained by an independent protocol (\Cref{protocol-equality,new-protocol-equality}).
The first bound improves the previous best known upper bound of $O(r^2 \log N)$ qubits. The second is the first sublinear-in-$N$ bound that is independent of the path length $r$ at every vertex. The relay-based construction of Hasegawa et al.~\cite{hasegawa2024power} removes the distance dependence by storing complete $\Theta(N)$-bit descriptions at designated relay vertices; these vertices determine its worst-case certificate size. Our second protocol compresses the certificates at all vertices, including potential relays.

The two improvements have different applications. The distance-independent bound gives \Cref{thm:quan-upper-sym} for unrestricted symmetry. The $O(r\log N)$ bound improves the radius dependence for $\symk$ from $O(nD^2\log n)$ to $O(nD\log n)$, and the central-path contribution for $\treeq_\mu$ from $O(\mu^2\log n)$ to $O(\mu\log n)$.

We finally show that quantum verification cannot make the local information cost of equality arbitrarily small.

\begin{restatable}[Quantum lower bound]
    {theorem}{equalitylower}
\label{thm:equality-lower}

Consider a one-round $\mathsf{dQMA}$ protocol for equality of $N$-bit
strings held at the endpoints of a path with at least three edges.
Suppose its fixed completeness $c$ and soundness error $s$ satisfy
$2c-1>s$. Then the protocol requires a certificate of
$\Omega(\log N)$ qubits at some node, even if the certificates are
arbitrarily entangled and the messages have unrestricted size.
\end{restatable}

The previously known bounds are
$\Omega((\log N)^{1/4-\varepsilon})$ and
$\Omega((\log N)^{1/2-\varepsilon}/r^{1+\varepsilon})$ on the
total proof and message size over all nodes, obtained by
simulating entangled protocols with separable ones
\cite{hasegawa2024power}, together with an $\Omega(\log N)$ bound
that applies to separable proofs only.

One round of quantum verification therefore outperforms $\mu$ rounds of classical randomized verification. Table~\ref{tab:results} summarizes the bounds. 

\begin{table}[t]
\centering
\small
\renewcommand{\arraystretch}{1.35}
\begin{tabularx}{\textwidth}{@{}lYY@{}}
\toprule
Property & One-round $\dQMA$: qubits per node and message & $\mu$-round $\dMA$: certificate bits per node \\
\midrule
$\treeq_\mu$ & $O(\mu\log n)$ & $\Omega(n/(\mu\log^2 n))$; fixed $\mu\ge1$ \\
$\sym$ & $\widetilde O(n^{5/3})$ & $\Omega(n^2/\mu)$; $\mu=o(n)$ \\
$\symk$ & $O(nD\log n)$ & $\Omega(n^2/\mu)$; $D\ge5$, $\mu\le D-4$, $\mu=o(n)$ \\
$\eqtwo$ & $O(\min\{r\log N,\,N^{2/3}(\log N)^{5/3}\})$ & $\Omega(N/\mu)$ if $r\ge2\mu+1$~\cite[Theorem~9]{fraigniaud2021distributed} \\
\bottomrule
\end{tabularx}
\caption{Quantum upper bounds have perfect completeness and soundness error $1/3$. Classical lower bounds assume fixed completeness $c$ and soundness error $s$ with $2c-1>s$. Here $r$ is the number of path edges, and $N$ is the input-string length.}
\label{tab:results}
\end{table}

\subsection{Our technique}

Now we provide a high-level description of the techniques and new ideas used to obtain our results. We defer the formal presentation of our models to \Cref{sec:pre} and our protocols
to \Cref{sec:eq}, \Cref{sec:tree} and  \Cref{sec:sym}.
For the following overview, we only need a few
graph-theoretic notions.

Given two graphs $G$ and $H$, an isomorphism from $G$ to $H$
is a bijection $\varphi:V(G)\to V(H)$ such that, for all
$u,v\in V(G)$, the vertices $u,v$ are adjacent in $G$
if and only if $\varphi(u),\varphi(v)$ are adjacent in $H$.
We write $G\cong H$ when such an isomorphism exists.
For rooted trees $T_1,T_2$ with roots $r_1,r_2$,
a rooted isomorphism is a graph isomorphism
$\varphi:V(T_1)\to V(T_2)$ satisfying $\varphi(r_1)=r_2$.
The isomorphism class of a graph consists of all graphs
isomorphic to it; for rooted trees, we define the rooted
isomorphism class using rooted isomorphisms.
Accordingly, two graphs have the same isomorphism type
when they are isomorphic, and two rooted trees have the
same rooted isomorphism type when they are isomorphic
by an isomorphism mapping root to root.

\subsubsection{Equality problem on paths}

The complexity from \Cref{thm:equality-uniform} comes from two different protocols. We first explain the $O({r\log N})$ upper bound which follows from a sharper soundness analysis of the protocol of~\cite{hasegawa2024power}.

\begin{figure}[h]
    \centering
    \begin{tikzpicture}[
  x=.82cm,y=.82cm,
  every node/.style={font=\small},
  reg/.style={draw,fill=white,minimum width=.443cm,
    minimum height=.443cm,inner sep=0pt},
  test/.style={draw,rounded corners=2pt,fill=white,
    minimum width=1.279cm,minimum height=.443cm,inner sep=1pt},
  send/.style={->,draw=blue!65!black,line width=.85pt},
  local/.style={->,line width=.6pt}
]
\draw[rounded corners=3pt,draw=gray!40,fill=gray!1] (0.15,-1.14) rectangle (3.25,1.14);
\node at (1.7,1.47) {$v_{0}$};
\node[test] (test0) at (1.05,.4) {$|h_x\rangle\langle h_x|$};
\node[reg] (kept0) at (1.05,-.55) {};
\node[reg] (out0) at (2.35,-.55) {};
\draw[local] (kept0.north) -- (test0.south);
\draw[rounded corners=3pt,draw=gray!40,fill=gray!1] (3.95,-1.14) rectangle (7.05,1.14);
\node at (5.5,1.47) {$v_{1}$};
\node[test] (test1) at (4.85,.4) {SWAP};
\node[reg] (kept1) at (4.85,-.55) {};
\node[reg] (out1) at (6.15,-.55) {};
\draw[local] (kept1.north) -- (test1.south);
\draw[rounded corners=3pt,draw=gray!40,fill=gray!1] (9.55,-1.14) rectangle (12.65,1.14);
\node at (11.1,1.47) {$v_{r-1}$};
\node[test] (test2) at (10.45,.4) {SWAP};
\node[reg] (kept2) at (10.45,-.55) {};
\node[reg] (out2) at (11.75,-.55) {};
\draw[local] (kept2.north) -- (test2.south);
\begin{scope}[local bounding box=rightcontents]
\node[test,minimum width=1.8cm,anchor=west] (rightswap) at (13.45,.4) {SWAP test};
\node[anchor=west,inner sep=0pt] (rightor)
  at ([xshift=10mm]rightswap.east) {or};
\node[test,minimum width=1.8cm,anchor=west] (rightproj)
  at ([xshift=10mm]rightor.east) {$|h_y\rangle\langle h_y|$};
\node[reg] (rightreg) at ([yshift=-1.1cm]rightor.center) {};
\draw[local] ([xshift=-.098cm]rightreg.north) -- (rightswap.south);
\draw[local] ([xshift=.098cm]rightreg.north) -- (rightproj.south);
\node[font=\footnotesize,anchor=north] at ([yshift=-5mm]rightswap.south) {$1/2$};
\node[font=\footnotesize,anchor=north] at ([yshift=-5mm]rightproj.south) {$1/2$};
\end{scope}
\draw[rounded corners=3pt,draw=gray!40,fill=none]
  ([xshift=-3mm,yshift=3mm]rightcontents.north west)
  rectangle ([xshift=3mm,yshift=-3mm]rightcontents.south east);
\node[anchor=south] at ([yshift=5mm]rightcontents.north) {$v_r$};
\draw[send] (out0.east) to[out=0,in=180] (test1.west);
\coordinate (lastincoming) at ([xshift=-10mm]test2.west);
\node[anchor=east,inner sep=0pt,text=blue!65!black] (omitted)
  at ([xshift=-2mm]lastincoming) {$\cdots$};
\draw[send] (out1.east) to[out=0,in=180]
  ([xshift=-2mm]omitted.west);
\draw[send] (lastincoming) -- (test2.west);
\draw[send] (out2.east) to[out=0,in=180] (rightswap.west);
\end{tikzpicture}
    \caption{Graphical example of the execution of the protocol. Each square represents a register. $v_0$ checks if its register contains $\ket{h_x}$. Intermediate vertices perform the SWAP test. The test of $v_r$ is uniform randomly chosen.}
    \label{fig:placeholder}
\end{figure}
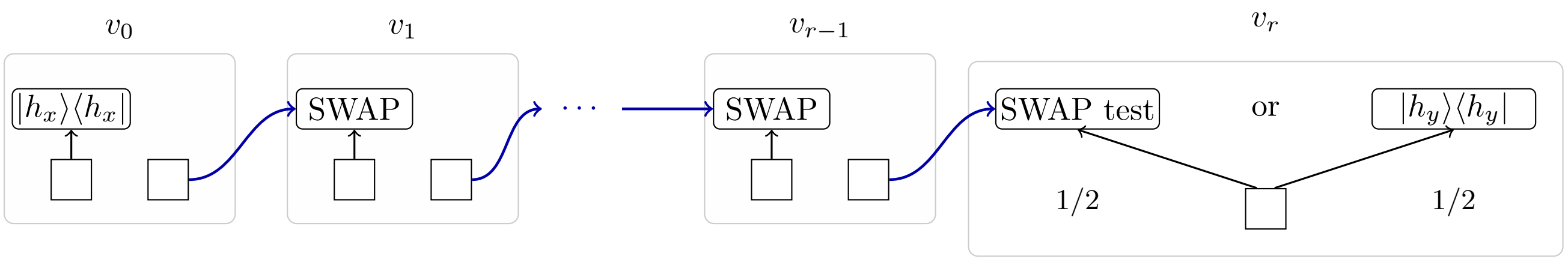

The main idea of that protocol is the following: An $N$-bit string $x$ can be represented by an $O(\log N)$-qubit state $|h_x\rangle$ so that distinct strings $|h_x\rangle$ and $|h_y\rangle$ have overlap bounded away from one \cite{buhrman2001quantum}. Such states, called \emph{quantum fingerprints}, allow strings to be compared without recovering them.
In the existing protocols~\cite{fraigniaud2021distributed,hasegawa2024power}, an honest prover supplies two copies of the same fingerprint along the path. Each vertex selects one of two registers uniformly at random, sends it to the right, then performs SWAP tests, while the endpoints check against their inputs. See \Cref{fig:placeholder}.

For unequal inputs on an $r$-edge path, the previous analysis of the protocol guarantees that at least one vertex rejects with probability $\Omega(1/r^2)$ per repetition~\cite{hasegawa2024power}. Hence, $O(r^2)$ repetitions are required for constant soundness. We improve this to \(\Omega(1/r)\), reducing the certificate size from \(O(r^2\log N)\) to \(O(r\log N)\).

Let us focus on cheating strategies of the malicious prover. The natural strategy for the prover is to change fingerprints in equal angles in small increments between two endpoints of the path. Thus, the angle difference between each consecutive certificates is only $1/r$, which makes the rejection probability $1/r^2$. The previous analysis used union bound to lower bound the probability that some node rejects, which is $\Omega(1/r^2)$.
However, since the proofs are not entangled in this case, the rejection events are effectively independent at each node, and the probability that at least one test will fail is $1/r$.  
This suggests that, the prover should {\it correlates} the rejection events to decrease the rejection probability. We show that even arbitrary entanglement cannot reduce rejection below \(\Omega(1/r)\).

Let \(|\Psi\rangle\) be a purification of the joint state after communication and before measurement, averaged over the random choices of which registers to send. The joint state of the two registers compared by the SWAP test at vertex \(v_j\) is
\[
\sigma_j=\operatorname{Tr}_{\neg j}|\Psi\rangle\langle\Psi|.
\]
Order this pair as the register received from \(v_{j-1}\), followed by the register retained at \(v_j\). Their individual states are \(\operatorname{Tr}_2\sigma_j\) and \(\operatorname{Tr}_1\sigma_j\), respectively. For unequal inputs $x\neq y$, sufficiently small overall rejection probability forces those endpoint registers to be close to the distinct fingerprints \(|h_x\rangle\) and \(|h_y\rangle\). Consequently, the sum of the trace distance ($\D(\cdot)$) of two registers on which SWAP tests are performed is
\begin{align}\label{eq:overview}
  \Omega(1)\le
\sum_{j=1}^{r}
\D\!\left(\operatorname{Tr}_2\sigma_j,\operatorname{Tr}_1\sigma_j\right).  
\end{align}

Condition on \(v_r\) choosing its SWAP test. The performed tests are then the fingerprint check at \(v_0\) and the SWAP tests at \(v_1,\ldots,v_r\). To keep track of which vertices reject together, label their joint outcomes by \(z\in\{0,1\}^{r+1}\), where \(z_i=0\) means acceptance at \(v_i\) and \(z_i=1\) means rejection. This gives an orthogonal decomposition of \(|\Psi\rangle = \sum_z \ket{\psi_z}\). Writing \(a_z=\|\psi_z\|\), we have
\[
q_\square
:=\Pr(\text{at least one vertex rejects})
=\sum_{z\ne0^{r+1}}a_z^2.
\]
The same decomposition gives
\begin{align}\label{eq:overview2}
    \sigma_j=
\sum_{z,z'}
\operatorname{Tr}_{\neg j}
\bigl(|\psi_z\rangle\langle\psi_{z'}|\bigr).
\end{align}
For each summand $\operatorname{Tr}_{\neg j}
\bigl(|\psi_z\rangle\langle\psi_{z'}|\bigr)$, if $z_i \neq z'_i$ for some $i \neq j$, the outcome of the test at $v_i$ on $|\psi_z\rangle$ and $|\psi_{z'}\rangle$ is distinct, hence their reduced states are orthogonal and $\operatorname{Tr}_{\neg j}
\bigl(|\psi_z\rangle\langle\psi_{z'}|\bigr)=0$. For \(z=z'\), applying \(\operatorname{Tr}_1\) or \(\operatorname{Tr}_2\) to the corresponding summand in~\eqref{eq:overview2} gives the same operator.
Thus, among the summands in~\eqref{eq:overview2}, only those indexed by strings \(z,z'\) that differ at \(j\) and agree everywhere else can make \(\operatorname{Tr}_2\sigma_j\) and \(\operatorname{Tr}_1\sigma_j\) different. These pairs are edges of the hypercube on \(\{0,1\}^{r+1}\).
Ultimately, together with~\eqref{eq:overview}, we obtain
\[
\begin{aligned}
\Omega(1)
\le
\sum_{j=1}^{r}
\D\!\left(\operatorname{Tr}_2\sigma_j,\operatorname{Tr}_1\sigma_j\right)
&\le
2\sum_{\{z,z'\}\text{ an edge}}a_za_{z'}
\\
&\le
2\sqrt{(r+1)q_\square}+(r+1)q_\square.
\end{aligned}
\]
Hence \(q_\square=\Omega(1/r)\), and so is the rejection probability.

\paragraph{Removing the dependence on distance.}
Now we explain the second $\tilde{O}({N^{2/3}})$ upper bound proved in~\Cref{thm:equality-uniform}.
Our starting point is to replace the large classical input strings with compact quantum states, which we call label states. In the protocol, the nodes only need to verify whether the represented classical values agree, rather than recover the represented strings from the states, so non-orthogonal states can be used. This idea is the same of the previous quantum protocol, with a standard $O(\log N)$-qubit fingerprints as label states. The label states associated with the input strings are compared in parallel along the path by the SWAP test. However, the SWAP test accepts two copies of every pure state, allowing the prover to change states gradually along a long path. Neighboring states can then be almost identical even when the endpoint states differ, explaining why the certificate size depends on \(r\).

We therefore incorporate the family of label states into the local test. The label states now have equal-magnitude amplitudes in a fixed basis, with signs determined by the represented bits:
\[
\ket{\phi_u}=\frac1{\sqrt L}\sum_{a=1}^L(-1)^{u_a}\ket a,
\qquad u\in\{0,1\}^L.
\]
Here $L=\tilde{\Theta}(N^{1/3})$ is a parameter which decides the length of encoded classical strings.
The new local test accepts exactly the linear combinations of identical pairs of label states, namely,
\[
W_L=\operatorname{span}
\{\ket{\phi_u}^{\otimes2}:u\in\{0,1\}^{L}\}.
\]
Each vertex receives a pair of registers and tests whether that pair lies in \(W_L\) before communication.
Every vertex except the right endpoint then sends its second register to its right neighbor, which performs another $W_L$-test to a new pair. Thus, the two stages test every adjacent pair in the sequence obtained by concatenating the vertices’ pairs along the path. After communication, the endpoints also check the first and last registers of this sequence against the label states specified by their inputs.

For soundness, the central task is to lower-bound the rejection probability when two endpoints hold different label states, independently of the path length. We analyze the protocol from the Hamiltonian viewpoint, which is a central methodological contribution of our protocol and analysis. Since the protocol checks every adjacent pair in the full register sequence, it can be viewed as the local Hamiltonian on a chain of qudits. More precisely, we can define a local Hamiltonian 
\[
H=\sum_i H_i
\]
where $H_i$ is the projector onto the rejection space of a local test applied to the $i$-th pair of registers. Here, input-dependent checks performed by two endpoinds are not included in $H$. For this Hamiltonian, we prove two properties.
First, the ground space of $H$ is spanned by tensor powers of label states. Second, we use a bound on the spectral gap (the minimum energy of a normalized state orthogonal to the ground space) to obtain an energy lower bound for unequal endpoint labels that is independent of the path length.

Each tensor power of a single label state is unchanged when the endpoint registers are exchanged. A state close to the ground space must therefore also change only slightly under this exchange. However, exchanging two different endpoint label states changes the state by a constant amount, because the label states assigned to distinct encoded values are designed to have overlaps bounded away from one. Hence a state satisfying both endpoint requirements for no-instance cannot be close to the ground space: it must have a constant weight outside it. The spectral-gap bound turns this constant weight into an energy lower bound independent of the path length. Consequently, the mismatch between the endpoints cannot be made arbitrarily cheap by spreading it over a longer chain.

The protocol applies the \(W_L\)-test to two kinds of pairs. Before communication, each vertex tests the two registers supplied to it by the prover. After communication, each vertex other than \(v_0\) tests the register received from its left neighbor together with its own retained first register. The endpoint checks are also performed after communication.

Let \(p>0\) be the probability that all tests accept, and let \(\rho\) be the normalized final state conditioned on this event. All tests after communication act on disjoint registers, so once one of these tests has accepted, it would still accept if repeated after the others. Consequently, \(\rho\) satisfies both endpoint checks, and for every $H_i$ corresponding to a test performed after communication, 
$
\operatorname{Tr}(H_i\rho)=0.
$
However, for a test performed before communication, \(\operatorname{Tr}(H_i\rho)\) can be positive even though that test accepted when it was performed. 

The energy of \(\rho\) is \(\operatorname{Tr}(H\rho)\). We will prove the following inequality where the endpoint conditions give the lower bound, while the detectability lemma~\cite{AnshuAradVidick2016} gives the upper bound:
\[
\widetilde{\Omega}(L^{-2})
\le
\operatorname{Tr}(H\rho)
=
\sum_i\operatorname{Tr}(H_i\rho)
\le
\frac{4(1-p)}p.
\]
Hence \(1-p=\widetilde{\Omega}(L^{-2})\), independently of the path length.

Finally, we encode each \(N\)-bit input using a Reed–Solomon code into \(K\) strings of \(\Theta(L)\) bits each, so that unequal inputs produce at least \(K/2\) of $K$ strings differ. To do so, we choose $N=\Theta(KL)$, $K=\tilde{\Theta}(L^2)$.
The prover uses $K$ label states from each of $L$-bit strings, and we perform the above protocol, $K$ times in parallel.
Since at least $K/2$ repetitions the two label states at endpoints differ, the corresponding test is rejected with probability $\tilde{\Omega}(L^{-2})$, and $K$ repetitions make the overall rejection probability $\Omega(1)$. Finally, the combined certificate is
\[
O(K\log L)=\widetilde O(N^{2/3}),
\]
independently of \(r\).

\paragraph{Imroved lower bound.}
We next explain the lower bound side. The known classical and quantum lower bounds for $\eqtwo$ on a path are based on the fooling set method, as used in previous work~\cite{fraigniaud2021distributed,hasegawa2024power}. In $\dQMA$ this is achieved by first combining one half of the honest proof for the input $(x,x)$ and one half of the honest proof for the input $(y,y)$, and then showing the combined new proof passes all tests with high probability on the mismatched input $(x,y)$, which is a contradiction. However, this argument crucially relies on the assumption that proofs have ``halves'', which means the argument only applies to separable proofs.

We instead use the causality of a single round of communication. Every node at distance at least 2 from $v_r$ cannot see the input $y$, so if the prover supplies the honest proof for the instance $(x,x)$ on the mismatched instance $(x,y)$, the nodes $v_0,\ldots,v_{r-2}$ accept with high probability. Soundness therefore forces $v_{r-1}$ or $v_r$ to reject with high probability. From outside the pair, these two nodes receive only their proof registers and the message from $v_{r-2}$, and everything else they use is produced by their own operations. Stopping the protocol just before they act, the reduced state $\sigma_x$ on those registers is independent of $y$, while their subsequent operations depend only on $y$, which is represented by a single POVM element $E_y$. Since $E_y$ must accept $\sigma_y$ due to completeness and must reject $\sigma_x$ due to soundness, the two states must be far apart in trace distance.

What remains is a counting problem; how small a dimension can hold $2^N$ quantum states that are pairwise far apart in trace distance? This is already answered in~\cite{hasegawa2024power}: it can be shown that the dimension must be at least polynomial in $N$, that is, the corresponding proof contains at least $\log N$ qubits.

\subsubsection{Exponential advantage}

Our first separation is given by a graph property that we call $\treeq_\mu$, and whose
members are \emph{trees}. To define $\treeq_\mu$, our idea is to start with an
instance of $\eqtwo$ on a path, with two input strings $z,z'\in\{0,1\}^N$, one input string at each
endpoint, and to replace each input string
by a tree attached to the corresponding endpoint, so that the
strings are encoded in the structure of the resulting graph that we denote
$T_{N,\mu}^{\mathrm{tr}}(z,z')$. We describe our encoding at a high level.

\paragraph*{Encoding.}
We focus on the equality problem $\eqtwo$, where $N$ is even
and each input string has Hamming weight exactly
$N/2$.\footnote{The Hamming weight of a binary string is
the number of positions containing $1$.}
Our first observation is that such a string is uniquely
determined by the numbers of zeros before the first one,
between consecutive ones, and after the last one.
More precisely, for every $z\in\{0,1\}^N$ of Hamming weight
$N/2$, there are unique values
$r_0,\dots,r_{N/2}\in\{0,\dots,N/2\}$ such that
\begin{equation}\label{eq:decom-z}
z=0^{r_0}\,1\,0^{r_1}\,1\cdots1\,0^{r_{N/2}}.
\end{equation}
These values satisfy $\sum_{i=0}^{N/2}r_i=N/2$,
since $z$ also contains exactly $N/2$ zeros.

We first describe the rooted tree that encodes a single
string $z$ satisfying the above conditions with decomposition given by $(r_0,\dots,r_{N/2})$.
We start with a root $h_z$ and attach the vertices
$a_0,\dots,a_{N/2}$ to it.
For each $i$, we attach $r_i$ leaves to $a_i$,
where $(r_0,\dots,r_{N/2})$ is the vector of $z$ satisfying \eqref{eq:decom-z}.

The names $a_0,\dots,a_{N/2}$ are only notation and are
not part of the input graph. An important feature of our encoding is the following: 
A tree encoding a string $z$ cannot be isomorphic to another tree encoding a string $z'\neq z$, i.e., our encoding must retain the order of the vectors $(r_0,\dots,r_{N/2})$:
two rooted encodings should be isomorphic if and only
if they represent the same string.

To obtain this property, we attach to each node $a_i$ a rooted
tree $\mathrm{Tag}(i)$, defined in \Cref{def:tag} (see \Cref{fig:treecomb}(a) for a graphical example).
The tag has a root $s_i$ with child $t_i$, from which $L=\Theta(\log N)$ paths of distinct lengths $1,\dots,L$ grow; the last vertex of path $b$ carries two extra leaves exactly when the $b$-th least significant bit of $i$ is $1$.
We show (\Cref{lem:tag}) that this rooted structure identifies the index $i$.
The corresponding value $r_i$ is then recovered by counting
the leaves attached to $a_i$.
Together, these observations allow us to recover the
encoded strings (\Cref{lem:isoT}).

Finally, to construct the graphs $T_{N,\mu}^{\mathrm{tr}}(z,z')$, we take
the rooted trees encoding $z$ and $z'$, call their roots
$h_L$ and $h_R$, and join these roots by a central path
of $\rho=2\mu+1$ edges, using new internal vertices. 
We denote the two rooted encodings by $T_L(z)$ and
$T_R(z')$, respectively. See \Cref{fig:treecomb}(b) for a graphical example.

This encoding has the following properties for each $z,z'\in\{0,1\}^N$:
\begin{itemize}
    \item  The encoded graph $T_{N,\mu}^{\mathrm{tr}}(z,z')$ is a tree (by construction).
    \item The shape of the tree $T_{N,\mu}^{\mathrm{tr}}(z,z')$ can be verified with a deterministic $\dMA$ protocol (Lemma~\ref{lem:shapeT}).
    \item The number of nodes of any tree $T_{N,\mu}^{\mathrm{tr}}(z,z')$ is $\Theta(N\log^2 N)$ (\Cref{fact:treecount}).
    \item In each tree, the pair of encoded strings $z$ and $z'$ can be recovered uniquely from the isomorphism class (\Cref{lem:isoT}).
\end{itemize}

 The fourth property means
that the tree structure alone determines the two encoded strings,
up to exchanging them, regardless of the names assigned to the
vertices. We therefore define $\treeq_\mu$ to consist of all graphs
isomorphic to $T_{N,\mu}^{\mathrm{tr}}(z,z)$ for some input string $z$. With this
definition,
\[
T_{N,\mu}^{\mathrm{tr}}(z,z')\in\treeq_\mu
\quad\Longleftrightarrow\quad
z=z'.
\]
Thus, checking membership in $\treeq_\mu$ on an encoded instance
is equivalent to checking equality of its two strings.
This equivalence is the basis of our separation. 

For the quantum upper bound, we combine the classical structural verification with the $O(r\log N)$-qubit equality protocol of \Cref{thm:equality-uniform} in a single communication round, during which the nodes exchange the classical information and the quantum registers in parallel. Since the quantum operations performed before communication do not require knowledge of $z$ and $z'$, the roots $h_L$ and $h_R$ can reconstruct these strings from the received classical messages and complete the corresponding local endpoint tests after reception, without any additional communication round.

Indeed, given an input graph $T$, the prover supplies $O(\log n)$-bit
classical certificates to allow the nodes to verify that there exist two strings $z,z'\in\{0,1\}^N$ such that $T\cong T_{N,\mu}^{\mathrm{tr}}(z,z')$ (Lemma~\ref{lem:shapeT}).
If this structural verification passes, then
$T\cong T_{N,\mu}^{\mathrm{tr}}(z,z')$ for the strings determined by the reported
run lengths. Moreover, the root of each tree can reconstruct the string being encoded in the tree. This classical verification requires only $\bigo{\log n}$ bits per vertex.\footnote{Here we use \Cref{fact:treecount}: The number of nodes $n$ of a tree $T_{N,\mu}^{\mathrm{tr}}(z,z')$ is $\Theta(N\log^2 N)$.}

The roots $h_L$ and $h_R$ can reconstruct the strings
$z$ and $z'$, respectively.
Along with the classical certificate of \Cref{lem:shapeT},
the prover supplies to the vertices in the central path  the
quantum certificates of the $O(\rho\log N)$-qubit protocol
from \Cref{thm:equality-uniform} to verify that $z=z'$.
Only the endpoint tests depend on these strings, so the
roots can reconstruct them during the communication round,
before performing these tests.

The central path has length $\rho=\Theta(\mu)$,
independent of the number of vertices in the tree.
Thus, each vertex in the central path receives
$O(\mu\log N)$ certificate qubits.
Together with the $O(\log n)$-bit classical certificates
and the relation $\log N=\Theta(\log n)$, this gives
a one-round $\dQMA$ protocol using
$O(\mu\log n)$ certificate qubits per vertex.

On a yes-instance, honest classical certificates and quantum
registers make every check accept with certainty. For
soundness, condition on the outcomes of the initial classical
measurements. If the structural or routing checks fail,
some node rejects. Otherwise, the input is isomorphic to
$T_{N,\mu}^{\mathrm{tr}}(z,z')$ for the reconstructed strings. If
$T\notin\treeq_\mu$, these strings must be distinct, and the
soundness guarantee of the equality protocol bounds the
probability that all nodes accept by $1/3$, even for
arbitrarily entangled quantum certificates.

For the classical lower bound, we adapt the fooling-set argument of Fraigniaud et al.~\cite{fraigniaud2021distributed} to our
family of encoded trees. The encoding ensures that combining
the left side of $T_{N,\mu}^{\mathrm{tr}}(z,z)$ with the right side of $T_{N,\mu}^{\mathrm{tr}}(z',z')$
gives a no-instance whenever $z\ne z'$. This allows us to follow an analogous
counting and splicing argument to apply to certificates on
the central path.

\subsubsection{Polynomial advantage for graph symmetry}

Our main technique to obtain a quantum advantage for the graph symmetry problem is to reduce the verification of the existence of a non-trivial automorphism to a series of equality problems on a path, which can then be verified using our improved quantum equality protocols from \Cref{thm:equality-uniform}.
\paragraph{Reducing symmetry to distributed equality.}
Recall that a bijection
$\alpha:V(G)\to V(G)$ is an automorphism exactly when
$\alpha(N(v))=N(\alpha(v))$ for every vertex $v$.
Thus, once a proposed bijection is fixed, preserving
adjacency can be checked by comparing two descriptions:
the images of the neighbors of $v$, and the actual
neighbors of its proposed image.
These descriptions are available at different vertices,
which may be far apart.

To perform these comparisons, the prover supplies each
vertex $v$ with two identifier values $\pi(v)$ and
$\sigma(v)$, intended to describe its image and inverse
image.
It also supplies a path starting at each vertex $v$.
The classical certificates allow the nodes to verify
these paths locally.
Write $\tau(v)$ for the endpoint of the path starting
at $v$.
At this stage, the paths are well defined, but we still
need to verify that their endpoints agree with the
proposed images and together define an automorphism.

We incorporate all these requirements into one equality
test per path.
At its starting vertex $v$ and endpoint $w=\tau(v)$,
respectively, we construct the strings
\[
\begin{aligned}
X_v&:=\operatorname{enc}\bigl(
\mathrm{id}(v),\pi(v),\deg(v),
(\pi(u))_{u\in N(v)}
\bigr),\\
Y_w&:=\operatorname{enc}\bigl(
\sigma(w),\mathrm{id}(w),\deg(w),
(\mathrm{id}(u))_{u\in N(w)}
\bigr).
\end{aligned}
\]
Here $\operatorname{enc}$ is an injective encoding
with sorted neighbor lists and padding to a common length.
Each endpoint can construct its own string after receiving
its neighbors' identifiers and classical certificate fields.

Suppose $X_v=Y_{\tau(v)}$ for every $v$.
The second fields give
$\pi(v)=\mathrm{id}(\tau(v))$, so each path ends at
the claimed image of its source.
The first fields give
$\sigma(\tau(v))=\mathrm{id}(v)$.
In particular, two distinct sources cannot have the same
endpoint, since that endpoint has only one value of
$\sigma$.
Thus $\tau$ is injective and, on the finite vertex set,
bijective.
Finally, equality of the degrees and neighbor lists gives
$\tau(N(v))=N(\tau(v))$ for every $v$.
Hence $\tau$ is an automorphism.
To ensure that it is nontrivial, we additionally certify
a spanning tree whose root $v^*$ satisfies
$\pi(v^*)\ne\mathrm{id}(v^*)$.

Conversely, any nontrivial automorphism $\alpha$ provides
certificates satisfying these conditions: use its images
and inverse images for $\pi$ and $\sigma$, choose paths
from $v$ to $\alpha(v)$, and root the spanning tree at
a vertex moved by $\alpha$.
All the corresponding string equalities then hold.
Consequently, certifying symmetry reduces to verifying
the classical data and one equality instance along each
certified path.

The quantum part performs precisely these equality tests.
The prover supplies the quantum registers along the paths,
and all quantum operations before reception are independent
of the endpoint strings.
The nodes can therefore exchange their classical headers
and quantum registers in the same round, construct the
strings after reception, and then complete the tests that
depend on them.
For $\symk$, we additionally certify the radius bound.
This ensures that the paths can be chosen with length
at most $2D$, allowing us to use the fingerprint equality
protocol with a smaller cost.

The common string length is $N_*=O(n\log n)$, and at most $n$ comparison paths pass through any vertex. Thus the distance-independent cost is multiplied by at most $n$, giving $n\cdot\widetilde O(N_*^{2/3})=\widetilde O(n^{5/3})$. Under a radius bound, the fingerprint cost is $O(D\log N_*)$ per path, hence $O(nD\log n)$ overall. This path congestion explains both the need for compression at every vertex in the unrestricted case and the benefit of short paths in the bounded-radius case.

For the classical lower bounds in \Cref{thm:BC}, we use the graphs $B_\rho(H,H')$ of \Cref{def:sym-family}, obtained by joining two $k$-vertex graphs $H,H'\in\mathcal F_k$ by a path of $\rho\in\{2\mu+2,2\mu+3\}$ edges, where $\mathcal F_k$ consists of pairwise non-isomorphic asymmetric graphs of diameter at most $2$ and minimum degree at least $3$. An Erd\H{o}s--R\'enyi counting argument gives $|\mathcal F_k|\ge 2^{(1-o(1))k^2/2}$ (\Cref{fact:count}), while asymmetry and the degree distinction between the attached graphs and the path ensure that $B_\rho(H,H')$ has a nontrivial automorphism exactly when $H=H'$ (\Cref{lem:pin}). Moreover, its radius is at most $\lceil\rho/2\rceil+2\le\mu+4$, so the same family applies to $\symk$ whenever $\mu\le D-4$ (\Cref{fact:ecc}). Then we apply again a fooling-set argument to show that in any $\mu$-round protocol the honest certificate tuples on $2\mu$ central-path vertices must distinguish all instances $B_\rho(H,H)$, since otherwise splicing two accepting assignments would violate soundness on some $B_\rho(H,H')$ with $H\ne H'$ (\Cref{claim:sym-certificates}); counting these tuples forces $\Omega(k^2/\mu)=\Omega(n^2/\mu)$ certificate bits at some node, using $n=2k+\rho-1$ and $\mu=o(n)$.

\subsection{Organization of the paper}

In \Cref{sec:pre} we define the model and notations used throughout the paper. In \Cref{sec:eq} we describe the improved upper and lower bounds for the equality problem $\eqtwo$. Then in \Cref{sec:tree,sec:sym} we prove \Cref{thm:treeq-separation} and \Cref{thm:quan-upper-sym}, respectively.

\section{Preliminaries}\label{sec:pre}

\paragraph{Graphs.}
For a positive integer $k$, write $[k]:=\{1,\dots,k\}$.
All graphs are finite, simple, undirected, and connected.
For a graph $G=(V,E)$, we write $n=|V|$,
$N_G(v)=\{u\in V:\{u,v\}\in E\}$, and
$\deg_G(v)=|N_G(v)|$.
The distance $d_G(u,v)$ is the number of edges in a shortest
path between $u$ and $v$. The eccentricity of a vertex $v$,
the radius of $G$, and its diameter are, respectively,
\[
    \operatorname{ecc}_G(v)=\max_{u\in V}d_G(u,v),
    \qquad
    \operatorname{rad}(G)=\min_{v\in V}\operatorname{ecc}_G(v),
    \qquad
    \operatorname{diam}(G)=\max_{v\in V}\operatorname{ecc}_G(v).
\]
In particular, $\operatorname{rad}(G)\le k$ means that some
vertex is within distance $k$ of every vertex, and implies
$\operatorname{diam}(G)\le 2k$.
A path $v_0,\ldots,v_r$ has $r$ edges and $r+1$ vertices.

A tree is a connected acyclic graph.
A spanning tree of $G$ is a tree with vertex set $V$ and
edge set contained in $E$.
In a rooted tree, the parent of a nonroot vertex is its
neighbor on the unique path to the root.
The depth of a vertex is its distance from the root, and
the depth of the rooted tree is the maximum vertex depth.

For rooted trees $T_1,T_2$ with roots $r_1,r_2$, a rooted isomorphism is a graph isomorphism $\varphi:V(T_1)\to V(T_2)$ satisfying $\varphi(r_1)=r_2$; we write $T_1\cong_{\mathrm{rt}}T_2$ when such an isomorphism exists.

An automorphism of $G$ is an isomorphism from $G$ to itself.
It is nontrivial if it moves at least one vertex.
We call a graph symmetric if it admits a nontrivial
automorphism, and asymmetric otherwise.
A graph property is a class of graphs closed under
isomorphism. We identify a property with the class of
graphs satisfying it.

\paragraph{Quantum computing.}
We follow mostly standard notation from quantum computing; see~\cite{nielsen2010quantum} for a thorough review. All Hilbert spaces are finite-dimensional. A quantum state is represented by a positive semidefinite operator of trace one. For vectors, $\|\ket\psi\|:=\sqrt{\langle\psi|\psi\rangle}$. For an operator $A$, its operator norm and trace norm are
\[
\|A\|:=\sup_{\|\ket\psi\|=1}\|A\ket\psi\|,
\qquad
\|A\|_1:=\tr\sqrt{A^\dagger A}.
\]
For Hermitian operators, $A\succeq B$ (equivalently, $B\preceq A$) means that $A-B$ is positive semidefinite.

For density operators on the same space, the trace distance is
\[
\D(\rho,\sigma):=\frac12\|\rho-\sigma\|_1
=\max_{0\preceq E\preceq I}\tr(E(\rho-\sigma)).
\]
A quantum channel is a completely positive trace-preserving map $\Lambda$. Trace distance is contractive under channels: $\D(\Lambda(\rho),\Lambda(\sigma))\le\D(\rho,\sigma)$. In particular, partial trace discards registers and is a quantum channel; the trace norm also satisfies $\|\tr_B X\|_1\le\|X\|_1$. The dual map $\Lambda^*$ is defined by $\tr(E\Lambda(X))=\tr(\Lambda^*(E)X)$.

We use root-fidelity
\[
\F(\rho,\sigma):=\|\sqrt\rho\sqrt\sigma\|_1
=\tr\sqrt{\sqrt\rho\,\sigma\sqrt\rho}.
\]
Uhlmann's theorem identifies it with the maximum absolute inner product of purifications of $\rho$ and $\sigma$ on a common sufficiently large auxiliary space. In particular, any such purifications have overlap at most $\F(\rho,\sigma)$. The Fuchs--van de Graaf inequalities state
\[
1-\F(\rho,\sigma)\le\D(\rho,\sigma)
\le\sqrt{1-\F(\rho,\sigma)^2}.
\]
For a pure state $\ket\psi$, $\F(\rho,\ket\psi\bra\psi)^2=\langle\psi|\rho|\psi\rangle$.

A two-outcome POVM is specified by effects $\{E,I-E\}$ with $0\preceq E\preceq I$, and its first outcome has probability $\tr(E\rho)$. For an orthogonal projector $P$, the non-destructive projective measurement $\{P,I-P\}$ leaves the state in $P\rho P/\tr(P\rho)$ conditioned on its first outcome. For a pure state this is $P\ket\psi/\|P\ket\psi\|$; states already in the image of $P$ are unchanged.

For an $N$-bit string $X$, a quantum fingerprint $\ket{h_X}$ has $O(\log N)$ qubits and satisfies $|\langle h_X|h_Y\rangle|^2\le1/2$ whenever $X\ne Y$. Its precise construction is not needed here (see~\cite{buhrman2001quantum}). The SWAP operator is $S(\ket a\otimes\ket b)=\ket b\otimes\ket a$, and $\Pi_\pm:=(I\pm S)/2$ project onto the symmetric and antisymmetric subspaces. The SWAP test accepts on $\Pi_+$ and rejects on $\Pi_-$, with rejection probability $\tr(\Pi_-\rho)$ on a joint state $\rho$. On $\ket\psi\otimes\ket\phi$, its acceptance probability is $(1+|\langle\psi|\phi\rangle|^2)/2$. Thus equal fingerprints are accepted with certainty, and distinct ones with probability at most $3/4$.

\subsection{Local certification models}\label{sec:definition}

An instance is a pair $I=(G,\mathbf{x})$, where $G=(V,E)$
is a connected graph with $n=|V|$ nodes and
$\mathbf{x}=(x_v)_{v\in V}$ is an assignment of classical
local inputs.
A distributed language $\mathcal L$ is a set of instances.
For a graph property, there are no additional inputs,
and membership depends only on the isomorphism class of $G$.

Each node $v$ knows its local input and its own distinct identifier $\mathrm{id}(v)\in[U]$, where $U=n^{O(1)}$ is a common identifier bound. Its incident edges have distinct local port numbers $1,\dots,\deg(v)$; the port numbers at opposite ends of an edge need not agree. A node can address an outgoing message by its local port and identify the local port on which each incoming classical or quantum message arrives. Our upper bounds use known $n$, $U$, and the common problem parameters. Nodes are not initially given the global topology or inputs held elsewhere. The lower bounds hold even with these parameters known.

\begin{definition}[$\dMA$ and $\dQMA$ protocols~\cite{fraigniaud2021distributed}]
\label{def:dma-dqma}
A $\mu$-round distributed Merlin--Arthur protocol for
$\mathcal L$ is described as follows.

An untrusted prover, Merlin, knows the entire instance
and the identifier and port assignments and supplies certificates
before the verifiers communicate.
The certificates and verification operations depend on
the model:
\begin{itemize}
    \item In a classical protocol ($\dMA$), Merlin supplies
    a classical certificate $z_v$ to each node $v$.
    The nodes perform randomized classical computation
    and exchange classical messages.

    \item In a quantum protocol ($\dQMA$), Merlin supplies
    an arbitrary joint quantum state $\rho$ on
    \[
        \mathcal H=\bigotimes_{v\in V}\mathcal H_v,
    \]
    with node $v$ receiving the register $\mathcal H_v$.
    The state may be entangled across nodes.
    The nodes may perform arbitrary local quantum
    operations, including measurements, and exchange
    quantum messages. Classical fields are allowed as qubits in the computational basis, measured before other local operations or routing.
\end{itemize}
For fixed size parameters, certificate registers have prescribed finite size. When measured classical records select roles or active register lengths, our upper-bound protocols reserve a fixed maximum register budget and use only the prescribed active slots, ignoring padding. Malformed records or incompatible register formats cause rejection. These budgets apply also on no-instances.

After receiving their certificates, the nodes execute
$\mu$ synchronous communication rounds along the edges
of $G$, with no further interaction with Merlin.
In each round, every node prepares and sends its outgoing
messages before receiving the messages sent to it in that
round, and may then perform local computation. Thus a single round has three phases: local preparation (including measurement of classical certificate fields), simultaneous message exchange, and local computation and measurements after reception. Operations before reception may depend only on information already available locally.
Local computation, local workspace, and message sizes
are unrestricted. Therefore, verification can be viewed as a $\mu$-round (classical or quantum) LOCAL algorithm with private randomness and no prior shared entanglement beyond the supplied quantum certificates.
Our classical lower bounds also hold when the verifiers additionally share public randomness, sampled after the certificates have been supplied.

At the end of verification, each node outputs accept or
reject, and the network accepts if and only if every node
accepts. Let $\mathsf{Acc}$ denote the event that the network accepts.
The protocol has completeness $c$ and soundness error $s$,
where $0\le s<c\le1$, if, for every instance $I$,
\[
\begin{aligned}
    I\in\mathcal L
    &\quad\Longrightarrow\quad
    \exists\,\Gamma:
    \Pr[\mathsf{Acc}\mid I,\Gamma]\ge c,\\
    I\notin\mathcal L
    &\quad\Longrightarrow\quad
    \forall\,\Gamma:
    \Pr[\mathsf{Acc}\mid I,\Gamma]\le s.
\end{aligned}
\]
Here, $\Gamma$ ranges over all certificates of the
prescribed size: classical certificate assignments
in $\dMA$ and arbitrary joint quantum states in $\dQMA$.
Probabilities are taken over the verifiers' randomness
and, in the quantum model, their measurement outcomes.

The certificate size is the maximum number of certificate
bits in $\dMA$, or qubits in $\dQMA$, per node, in the worst case over instances with the relevant size parameters fixed (these parameters include $n$ and, for problems involving
input strings, their length $N$). The message size is the maximum number of bits or qubits in one directed edge message in any round. Classical fields count toward the corresponding message and certificate sizes. All upper bounds below bound both resources, although message size is unrestricted in the model and in our lower bounds.
\end{definition}

\paragraph{Distributed equality.}
In the distributed equality problem $\eqn$, a set of
$t\ge2$ designated nodes, called terminals, hold strings
$x_1,\ldots,x_t\in\{0,1\}^N$.
The task is to verify whether
\(
    x_1=x_2=\cdots=x_t.
\)
Nonterminal nodes hold no input.
In the path setting, the network is a path $P=v_0,\ldots,v_r$,
the two terminals $v_0,v_r$ are its endpoints, and their inputs
are denoted by $x$ and $y$.
We use $N$ for the length of each input string and $n$
for the number of network nodes.

\paragraph{Asymptotic notation.}
The notations $\widetilde O$ and $\widetilde \Omega$ hide polylogarithmic factors in the relevant size parameter: in $n$ for bounds in the number of nodes, and in $N$ for bounds in the input-string length. Unless explicitly written as $\ln$, logarithms are to base $2$.

\section{Distributed equality problem}\label{sec:eq}

In this Section we prove \Cref{thm:equality-uniform} by giving two independent protocols achieving the corresponding bounds.

\subsection{Improved analysis of the equality protocol}
In this section we prove the $O(r\log N)$ upper bound of~\Cref{thm:equality-uniform}. This result is proved by a refined analysis of the previous protocols used in~\cite{fraigniaud2021distributed,hasegawa2024power}. Let $m\in \Theta(r)$ be the repetition parameter to be fixed afterwards.
We first describe our protocol, which is almost identical to the one from~\cite{hasegawa2024power}.

\begin{tcolorbox}[code={\refstepcounter{protocolcount}}, title={Protocol \theprotocolcount}, enhanced, breakable=true, label={protocol-equality}]
\begin{description}
\item[Certificate.] 
For each repetition $k\in[m]$, each node $v_j$ with $0\le j\le r-1$ receives an ordered pair of blocks $(\mathsf{A}^{(k)}_{j},\mathsf{B}^{(k)}_{j})$, and $v_r$ receives a single block $\mathsf{A}^{(k)}_{r}$. 
On a yes-instance the honest certificate places every individual block in the state $|h_x\rangle$, so each pair is in $|h_x\rangle^{\otimes 2}$ and the single block at $v_r$ is in $|h_x\rangle$.
\item[Communication.] 
Independently for each repetition $k$: each $v_j$ with $0\le j\le r-1$ flips a private fair coin $s^{(k)}_{j}$ and, if $s^{(k)}_{j}=1$, swaps the two blocks of its pair. 
This is the \emph{symmetrization} step, which we import from the improved equality protocol of Hasegawa, Kundu and Nishimura~\cite{hasegawa2024power}. 
Concretely, the node either applies the unitary that exchanges the two $\ell$-qubit blocks, or does nothing, each with probability $\tfrac12$. 
This is equivalently implemented by preparing an ancilla qubit in $|+\rangle$, applying the swap controlled on the ancilla, and discarding the ancilla. 
The node then sends the post-swap register $\tilde{\mathsf{B}}^{(k)}_{j}$ to $v_{j+1}$ and keeps the other register $\tilde{\mathsf{A}}^{(k)}_{j}$.
\item[Tests.] 
For each repetition $k$: 
\begin{description}
    \item[Test $T_0$:] Node $v_0$ measures $\{\ket{h_x}\bra{h_x},I-\ket{h_x}\bra{h_x}\}$ on $\tilde{\mathsf{A}}^{(k)}_{0}$ and accepts on the first outcome. 
    \item[Test $T_j$:] Each intermediate $v_j$, $1\le j\le r-1$, performs the swap test on $(\tilde{\mathsf{B}}^{(k)}_{j-1},\tilde{\mathsf{A}}^{(k)}_{j})$. 
    \item[Test $T_r$ and $T_{r+1}$:] Node $v_r$ flips a private fair coin $t^{(k)}$ and performs the swap test on $(\tilde{\mathsf{B}}^{(k)}_{r-1},\mathsf{A}^{(k)}_{r})$ if $t^{(k)}=0$ ($T_r$), or measures $\{\ket{h_y}\bra{h_y},I-\ket{h_y}\bra{h_y}\}$ on $\mathsf{A}^{(k)}_{r}$ if $t^{(k)}=1$ ($T_{r+1}$). It rejects iff the performed test fails.
\end{description}
\end{description}
\end{tcolorbox}

Per node this uses $2m\ell$ proof qubits and $m\ell$ message qubits, where $m$ is the repetition parameter and $\ell$ is the size of the fingerprints. By choosing $m\in\bigo{r}$ and using the fact that the fingerprints have size $O(\log N)$, we conclude that it uses $O(r\log N)$ qubits each for $m=\Theta(r)$. Completeness is perfect and the analysis is trivial. What remains to be done is thus the proof of soundness.

\subsubsection*{Soundness}

We show that when $x\neq y$, the probability that some node rejects is $\Omega(1/r)$ in each repetition $k$ of Protocol~\ref{protocol-equality}, for any (possibly entangled) quantum proofs. Before that, we first prove the following lemma.

\begin{lemma}

\label{lem:swap}
Let $\rho$ be a bipartite state on two registers $\mathsf{A}$ and $\mathsf{B}$ with $\dim \mathsf{A}=\dim \mathsf{B}$, let
$\Pi_{\pm}$ be a projector onto the symmetric and antisymmetric
subspaces, and set the rejection probability of the swap test $\alpha:=\tr(\Pi_{-}\rho)$ and
$\kappa:=\|\Pi_{+}\rho\Pi_{-}\|_{1}$. Then
\begin{enumerate}
    \item $\D(\rho_{A},\rho_{B})\le2\kappa$
    \item $\kappa\le\sqrt{\alpha(1-\alpha)}\le\sqrt{\alpha}$
\end{enumerate}
where $\rho_{A},\rho_{B}$ are the reduced states on registers $\mathsf{A}$ and $\mathsf{B}$ and $\D$ is the trace distance.
\end{lemma}

\begin{proof}
Let $S=\Pi_{+}-\Pi_{-}$ be the swap operator. Then we have
\[
\rho_{A}-\rho_{B}
=\tr_{B}\bigl(\rho - S\rho S\bigr)
=2\tr_{B}\bigl(\Pi_{+}\rho \Pi_{-}
+\Pi_{-}\rho \Pi_{+}\bigr),
\]
using $\tr_{B}(S\rho S)=\tr_{A}(\rho)$ and $S\Pi_{\pm}=\pm\Pi_{\pm}$. The partial trace does not increase
$\|\cdot\|_{1}$, so 
\begin{align*}
    \D(\rho_A,\rho_B) = \frac12\|\rho_{A}-\rho_{B}\|_{1} &\le (\| \Pi_{+}\rho \Pi_{-} \|_1 + \|\Pi_{-}\rho \Pi_{+}\|_1)   \\ 
    &= 2\kappa.
\end{align*}

For a pure state $\rho = |\psi\rangle\langle\psi|$, $\Pi_{+}\ket{\psi}\bra{\psi}\Pi_{-}$ is a rank-1 operator and its trace norm is 
\begin{align*}
    \kappa &= \|\Pi_{+}\ket{\psi}\bra{\psi}\Pi_{-}\|_{1}\\ &= \|\Pi_+\ket{\psi}\|\cdot \|\Pi_-\ket{\psi}\| \\ &= \sqrt{\alpha(1-\alpha)}
\end{align*}
For a mixed state $\rho = \sum_k p_k \ket{\psi_k}\bra{\psi_k}$ and $\alpha_k = \|\Pi_- \ket{\psi_k} \|^2$ we have $\alpha = \sum_k p_k\alpha_k$ by linearity. Now similarly to the case of pure states,

\begin{align*}
    \kappa = \| \Pi_+ \rho \Pi_- \|_1 &= \| \sum_k p_k \Pi_+ \ket{\psi_k}\bra{\psi_k} \Pi_- \|_1 \\
    &\le \sum_k p_k \| \Pi_+ \ket{\psi_k}\bra{\psi_k} \Pi_- \|_1 \\
    &\le \sum_k p_k \| \Pi_+ \ket{\psi_k} \| \cdot \|\Pi_- \ket{\psi_k}\| \\
    &= \sum_k p_k \sqrt{\alpha_k(1-\alpha_k)}.
\end{align*}
The concavity of $\alpha\mapsto\sqrt{\alpha(1-\alpha)}$ gives $\kappa \le \sqrt{\alpha (1-\alpha)}$.
\end{proof}

The next proposition improves the $\Omega(1/r^{2})$ lower bound for the rejection probability in the earlier protocols in~\cite{fraigniaud2021distributed,hasegawa2024power} to $\Omega(1/r)$.

The distinction is that we keep the tests' joint rejecting weight. Separate estimates $\D(\mu_{j-1},\mu_j)\le2\kappa_j\le2\sqrt{\alpha_j}$ and Cauchy--Schwarz yield only $\sum_j\alpha_j=\Omega(1/r)$; relating this sum to the probability of any rejection can lose another factor of $r$. The hypercube decomposition below instead bounds $\sum_j\kappa_j$ directly by the total weight outside the all-accepting subspace, avoiding this second loss.

\begin{lemma}[Soundness of a single repetition]\label{prop:single}
Suppose $x\ne y$. Then for every state of the blocks of a fixed
repetition $k\in [m]$, arbitrarily entangled internally and with the rest of
the proof, the probability that all tests of repetition $k$ accept is
at most $1-\gamma$, where $\gamma:=\tfrac1{70(r+1)}$.
\end{lemma}

\begin{proof}
 We focus on a single repetition $k$ of Protocol~\ref{protocol-equality} and drop the superscript $k$ for simplicity. Before the one-round communication, each node $v_j$ ($0\le j \le r-1$) holds a pair of registers $(\mathsf{A}_j, \mathsf{B}_j)$ and $v_r$ holds $\mathsf{A}_r$. After the symmetrization step we write the registers $(\tilde{\mathsf{A}}_j, \tilde{\mathsf{B}}_j)$, and $\tilde{\mathsf{B}}_j$ is sent to $v_{j+1}$. The reduced states on the two registers $\tilde{\mathsf{A}}_j,\tilde{\mathsf{B}}_j$ are identical, and we denote both by $\mu_j$. Indeed, if $\rho_{\mathsf A_j}$ and $\rho_{\mathsf B_j}$ are their marginals before symmetrization, then $\mu_j=(\rho_{\mathsf A_j}+\rho_{\mathsf B_j})/2$. This identity makes the trace-distance bounds telescope along the path. Let $P_x:=\ket{h_x}\bra{h_x}$ and $P_y:=\ket{h_y}\bra{h_y}$. Let $\nu$ be the reduced state of $\mathsf{A}_{r}$, let
$\alpha_{0}:=1-\tr(P_{x}\mu_{0})$ and
$\alpha_{r+1}:=1-\tr(P_{y}\nu)$, and for $1\le j\le r$ let
$\kappa_j = \| \Pi_+ \sigma_j \Pi_- \|_1$ be the quantity of Lemma~\ref{lem:swap} performed by the node $v_j$ on the reduced state $\sigma_j$ on $(\tilde{\mathsf{B}}_{j-1},\tilde{\mathsf{A}}_j)$. Then by definition, the reduced states of $\sigma_j$ on each of the two registers are $(\mu_{j-1},\mu_{j})$ for $j\le r-1$ and $(\mu_{r-1},\nu)$ for $j=r$. 

Let $q$ be the probability that some test (of the fixed repetition $k$) rejects. Since the test $T_{0}$ is
always performed and the test $T_{r+1}$ is performed with probability
$\tfrac12$,
$q\ge\max\bigl(\alpha_{0},\tfrac12\alpha_{r+1}\bigr)$. If $\alpha_{0}\ge\tfrac1{100}$ or $\alpha_{r+1}\ge\tfrac1{70}$,
then $q\ge\tfrac1{140}\ge\gamma$ for every $r\ge1$, and we
are done. Hence assume that $\alpha_{0}<\tfrac1{100}$ and
$\alpha_{r+1}<\tfrac1{70}$.

By Lemma~\ref{lem:swap} and the triangle inequality,
\begin{align*}
    \D(\mu_{0},\nu) &\le \sum_{j=1}^{r-1} \D(\mu_{j-1},\mu_j) + \D(\mu_{r-1},\nu)\\
    &\le 2\sum_{j=1}^{r}\kappa_{j},
\end{align*}
while testing the observable $\ket{h_y}\bra{h_y}$ against the pair $(\mu_{0},\nu)$ gives
\[
\D(\mu_{0},\nu)\ \ge\ \tr(\ket{h_y}\bra{h_y}\cdot \nu)-\tr(\ket{h_y}\bra{h_y}\cdot \mu_{0})
\ \ge\ \Bigl(1-\tfrac1{100}\Bigr)
-\Bigl(\tfrac12+\sqrt{\alpha_{0}}\Bigr)\ \ge\ 0.39,
\]
using $\tr(\ket{h_y}\bra{h_y}\cdot \mu_{0})\le\tr(\ket{h_y}\bra{h_y}\cdot \ket{h_x}\bra{h_x})+\D(\mu_{0},\ket{h_x}\bra{h_x})$, $\tr(\ket{h_y}\bra{h_y}\cdot \ket{h_x}\bra{h_x})\le\tfrac12$ for $x\ne y$, and the
Fuchs--van de Graaf inequality $\D(\mu_{0},\ket{h_x}\bra{h_x})\le \sqrt{\alpha_{0}}$. Hence $\sum_{j=1}^{r}\kappa_{j}\ge0.19$.

For each node, we define two orthogonal projectors (good and bad) defined by the performed test.
Specifically,
\begin{itemize}
    \item $Q^{(0)}_+ = \ket{h_x}\bra{h_x}$ and $Q^{(0)}_- = I-\ket{h_x}\bra{h_x}$
    \item $Q^{(j)}_+ = \Pi_+$ and $Q^{(j)}_- = \Pi_-$ for $1 \le j \le r$
\end{itemize}
where $Q^{(j)}_+ + Q^{(j)}_- = I$ and $Q^{(j)}_+ Q^{(j)}_- = 0$ for all $0 \le j \le r$. Let $\ket{\Psi}$ be a purification of the post-symmetrization proof state, including the untouched control ancillas and any other purifying registers. The identity factor below acts on those ancillary registers. The test register sets $\tilde{\mathsf A}_0$, $(\tilde{\mathsf B}_{j-1},\tilde{\mathsf A}_j)$ for $1\le j<r$, and $(\tilde{\mathsf B}_{r-1},\mathsf A_r)$ are pairwise disjoint. Conditioned on the independent endpoint coin being $t^{(k)}=0$, the tests $T_0,\dots,T_r$ are exactly these commuting projective measurements. For each $z =z_0z_1\cdots z_r\in \{0,1\}^{r+1}$, we write
\begin{align*}
    \ket{\psi_z} = (Q^{(0)}_{z_0} \otimes Q^{(1)}_{z_1} \otimes \cdots \otimes Q^{(r)}_{z_r} \otimes I ) \ket{\Psi}
\end{align*}
where in this relation $Q^{(j)}_{z_j}$ is $Q^{(j)}_{+}$ for $z_j=0$ and $Q^{(j)}_{-}$ for $z_j=1$.
Define $a_z:=\|\ket{\psi_z}\|$. The projectors defining these vectors are mutually orthogonal and sum to the identity, so
\[
\sum_{z\in\{0,1\}^{r+1}}a_z^2=\|\ket{\Psi}\|^2=1.
\]
By definition we have
\[
\kappa_j = \| Q^{(j)}_+ \sigma_j Q^{(j)}_- \|_1
\]
and
\[
\sigma_j = \mathrm{tr}_{\neg j} (\ket{\Psi}\bra{\Psi})
\] where $\mathrm{tr}_{\neg j}$ traces out all registers except the registers used for the test of node $v_j$. Using $\ket{\Psi}=\sum_z \ket{\psi_z}$,
\[
(Q^{(j)}_+\otimes I)\ket{\Psi}\bra{\Psi} (Q^{(j)}_-\otimes I) = \sum_{z:z_j=0}\sum_{z':z'_j=1} \ket{\psi_z}\bra{\psi_{z'}}.
\]
Thus we have
\[
Q^{(j)}_+ \sigma_j Q^{(j)}_- = \sum_{z:z_j=0} \sum_{z':z'_j=1} \mathrm{tr}_{\neg j} (\ket{\psi_z}\bra{\psi_{z'}})
\]
Next we show that for $z$, $z'$ with Hamming distance at least $2$, $\mathrm{tr}_{\neg j}(\ket{\psi_z}\bra{\psi_{z'}})=0$: there is a position $i\ne j$ at which $z$ and $z'$ differ. Since this position is traced out, $Q^{(i)}_{z_i}Q^{(i)}_{z'_i}=0$ and

\begin{align*}
\mathrm{tr}_{\neg j}(\ket{\psi_z}\bra{\psi_{z'}}) &=
\mathrm{tr}_{\neg j}\left(\left(Q^{(i)}_{z_i}\otimes I\right)\ket{\psi_z}\bra{\psi_{z'}} \left(Q^{(i)}_{z'_i}\otimes I\right)\right)\\ &= 
\mathrm{tr}_{\neg j}\left(\left(Q^{(i)}_{z_i}Q^{(i)}_{z'_i}\otimes I\right)\ket{\psi_z}\bra{\psi_{z'}} \right)\\
&=0.
\end{align*}
Therefore, we have
\begin{align*}
    Q^{(j)}_+ \sigma_j Q^{(j)}_- = \sum_{z:z_j=0}  \mathrm{tr}_{\neg j} (\ket{\psi_z}\bra{\psi_{z\oplus e_j}}).
\end{align*}
Here $e_j$ is a vector whose $j$-th element is $1$ and the other elements are $0$. By contractivity of the trace norm under partial trace,
\[
\bigl\|\mathrm{tr}_{\neg j}(\ket{\psi_z}\bra{\psi_{z\oplus e_j}})\bigr\|_1
\le\bigl\|\ket{\psi_z}\bra{\psi_{z\oplus e_j}}\bigr\|_1
=a_z a_{z\oplus e_j}.
\]
Therefore,

\begin{align*}
    \kappa_j &= \| Q^{(j)}_+ \sigma_j Q^{(j)}_- \|_1 \\
    &\le \sum_{z:z_j=0} \| \mathrm{tr}_{\neg j} (\ket{\psi_z}\bra{\psi_{z\oplus e_j}}) \|_1 \\
    &\le \sum_{z:z_j=0} a_z \cdot a_{z\oplus e_j}.
\end{align*}
Summing up over all $1\le j \le r$, we have
\begin{align*}
    \sum_{j=1}^r \kappa_j \le \sum_{\{z,z'\}:\text{ an edge of the hypercube }\{0,1\}^{r+1}} a_z a_{z'} 
\end{align*}

Define $q_{\square}:=1-a_{\mathbf0}^{2}=\sum_{z\ne\mathbf0}a_{z}^{2}$ where $\mathbf0 = 0^{r+1}$ is all-zero vector. This is exactly the probability that some test among $T_0,\dots,T_r$ rejects conditioned on $t^{(k)}=0$. Since this coin outcome has probability $1/2$, the unconditional rejection probability satisfies $q\ge q_\square/2$. Now our goal is to bound $q_{\square}$. To this end, we divide 
\[
\sum_{\{z,z'\}:\text{ an edge of the hypercube }\{0,1\}^{r+1}} a_z a_{z'} 
= \underbrace{\sum_{i=0}^r  a_{\mathbf0} a_{e_i}}_{\text{(I)}}  +  \underbrace{\sum_{\{z,z'\}:\text{ an edge of the hypercube }\{0,1\}^{r+1}, z \neq \mathbf0, z'\neq \mathbf0} a_z a_{z'}}_{\text{(II)}}
\]

By Cauchy-Schwarz
\[
\text{(I)}=\sum_{i=0}^r a_{\mathbf0} a_{e_i}\le \sqrt{
\left(\sum_{i=0}^r 1^2 \right)\cdot
\sum_{i=0}^r a_{e_i}^2
}
=\sqrt{(r+1)\cdot \sum_{i=0}^r a_{e_i}^2}
\le \sqrt{(r+1)q_{\square}}.
\]

For (II), using $2a_za_{z'}\leq a_z^2+a_{z'}^2$, we obtain
\[
\begin{aligned}
\mathrm{(II)}
&\leq
\frac12
\sum_{\substack{
\{z,z'\}:\,\text{an edge of }\{0,1\}^{r+1}\\
z\neq\mathbf{0},\ z'\neq\mathbf{0}
}}
(a_z^2+a_{z'}^2)\\
&=\frac12 \sum_{z\neq \mathbf0} \mathrm{deg}(z)\cdot a_z^2 \\
&\leq
\frac{r+1}{2}\sum_{z\neq\mathbf{0}}a_z^2 \\
&=
\frac{r+1}{2}q_{\square}.
\end{aligned}
\]
The second inequality holds because the degree of each vertex $z$ is at most
$r+1$.
To summarize,
\[
0.19 \le \text{(I)} + \text{(II)} \le \sqrt{(r+1)\cdot q_\square} + \frac12 (r+1)\cdot q_\square.
\]
With $u:=\sqrt{(r{+}1)q_{\square}}$, the
constraint $u+u^{2}/2\ge0.19$ gives $u\ge0.174$, i.e.\
$q_{\square}\ge\tfrac1{35(r+1)}$ and
$q\ge\tfrac1{70(r+1)}\ge\gamma$.
\end{proof}

\begin{lemma}\label{lem:eq-fingerprint}
With $m:=\lceil70(r+1)\ln3\rceil=\Theta(r)$ repetitions, the protocol
has perfect completeness and soundness error at most $1/3$ against
arbitrary entangled certificates, with $O(r\log N)$ qubits of
proof and of messages per node.
\end{lemma}

\begin{proof}
Completeness obviously holds. For $x\ne y$, let $M_k$ be the effective accepting POVM element of repetition $k$, averaged over its private coins. Lemma~\ref{prop:single} bounds $\tr(M_k\sigma)\le1-\gamma$ for every state $\sigma$, so $\|M_k\|\le1-\gamma$. The repetitions use disjoint registers and independent coins; hence their joint accepting effect is $\bigotimes_{k=1}^m M_k$. For any certificate state $\rho$, including states entangled across repetitions,
\[
\tr\!\left[\left(\bigotimes_{k=1}^mM_k\right)\rho\right]
\le\prod_{k=1}^m\|M_k\|
\le(1-\gamma)^m\le e^{-\gamma m}\le\tfrac13.
\]
\end{proof}

\subsection{A new equality protocol}\label{sec:new-equality-protocol}

In this section, we prove the $\widetilde{O}(N^{2/3})$ upper bound of \Cref{thm:equality-uniform}. In the protocol used to prove the $O(r\log N)$ upper bound, the $O(r)$ repetitions are inevitable, as it is known that there is a concrete cheating strategy for $o(r)$ repetitions \cite[Section~3]{fraigniaud2021distributed}. Therefore, to obtain a new protocol without dependence on $r$, we need to design a new protocol.

\paragraph{Roadmap.}
The full symmetric subspace accepts $\ket\psi^{\otimes2}$ for every pure $\ket\psi$, permitting gradual rotation along a path. Our subspace $W_L$ has dimension $1+\binom L2$, rather than $\binom{L+1}2$, and imposes equality of the diagonal amplitudes in addition to symmetry. These constraints force the chain's zero-energy space to be the span $S_t$ of tensor powers of the finitely many phase labels. They prevent arbitrary pure-state powers from remaining accepting along the chain.

The proof uses two length scales. The three-register estimate first yields the inverse-square bound $H_t\succeq4t^{-2}(I-G_t)$. Grouping $m=\Theta(L\log L)$ registers into a block makes distinct label powers nearly orthogonal: their off-diagonal Gram-matrix row sums are at most $1/10$. The resulting overlapping-block gap exceeds $3/4$, so a Knabe bound promotes the $\Omega(1/m^2)$ estimate on a pair of blocks to the same lower bound on arbitrarily long chains. The detectability lemma then gives rejection probability $\Omega(1/m^2)$ for unequal endpoint labels.

This also explains the parameters below. Taking $K=\Theta(m^2)$ Reed--Solomon evaluations amplifies rejection to a constant. Each evaluation carries $\Theta(L)$ classical bits, so $N=\Theta(KL)=\widetilde\Theta(L^3)$, whereas each vertex receives $O(K\log L)=\widetilde O(L^2)$ qubits. Thus the exponent $2/3$ comes from the ratio between these cubic input and quadratic certificate scales.

\subsubsection{Encoding of classical strings}

We first describe a fixed encoding of a classical input $X\in\{0,1\}^N$ into $K$ quantum states of $O(\log L)$ qubits, which is based on the standard Reed--Solomon encoding. For simplicity, assume that $N=KL/6$, where $L$ is divisible by $3$ and sufficiently large, and set
\[
m:=\left\lceil\frac L2\ln L+\frac{6L}{5}\right\rceil,
\qquad
K:=2\left\lceil300(\ln3)m^2\right\rceil
=\Theta\!\left(L^2(\log L)^2\right).
\]
Thus $K$ is even, $K\ge600(\ln3)m^2$, and, for sufficiently large $L$, $K\le2^{L/3}$. General input lengths are handled by padding both inputs with zeros to the next admissible length; finitely many smaller lengths can be handled directly.
These parameters give $N=\Theta(L^3(\log L)^2)$ and $L=\Theta(N^{1/3}(\log N)^{-2/3})$. The certificates have $2K\lceil\log L\rceil$ qubits per node, hence size
\[
O\!\left(L^2(\log L)^3\right)
=O\!\left(N^{2/3}(\log N)^{5/3}\right).
\]

Given a string $X\in\{0,1\}^N$, divide $X$ into $K/2$ consecutive blocks of length $L/3$. We regard each block as an element of the finite field with $2^{L/3}$ elements. That is, the input $X$ is a sequence of such elements $q_0,\ldots,q_{K/2-1}$. Next, use these $K/2$ field elements as the coefficients of a polynomial
\[
f_X(t):= \sum_{i=0}^{K/2-1} q_i t^i
\]
of degree at most $K/2-1$. Fix $K$ distinct field elements
$g_1,\ldots,g_K$, and evaluate $f_X$ at these points:
\[
f_X(g_1),\ldots,f_X(g_K).
\]
If $X\neq Y$, then $f_X-f_Y$ is a nonzero polynomial of degree at
most $K/2-1$ and can vanish at no more than
$K/2-1$ points. Consequently,
$
f_X(g_b)\neq f_Y(g_b)
$
for at least $K/2$ values of $b$.

We next fix $2^{L/3}$ binary strings of length $L$ such that any
two distinct chosen strings differ in at least $L/8$ and at most
$7L/8$ positions (such a collection can be obtained greedily), and

fix a one-to-one correspondence between the $2^{L/3}$ field
elements and these selected strings. For each $b$, let $z_b(X)$
be the $L$-bit label corresponding to $f_X(g_b)$.

In this way, we obtain $K$ strings $z_1(X),\ldots,z_K(X)\in \{0,1\}^L$ from an $N$-bit string $X$. By construction, if $X\neq Y$, then for at least $K/2$ positions $b$, the two labels $z_b(X)$ and $z_b(Y)$ have Hamming distance between $L/8$ and $7L/8$.
For completeness, let $H_2(p):=-p\log_2p-(1-p)\log_2(1-p)$. For each selected string, fewer than
\[
2\sum_{j<L/8}\binom Lj\le2^{1+H_2(1/8)L}<2^{0.55L}
\]
strings are excluded by the two forbidden distance ranges, for sufficiently large $L$, since $H_2(1/8)<0.55$. Greedy selection therefore yields at least $2^{0.45L}\ge2^{L/3}$ strings. We now define the corresponding label states.

\begin{definition}[Label states]\label{def:label-state}
For every $L$-bit string $u=u_1\cdots u_L\in\{0,1\}^L$, define
\[
\ket{\phi_u}
=
\frac{1}{\sqrt L}
\sum_{x=1}^L(-1)^{u_x}\ket{x},
\]
and let
\[
W_L
=
\operatorname{span}
\left\{
\ket{\phi_u}\otimes\ket{\phi_u}
:
u\in\{0,1\}^L
\right\}.
\]
Each physical register has $\lceil\log L\rceil$ qubits. We embed $\mathbb C^L$ as the span of its first $L$ computational basis vectors, labelled $\ket1,\dots,\ket L$. Let $\Pi_L$ be the orthogonal projector onto $W_L$, extended by zero on the rest of the two-register physical space. Thus an accepting pre-communication measurement removes every component outside $\mathbb C^L\otimes\mathbb C^L$, justifying the subsequent chain analysis on $L$-dimensional registers. The definition of $W_L$ uses all $L$-bit strings, including those not used by the encoding.
\end{definition}

The identity
\[
\langle\phi_u|\phi_v\rangle=1-\frac{2\,\operatorname{dist}(u,v)}{L}
\]
shows why both distance restrictions matter: complementing all bits negates a label state. At each of the $K/2$ positions for which $z_b(X)$ and $z_b(Y)$ differ,
\[
\left|
\left\langle
\phi_{z_b(X)}
\middle|
\phi_{z_b(Y)}
\right\rangle
\right|
\leq \frac34.
\]

\paragraph*{Protocol.}

Given these definitions, we describe our protocol. 

\begin{tcolorbox}[code={\refstepcounter{protocolcount}}, title={Protocol \theprotocolcount}, enhanced, breakable=true, label={new-protocol-equality}]
\begin{enumerate}
\item
\textbf{Certificates and local preparation.}
For every node $v_i$ and every index $b\in [K]$, the prover supplies a
certificate register $\A_{i,b}\otimes \B_{i,b}$, each containing $\lceil \log L\rceil$ qubits. The node performs the non-destructive projective measurement $\{\Pi_L,I-\Pi_L\}$ on $\A_{i,b}\otimes\B_{i,b}$ and records rejection on the second outcome. The accepting outcome projects onto $W_L$ and leaves every state already in $W_L$ unchanged.

\item
\textbf{Communication.}
Each node $v_i$ with $i<r$ keeps all registers $\A_{i,b}$ and sends
all registers $\B_{i,1},\ldots,\B_{i,K}$ to $v_{i+1}$.  The right
endpoint $v_r$ keeps its register $\B_{r,b}$.
After the communication, $v_i$ holds $\B_{i-1,b}\A_{i,b}$ for every
$1\leq i\leq r$. 

\item
\textbf{Measurements and acceptance.}
For every $b\in [K]$, the following measurements are performed, where accepting projectors are defined in Definition~\ref{def:label-state} (the rejecting projector is the orthogonal complement of the corresponding accepting projector).
\begin{center}
\renewcommand{\arraystretch}{1.35}
\begin{tabular}{c@{\qquad}c@{\qquad}c}
\textbf{node} & \textbf{register(s)} & \textbf{accepting projector} \\
\hline
$v_0$ & $\A_{0,b}$ & $\ket{\phi_{z_b(X)}}\bra{\phi_{z_b(X)}}$ \\
$v_i\ (1\leq i\leq r)$ & $\B_{i-1,b}\A_{i,b}$ & $\Pi_L$ \\
$v_r$ & $\B_{r,b}$ & $\ket{\phi_{z_b(Y)}}\bra{\phi_{z_b(Y)}}$
\end{tabular}
\end{center}

A node accepts if every measurement assigned to it, for every
$b\in\{1,\ldots,K\}$, has the accepting outcome.  The protocol accepts if and only if every node accepts. 
\end{enumerate}
\end{tcolorbox}

At the right endpoint $v_r$, the received register $\B_{r-1,b}$ and the local registers $\A_{r,b},\B_{r,b}$ serve two distinct tests: the link test on $\B_{r-1,b}\A_{r,b}$ and the endpoint test on $\B_{r,b}$. Both are performed. Steps~1 and~2 are independent of $X$ and $Y$; only the final endpoint tests in Step~3 use those strings. This permits a single-round composition with classical input reconstruction.

\paragraph{Perfect completeness.}
When $X=Y$, the two endpoints have the same label at
every index $b$. For each node and each $b$, the prover directly supplies the state in $W_L$
\[
\ket{\phi_{z_b(X)}}\otimes\ket{\phi_{z_b(X)}}.
\]
The resulting state for this index is
\[
\ket{\phi_{z_b(X)}}^{\otimes(2r+2)},
\]
so every measurement accepts with probability one.

\subsubsection{Soundness for one label index}
We write $z=z_b(X)$, $w=z_b(Y)$ for $z\neq w$ for a fixed index $b\in [K]$ and analyze Protocol~\ref{new-protocol-equality} for this $b$.

Set $m= \lceil \frac L2\ln L + \frac{6L}5 \rceil$. We prove the following.

\begin{lemma}\label{lem:one-label-soundness}
    For the distinct labels $z=z_b(X)$, $w=z_b(Y)$, every certificate in Protocol~\ref{new-protocol-equality} is rejected by at least one node with probability at least
    \[
    \frac1{129m^2} = \Omega\left(  \frac1{L^2(\log L)^2}\right).
    \]
\end{lemma}

\paragraph{The chain Hamiltonian}
To prove Lemma~\ref{lem:one-label-soundness}, we first study the sum
of the rejection probabilities of the tests of intermediate nodes in Protocol~\ref{new-protocol-equality} whose accepting projector is $\Pi_L$.

At the fixed label index $b$, $\A_i$, $\B_i$ denote the registers $\A_{i,b}$ and $\B_{i,b}$ at vertex $v_i$ for simplicity. Set $n=2r+2$ (only in this chain analysis, $n$ denotes the number of registers rather than graph vertices) and identify these $n$ registers, in order, with
\[
\A_0,\B_0,\A_1,\B_1,\ldots,\A_r,\B_r.
\]
For each integer $t\ge2$, consider an ordered chain of $t$ registers, each with Hilbert space $\mathbb C^L$, and define
\[
H_t=\sum_{j=1}^{t-1}(I-\Pi^{j,j+1})
\quad\text{on }(\mathbb C^L)^{\otimes t}.
\]
Here $\Pi^{j,j+1}$ acts as $\Pi_L$ on registers $j$ and $j+1$
and as the identity on all other registers. We use the same
superscript convention specifying the registers to be considered for other operators.
Under this notation,
\[
H_n
=
\sum_{i=0}^{r}(I-\Pi^{\A_i\B_i})
+
\sum_{i=1}^{r}(I-\Pi^{\B_{i-1}\A_i}).
\]
The first sum corresponds to the measurements before communication,
and the second to those after communication in Protocol~\ref{new-protocol-equality}.
The two endpoint measurements are not included in $H_n$.

We may also apply $H_t$ to any $t$ consecutive registers in the above
ordering, numbering the registers within that interval from $1$ to $t$
from left to right. Therefore, for a quantum state $\ket{x}$, $\langle x|H_t\ket{x}$ is the sum of the rejection probabilities when each adjacent-pair test is applied separately to $\ket{x}$. 
In the language of Hamiltonian complexity, $H_t$ is a \textit{frustration-free local Hamiltonian} on a chain of qudits.

\paragraph*{The accepting space.}
Let
\[
S_t=\mathrm{span}\{\ket{\phi_u}^{\otimes t}: u\in \{0,1\}^L\}.
\]
Every vector $\ket{\phi_u}^{\otimes t}$ is accepted by every test, and thus a ground state of $H_t$, meaning $S_t \subseteq \mathrm{ker}H_t$. 
The following lemma shows the two spaces are identical.
\begin{lemma}\label{lem:ground-space}
    For every $t\ge 2$, $\mathrm{ker}H_t = S_t$. Moreover, every vector in $S_t$ is fixed by every permutation of the registers.
\end{lemma}
\begin{proof}
Since $ S_t \subseteq \mathrm{ker}H_t$ is obvious, we focus on proving $\mathrm{ker}H_t \subseteq S_t$.

First, observe that
\[
\ket{\Omega}=\frac1{\sqrt{L}}\sum_a \ket{aa},\qquad \ket{s_{ab}}=\frac{\ket{ab}+\ket{ba}}{\sqrt{2}} \quad (1\le a<b\le L)
\]
is an orthonormal basis of $W_L$:
From the definition of label states,
\[
\ket{\phi_u}^{\otimes 2}=\frac1{\sqrt{L}}\ket{\Omega} + \frac{\sqrt{2}}L \sum_{a<b}(-1)^{u_a+u_b}\ket{s_{ab}},
\]
hence every state in $W_L$ is spanned by this basis.
On the other hand, averaging over all strings $u$ gives
\[
\frac1{2^{L}}\sum_{u\in \{0,1\}^L}\ket{\phi_u}^{\otimes 2} = \frac1{\sqrt{L}}\ket{\Omega}, \qquad \frac1{2^L}\sum_u (-1)^{u_a+u_b}\ket{\phi_u}^{\otimes 2}=\frac{\sqrt{2}}{L}\ket{s_{ab}}
\]
hence $\ket{\Omega}, \ket{s_{ab}} \in W_L$.

Let $\ket{1},\ldots,\ket{L}$ be an orthonormal basis of $\mathbb{C}^L$, and write an arbitrary state in $(\mathbb{C}^L)^{\otimes 2}$ as $\sum_{a,b=1}^Ld_{ab}\ket{ab}$. Suppose that this state is in $W_L$ and write
\[
\sum_{a,b}d_{ab}\ket{ab}=\alpha \ket{\Omega} + \sum_{a<b}\beta_{ab}\ket{s_{ab}}.
\]
Comparing the coefficients gives
\[
d_{aa}=\frac{\alpha}{\sqrt{L}} \quad (1\le a \le L), \qquad 
d_{ab}=\frac{\beta_{ab}}{\sqrt{2}} \quad (a<b),
\]
and in particular, for all $a\neq b \in [L]$,
\begin{align}\label{eq:W-condition}
    d_{aa}=d_{bb},\quad d_{ab}=d_{ba}.
\end{align}
Conversely a state $\sum_{a,b=1}^Ld_{ab}\ket{ab}$ satisfying \eqref{eq:W-condition} equals $\sqrt L\,d_{11}|\Omega\rangle+\sqrt2\sum_{a<b}d_{ab}|s_{ab}\rangle$ and thus lies in $W_L$.

Next, write 
\[|x\rangle=\sum_{\mathbf a = (a_1,\cdots,a_t)\in[L]^t}x(\mathbf a)|\mathbf a\rangle = \sum_{\mathbf a = (a_1,\cdots,a_t)\in[L]^t}x(\mathbf a)|a_1\rangle\otimes \cdots \otimes \ket{a_t} \in \mathrm{ker} H_t.
\] 
Since $H_t=\sum_j(I-\Pi^{j,j+1})$ is a sum of
positive semidefinite operators, $|x\rangle\in\ker H_t$
if and only if $\Pi^{j,j+1}|x\rangle=|x\rangle$
for every $j$, that is, if and only if for every $j$ and every fixed choice of the entries other than $a_j,a_{j+1}$,
the vector $\sum_{a_j,a_{j+1}}x(\mathbf a)|a_ja_{j+1}\rangle$ lies in $W_L$. 
By \eqref{eq:W-condition},
this holds if and only if 
\begin{align}\label{eq:operation}
    x(\cdots,a_j,a_{j+1},\cdots)=x(\cdots,a_{j+1},a_{j},\cdots),\quad x(\cdots,a,a,\cdots)=x(\cdots,b,b,\cdots).
\end{align}

For $\mathbf a\in[L]^t$, let $p(\mathbf a)\in\{0,1\}^L$ be the vector whose $c$-th entry
is the number of occurrences of the symbol $c$ in $\mathbf a$, modulo two. 
Then, the coefficient $x(\mathbf a)$ is determined by $p(\mathbf a)$: for every symbol $c$, keep one instance of $c$ that appears an odd number of times in $\mathbf a$, group the remaining into pairs, replace all of them with \(11\) by the two operations of \eqref{eq:operation}, and arrange the remaining symbols in increasing order. The sequence obtained by this operation is determined solely by \(p(\mathbf a)\) and \(t\).

Then, for
\[
\ket{v_{p'}}=\sum_{\mathbf a :p(\mathbf a)=p'}\ket{a_1\cdots a_t} \quad (p'\in \{0,1\}^L),
\]
$\ket{x}\in \mathrm{span}\{\ket{v_{p'}}\}_{p'\in\{0,1\}^L}$.
On the other hand, by the definition of label states,
\[
\ket{\phi_u}^{\otimes t} = \frac1{\sqrt{L^t}}\sum_{p'\in\{0,1\}^L}(-1)^{u\cdot p'}\ket{v_{p'}}.
\]
Therefore,
\begin{align*}
    \frac{\sqrt{L^t}}{2^L} \cdot \sum_{u\in \{0,1\}^L} (-1)^{u\cdot p'}\ket{\phi_u}^{\otimes t}
    &= \frac1{2^L}\sum_{u,p''\in\{0,1\}^L}(-1)^{u\cdot p''+u\cdot p'}\ket{v_{p''}}\\
    &=\ket{v_{p'}} + \frac1{2^L} \sum_{u,p''\in\{0,1\}^L:p''\neq p'}(-1)^{u\cdot (p''+p')}\ket{v_{p''}}\\
    &=\ket{v_{p'}}.
\end{align*}
Here, the last equality used the following. For each $p''\neq p'$, choose an arbitrary index $k$ such that $p''_k\neq p'_k$. 
Since $p''_k\ne p'_k\in\{0,1\}$, we have $p''_k+p'_k\equiv1\pmod2$, which simplifies the coefficient of $u_k$ in the exponent. Then
\begin{align*}
    \sum_{u} (-1)^{u\cdot(p''+p')} &= \sum_{u_i\in\{0,1\}:i\neq k} \sum_{u_k\in\{0,1\}} (-1)^{\sum_{i\neq k}u_i(p''_i+p'_i) + u_k}\\
    &= \sum_{u_i\in\{0,1\}:i\neq k} (-1)^{\sum_{i\neq k}u_i(p''_i+p'_i)} + (-1)^{\sum_{i\neq k}u_i(p''_i+p'_i)+1}\\
    &=0.
\end{align*}
Therefore, we have 
\[
\ket{x}\in \mathrm{span}\{\ket{v_{p'}}\}_{p'}=\mathrm{span}\{\ket{\phi_u}^{\otimes t}\}_u,
\]
meaning $\mathrm{ker}H_t \subseteq S_t$.  

Finally, every $|\phi_u\rangle^{\otimes t}$ is fixed by every permutation of the registers, hence so is every vector in $S_t$.
\end{proof}

\paragraph*{A spectral gap independent of the path length.}
Let $G_t$ be the orthogonal projector onto $S_t$.
For the proof of Lemma~\ref{lem:one-label-soundness}, we use the following lemma showing a spectral gap of the Hamiltonian that is independent of the path length $r$. Its proof appears later in this subsection, under \emph{Spectral gap of the Hamiltonian}.
\begin{lemma}[Spectral gap]
\label{lem:spectral-gap}
For every integer $s\ge 2$, with $m$ as in the statement of Lemma~\ref{lem:one-label-soundness},
\[
H_{sm}\succeq\frac1{4m^2}(I-G_{sm}).
\]
\end{lemma}

We next combine this bound with the endpoint conditions.

\begin{lemma}\label{lem:property1}
For the distinct labels $z,w$ fixed above, let $t\geq2$ and let
$\ket{x}\in(\mathbb{C}^L)^{\otimes t}$ satisfy
\[
\begin{aligned}
\bigl(\ket{\phi_z}\bra{\phi_z}\otimes I^{\otimes(t-1)}\bigr)
\ket{x}&=\ket{x},\\
\bigl(I^{\otimes(t-1)}\otimes\ket{\phi_w}\bra{\phi_w}\bigr)
\ket{x}&=\ket{x}.
\end{aligned}
\]
Then
\[
\bra{x}H_t\ket{x}
\geq
\frac{\|\ket{x}\|^2}{32m^2}.
\]
\end{lemma}

\begin{proof}
First suppose that $t=sm$ for an integer $s\geq2$.
Let $F$ exchange the first and last registers.
Every vector in $S_t$ is fixed by $F$, so the orthogonal projector
$(I-F)/2$ vanishes on $S_t$. Hence, for any vector $\ket{v}$, 
\[
\begin{aligned}
\bra{v}\frac{I-F}{2}\ket{v}
&=
\left\|\frac{I-F}{2}\ket{v}\right\|^2\\
&=
\left\|\frac{I-F}{2}(I-G_t)\ket{v}\right\|^2\\
&\leq
\|(I-G_t)\ket{v}\|^2\\
&=
\bra{v}(I-G_t)\ket{v}.
\end{aligned}
\]
The inequality holds because an orthogonal projection cannot
increase the norm. We now have the inequality
\[
\frac{I-F}{2}\preceq I-G_t.
\]
The endpoint conditions imply
\[
\ket{x}
=
\ket{\phi_z}\otimes\ket{\chi}\otimes\ket{\phi_w}
\]
for some vector $\ket{\chi}$ on the middle $t-2$ registers.
Consequently,
\[
\bra{x}F\ket{x}
=
|\langle\phi_z|\phi_w\rangle|^2\,\|\ket{x}\|^2.
\]
Combining it with the above inequality and the spectral-gap bound of Lemma~\ref{lem:spectral-gap} gives
\[
\begin{aligned}
\bra{x}H_t\ket{x}
&\geq
\frac{1}{4m^2}\bra{x}(I-G_t)\ket{x}\\
&\geq
\frac{1}{4m^2}\bra{x}\frac{I-F}{2}\ket{x}\\
&=
\frac{1-|\langle\phi_z|\phi_w\rangle|^2}{8m^2}
\,\|\ket{x}\|^2\\
&\geq
\frac{\|\ket{x}\|^2}{32m^2},
\end{aligned}
\]
where the last inequality uses
$|\langle\phi_z|\phi_w\rangle|\leq3/4$. This completes the proof when $t=sm$.

For arbitrary $t$, choose $t'=sm>t$ with $s\geq2$ and set
\[
\ket{\widetilde{x}}
=
\ket{x}\otimes\ket{\phi_w}^{\otimes(t'-t)}.
\]
The last register of $\ket{x}$ is fixed to $\ket{\phi_w}$.
Thus every newly added adjacent pair, including the pair joining
the original and appended registers, is in the state
$\ket{\phi_w}\otimes\ket{\phi_w}$ and is accepted by $\Pi_L$.
Therefore
\[
\bra{\widetilde{x}}H_{t'}\ket{\widetilde{x}}
=
\bra{x}H_t\ket{x},
\qquad
\|\ket{\widetilde{x}}\|^2=\|\ket{x}\|^2.
\]
The vector $\ket{\widetilde{x}}$ satisfies the same endpoint
conditions, so applying the case already proved at length $t'$
gives the claim.
\end{proof}

\paragraph{Proof of Lemma~\ref{lem:one-label-soundness}}
Let
\[
H'
=
H_n
+\bigl(I-\ket{\phi_z}\bra{\phi_z}\bigr)^{\A_0}
+\bigl(I-\ket{\phi_w}\bra{\phi_w}\bigr)^{\B_r}
\]
be the sum of all rejecting projectors at the fixed label index $b$.
We use the detectability lemma in the following form.
\begin{proposition}[Detectability lemma, \cite{AnshuAradVidick2016}]\label{lem:detectability}
    Let $R_1,\ldots, R_M$ be projectors and $H=\sum_k R_k$, and assume that each $R_k$ commutes with all but at most $g$ of the others. For a unit vector $\ket{\psi}$, let $\ket{\xi}=\prod_k(I-R_k)\ket{\psi}$, where the product is taken in arbitrary order. If $\ket{\xi}\neq 0$ and $\epsilon:=\bra{\xi}H\ket{\xi}/\|\ket{\xi}\|^2$, then
    \[
    \|\ket{\xi}\|^2 \le \frac1{1+\epsilon/g^2}.
    \]
\end{proposition}

Using Lemma~\ref{lem:property1} and the detectability lemma, we can prove Lemma~\ref{lem:one-label-soundness}. 

\begin{proof}[Proof of Lemma~\ref{lem:one-label-soundness}.]
    The certificate may be entangled with the registers of the other label indices. Tracing these registers out leaves a mixed state on the $2r+2$ registers of the index $b$, and since the acceptance probability is linear in the state, it suffices to consider a pure state $\ket{\psi}$ on these registers. Let
    \[
    \begin{aligned}
    \ket{\phi}:=
    (\ket{\phi_z}\bra{\phi_z})^{\A_0}
    (\ket{\phi_w}\bra{\phi_w})^{\B_r}
    \cdot
    \left(\prod_{i=1}^{r}\Pi^{\B_{i-1}\A_i}\right)
    \left(\prod_{i=0}^{r}\Pi^{\A_i\B_i}\right)
    \ket{\psi}.
    \end{aligned}
    \]
    The rightmost product is applied before communication and the
    remaining factors after communication. The acceptance probability
    is therefore $\|\ket{\phi}\|^2$. If $\ket{\phi}=0$, the protocol rejects
    with certainty.

    Otherwise, all projectors applied after communication act on
    disjoint registers, so both endpoint projectors fix $\ket{\phi}$. Their rejecting projectors therefore annihilate $\ket{\phi}$, and Lemma~\ref{lem:property1} with $t=n$ gives
\[
\frac{\bra{\phi}H'\ket{\phi}}{\|\ket{\phi}\|^2}
=
\frac{\bra{\phi}H_n\ket{\phi}}{\|\ket{\phi}\|^2}
\geq
\frac{1}{32m^2}.
\]
Each term of $H'$ shares a register with at most two other terms,
and hence commutes with all but at most two of them.
Proposition~\ref{lem:detectability} with $g=2$ therefore yields
    \[
    \|\ket{\phi}\|^2 \le \frac1{1+1/(128m^2)},
    \]
    so the protocol rejects with probability at least $1/(128m^2+1)\ge1/(129m^2)$.
\end{proof}

\paragraph{Spectral gap of the Hamiltonian}\label{sec:spectral-gap}
Now we move to the proof of the spectral gap shown in Lemma~\ref{lem:spectral-gap}. To do so, we use projectors on pairs of blocks, following earlier arguments~\cite{FNW92,Nach96} for proving spectral gaps of one-dimensional frustration-free Hamiltonians. We fix $s\ge2$, consider the $sm$  registers on which $H_{sm}$ acts, and partition the registers into $s$ consecutive blocks of $m$ registers each. For $1\le j<s$, let $G^{j,j+1}_{2m}$ be the projector onto $S_{2m}$ on the $2m$ registers of these two blocks, and let
\[
\gamma_j \;=\; I-G^{j,j+1}_{2m}\qquad(1\le j<s).
\]
The two ingredients of the proof are the following claims. Their proofs are given in the appendix. The first uses the three-site gap $1/2$ and a weighted sum of local terms to obtain an inverse-square finite-chain estimate. At this scale the three-site bound alone does not give a length-independent gap; the second claim supplies the stronger overlapping-block estimate needed after coarse-graining.
 
\begin{claim}\label{cl:A}
For every integer $t\ge2$, $H_t\succeq\dfrac{4}{t^2}\,(I-G_t)$.
\end{claim}
 
\begin{claim}\label{cl:B}
For $1\le j\le s-2$, the operator $\gamma_j+\gamma_{j+1}$ has no eigenvalue in the open interval $(0,3/4)$.
\end{claim}
For the second claim, Lemma~\ref{lem:gram} makes distinct label powers on an $m$-register block nearly orthogonal. The resulting three-block projection estimate, \eqref{eq:claim-b-bound}, gives a gap at least $9/11>3/4$ for $\gamma_j+\gamma_{j+1}$. Thus the block size $m=\Theta(L\log L)$ raises the local estimate above the threshold in the following Knabe bound.

We prove Lemma~\ref{lem:spectral-gap} using Claim~A and
the following variant of Knabe's bound~\cite{Kn88}.
Claim~B gives the condition on $\gamma_j+\gamma_{j+1}$
needed to apply this bound to
$\sum_{j=1}^{s-1}\gamma_j$. 
\begin{proposition}\label{prop:knabe}
Let $\gamma_1,\dots,\gamma_k$ be projectors such that $\gamma_i$ and $\gamma_j$ commute whenever $|i-j|\ge2$, and suppose that $\gamma_i+\gamma_{i+1}$ has no eigenvalue in the open interval $(0,3/4)$ for $1\le i<k$. Then $\sum_{i=1}^k\gamma_i$ has no eigenvalue in the open interval $(0,1/2)$.
\end{proposition}
 
\begin{proof}
The operator $\gamma_i+\gamma_{i+1}$ is positive semidefinite with no eigenvalue in $(0,3/4)$, so $(\gamma_i+\gamma_{i+1})^2\succeq\frac34(\gamma_i+\gamma_{i+1})$. As $(\gamma_i+\gamma_{i+1})^2=\gamma_i+\gamma_{i+1}+\gamma_i\gamma_{i+1}+\gamma_{i+1}\gamma_i$, this says
\begin{equation}\label{eq:beta}
\gamma_i\gamma_{i+1}+\gamma_{i+1}\gamma_i\;\succeq\;-\frac14\,(\gamma_i+\gamma_{i+1})\qquad(1\le i<k).
\end{equation}
Expanding the square,
\begin{align*}
    \left(\sum_{i=1}^k\gamma_i\right)^2
    &=\sum_{i=1}^k\gamma_i+\sum_{i=1}^{k-1}\bigl(\gamma_i\gamma_{i+1}+\gamma_{i+1}\gamma_i\bigr)+\sum_{\substack{1\le i,j\le k\\|i-j|\ge2}}\gamma_i\gamma_j \\
    &\succeq \sum_{i=1}^k\gamma_i-\frac14\sum_{i=1}^{k-1}\bigl(\gamma_i+\gamma_{i+1}\bigr)\\
    &\succeq \frac12\sum_{i=1}^k\gamma_i .
\end{align*}
Here, the first inequality applied \eqref{eq:beta} to the middle sum and the fact that the last sum is positive semidefinite ($\gamma_i\gamma_j$ is a projector whenever $|i-j|\ge2$). The second inequality holds because each $\gamma_i$ occurs at most twice in $\sum_{i=1}^{k-1}(\gamma_i+\gamma_{i+1})$.
 
Every positive eigenvalue $\lambda$ of $\sum_i\gamma_i$ therefore satisfies $\lambda^2\ge\lambda/2$, so $\lambda\ge1/2$.
\end{proof}

\begin{proof}[Proof of Lemma~\ref{lem:spectral-gap}]
Apply Claim~A with $t=2m$ to each pair of blocks $j$ and $j+1$. Note that the blocks $j$ and $j+1$ contain $2m$ registers numbered from $(j-1)m+1$ to $(j+1)m$. Claim A thus gives
\[
\sum_{i=(j-1)m+1}^{(j+1)m-1}
(I-\Pi^{i,i+1})
\succeq
\frac{4}{(2m)^2}(I-G_{2m}^{j,j+1})
=
\frac{1}{m^2}\gamma_j.
\]
Summing over $1\le j<s$,
\begin{align*}
\sum_{j=1}^{s-1}
\sum_{i=(j-1)m+1}^{(j+1)m-1}
(I-\Pi^{i,i+1})
\succeq
\frac1{m^2}\sum_{j=1}^{s-1}\gamma_j.
\end{align*}
In the left-hand side, each term $(I-\Pi^{i,i+1})$ appears at most twice, so
$
2H_{sm}\succeq \text{(LHS)}
$. 

Therefore, to complete the proof, it suffices to show
\begin{align}\label{eq:gamma-inequality}
    \sum_{j=1}^{s-1}\gamma_j
\succeq \frac12(I-G_{sm}).
\end{align}
The right-hand side vanishes on $S_{sm}$ and equals
$\frac12 I$ on $S_{sm}^{\perp}$.
It therefore suffices to show that
$\ker(\sum_{j=1}^{s-1}\gamma_j)=S_{sm}$
and that all positive eigenvalues of the sum are at least $1/2$.
The latter is easy to see: the projectors $\gamma_j$ and $\gamma_{j'}$ act on disjoint blocks when $|j-j'|\ge2$ and therefore commute, and by Claim~B the operator $\gamma_j+\gamma_{j+1}$ has no eigenvalue in $(0,\frac34)$. Proposition~\ref{prop:knabe} thus shows that every positive eigenvalue of $\sum_j\gamma_j$ is at least $1/2$.

We thus prove the kernel identity.
By Lemma~\ref{lem:ground-space}, $G_{2m}$ is the orthogonal projector onto
$S_{2m}=\ker H_{2m}$. Thus,
\[
\ker(I-G_{2m})=\ker H_{2m}
=\bigcap_{i=1}^{2m-1}\ker(I-\Pi^{i,i+1}).
\]
For each $1\le j<s$, we have
\[
\gamma_j = I-G_{2m}^{j,j+1}
=
I^{\otimes(j-1)m}\otimes(I-G_{2m})
\otimes I^{\otimes(s-j-1)m}.
\]
Together with the kernel identity above, this gives
\[
\ker\gamma_j
=
\bigcap_{i=(j-1)m+1}^{(j+1)m-1}
\ker(I-\Pi^{i,i+1}).
\]
Consequently, for every vector $\ket{x}$ on all $sm$ registers,
\[
\begin{aligned}
&\ket{x}\in\bigcap_{j=1}^{s-1}\ker\gamma_j\\
&\quad\Longleftrightarrow\quad
\begin{gathered}[t]
(I-\Pi^{i,i+1})\ket{x}=0
\quad\text{for every }1\le j<s
\text{ and every }(j-1)m+1\le i\le(j+1)m-1
\end{gathered}\\
&\quad\Longleftrightarrow\quad
(I-\Pi^{i,i+1})\ket{x}=0
\quad\text{for every }1\le i<sm\\
&\quad\Longleftrightarrow\quad
\ket{x} \in \ker H_{sm}.
\end{aligned}
\]
Again by Lemma~\ref{lem:ground-space}, this implies
\[
\bigcap_{j=1}^{s-1}\ker\gamma_j = \ker\left(\sum_{j=1}^{s-1}\gamma_j\right)=S_{sm}.
\]

\end{proof}

\subsubsection{Soundness analysis}
Set $\delta_0:=1/(300m^2)<1/(129m^2)$, using the one-label rejection bound of Lemma~\ref{lem:one-label-soundness}.
For each label index $b\in [K]$, let $Q_b$ be the product of the accepting projectors measured before communication, and let $P_b$ be the product of those measured after communication. Within either product, the factors act on disjoint registers, so both $P_b$ and $Q_b$ are orthogonal projectors. In particular, $P_b^\dagger P_b=P_b$. The effective accepting POVM element of first measuring $Q_b$ and then $P_b$ is
\[
A_b=(P_bQ_b)^\dagger(P_bQ_b)=Q_bP_bQ_b.
\]
The two appearances of $Q_b$ in this expression describe a single sequential measurement, rather than measuring it twice.
For a position with unequal labels $z_b(X)\neq z_b(Y)$, Lemma~\ref{lem:one-label-soundness} gives for every vector $\ket{x}$,
\[
\bra{x}A_b\ket{x}=\|P_bQ_b\ket{x}\|^2\le (1-\delta_0)\|Q_b\ket{x}\|^2\le (1-\delta_0)\|\ket{x}\|^2.
\]
Thus $\|A_b\|\le 1-\delta_0$ at each position $b$ such that $z_b(X)\neq z_b(Y)$.
Let $\rho$ be any certificate density operator on the overall registers and $A_{\mathsf{all}}=\bigotimes_{b=1}^K A_b$.
Then the acceptance probability of the entire protocol is
\begin{align*}
\mathrm{Pr}[\text{The protocol accepts}]&=\mathrm{Tr}(A_{\mathsf{all}}\rho) \\
&\le \|A_{\mathsf{all}}\|\\
&=\prod_{b=1}^K\|A_b\|\\
&\le (1-\delta_0)^{K/2}\\
&\le 1/3,
\end{align*}
where the second last inequality comes from the fact that for $X\neq Y$, at least $K/2$ positions have different labels. The last inequality follows from our choice of $K$, since
\[
(1-\delta_0)^{K/2}\le e^{-\delta_0K/2}
\le e^{-\ln3}=1/3.
\]

\subsection{Improved lower bound}
To prove our lower bound in Theorem~\ref{thm:equality-lower}, we first formalize quantum protocols on a path as follows.

\begin{definition}[one-round $\dQMA$ on a path]\label{def:model}
Let $v_0,\dots,v_r$ be nodes on a path; $v_0$ holds $x\in\{0,1\}^N$
and $v_r$ holds $y\in\{0,1\}^N$. A protocol specifies finite
register sizes and, for each node, input-dependent local
operations, as follows.
\begin{enumerate}
\item The prover prepares an arbitrary pure state $\xi$ on
$\bigotimes_{j=0}^{r}H_{j}$, where $H_j$ has $c(v_j)$ qubits, and
node $v_j$ receives $H_j$.
\item Simultaneously, each node applies an isometry to its
register, producing a kept part and outgoing message registers, and
sends each message to the corresponding neighbour; $m(v_j,v_{j+1})$
denotes the total qubits exchanged between $v_j$ and $v_{j+1}$. The
isometry of $v_0$ may depend on $x$, that of $v_r$ on $y$.
\item Each node measures its kept part together with its incoming
messages with a two-outcome POVM (accept/reject). The POVM of $v_0$
may depend on $x$, that of $v_r$ on $y$.
\end{enumerate}
The protocol has completeness $c$ and soundness $s$ for
$\eqtwo$ if: for $x=y$ some proof makes all nodes accept with
probability $\ge c$, and for $x\ne y$, every proof makes all nodes
accept with probability $\le s$.
\end{definition}

Fix a protocol as in Definition~\ref{def:model} with completeness
$c$ and soundness $s$, and for each $x$ fix an honest proof
$\xi_x$ for the instance $(x,x)$ accepted with probability
$\ge c$. We first observe the following.

\begin{lemma}\label{lem:locality}
On the instance $(x,y)$ with proof $\xi_x$, the joint outcome
distribution of the measurements of $v_0,\dots,v_{r-2}$ coincides
with that on the instance $(x,x)$ with the same proof.
\end{lemma}

\begin{proof}
The only $y$-dependent elements of the protocol are the isometry
and the POVM of $v_r$. The registers measured by $v_j$ for
$j\le r-2$ are its kept part and the messages from $v_{j\pm1}$;
none of these is acted on by $v_r$'s isometry, because in a
simultaneous round $v_{r-1}$'s outgoing message to $v_{r-2}$ is
produced from $H_{r-1}$ alone, before anything sent by $v_r$
arrives. Hence the state of those registers, and the measurements
applied to them, are identical in the two instances.
\end{proof}

Let \[
q_{\partial}\;:=\;c(v_{r-1})+c(v_{r})
+m(v_{r-2},v_{r-1})
\]
be the number of qubits of certificate and messages handled by the two
nodes nearest the endpoint holding $y$. Assume that the protocol runs on proof $\xi_x$ with left input $x$ and stop after the isometries of $v_0,\dots,v_{r-2}$ have been applied and
their messages sent, but before any operation of $v_{r-1}$ or
$v_r$. Let
\[
W\;:=\;H_{r-1}\otimes M_{r-2\to r-1}\otimes H_{r},
\]
of dimension $2^{c(v_{r-1})+m(v_{r-2},v_{r-1})+c(v_r)}
\le2^{q_\partial}$, and let $\sigma_x$ be the reduced state on $W$
at that moment. After this moment, all remaining operations of $v_{r-1}$ and $v_r$ act only on $W$ and on registers produced from it (the exchanged
messages come from $H_{r-1}$ and $H_r$, and the message $M_{r-1\to
r-2}$ is discarded), and are independent of $x$. Hence there is a POVM element
$0\le E_y\le\mathbf 1_W$, depending only on $y$ and the protocol,
with
\begin{equation}\label{eq:window}
\Pr[\,v_{r-1}\text{ and }v_r\text{ accept}\mid
\text{instance }(x,y),\ \text{proof }\xi_x\,]
=\tr\!\left[E_y\,\sigma_x\right],
\end{equation}
which decomposes the acceptance probability to $E_y$ (independent of $x$) and $\sigma_x$ (independent of $y$).
We now show that for distinct $x\neq y$, the reduced states $\sigma_x$ and $\sigma_y$ are $\Omega(1)$-far from each other in trace distance:
\begin{lemma}\label{lem:dist}
For all $x\ne y$,
$\ \tr[E_x\sigma_x]\ge c$ and $\ \tr[E_y\sigma_x]\le s+(1-c)$.
Consequently $\D(\sigma_x,\sigma_y)\ge\delta:=2c-1-s$.
\end{lemma}

\begin{proof}
The first inequality is completeness together with
\eqref{eq:window} at $y=x$: all nodes accepting implies the window
pair accepting. For the second, on instance $(x,y)$ with proof
$\xi_x$ let $A_{\le r-2}$ be the event that $v_0,\dots,v_{r-2}$ all
accept and $A_\partial$ the event that both $v_{r-1}$ and $v_r$ accept;
these are outcomes of measurements on disjoint registers, so their
joint distribution is well defined. Soundness gives
$\Pr[A_{\le r-2}\wedge A_\partial]\le s$, and
Lemma~\ref{lem:locality} with completeness gives
$\Pr[A_{\le r-2}]\ge c$. Hence
\[
\tr[E_y\sigma_x]=\Pr[A_\partial]
\le\Pr[A_{\le r-2}\wedge A_\partial]+\Pr[\neg A_{\le r-2}]
\le s+(1-c).
\]
Finally, for $x\ne y$,
$\D(\sigma_x,\sigma_y)\ge\tr[E_y\sigma_y]-\tr[E_y\sigma_x]
\ge c-s-(1-c)=\delta$.
\end{proof}

\begin{theorem}\label{thm:main}
Any one-round $\dQMA$ protocol for $\eqtwo$ on a path of length
$r\ge2$ with fixed completeness $c$ and soundness error $s$ satisfying
$\delta=2c-1-s>0$ has
\[
q_\partial \ge \Omega(\log N).
\]
Moreover, if $r\ge3$, then
\[
c(v_{r-2})+c(v_{r-1})+c(v_r)=\Omega(\log N),
\]
independently of the message sizes.
\end{theorem}

\begin{proof}
By Lemma~\ref{lem:dist}, the $2^{N}$ states $\{\sigma_x\}$ have pairwise trace distance at
least $\delta$ on $W$. Write $D:=\dim W$. 

Fix for each $x$ a purification $|\psi_x\rangle$ of $\sigma_x$ on
$W\otimes W'$, where $W'$ has dimension $D$, so that the
$|\psi_x\rangle$ are unit vectors in a space of dimension $D^{2}$.
By Uhlmann's theorem,
$|\langle\psi_x|\psi_y\rangle|\le\F(\sigma_x,\sigma_y)$, and by
the Fuchs--van de Graaf inequality and Lemma~\ref{lem:dist},
$\F(\sigma_x,\sigma_y)\le\sqrt{1-\delta^{2}}$ for all $x\ne y$.

We thus have $2^{N}$ unit vectors in dimension $D^{2}$ whose
pairwise inner products are at most $\sqrt{1-\delta^{2}}<1$ in
absolute value. By the lower bound argument of~\cite{hasegawa2024power}(see also~\cite{de2001quantum,buhrman2001communication}), we have 
\[
\log(D^2)=\Omega_\delta(\log N).
\]
Since $D\le2^{q_\partial}$, this proves the first assertion.

For the certificate-only assertion, assume $r\ge3$ and let $\tau_x$ be the reduced state of the honest proof $\xi_x$ on
\[
\widetilde W:=H_{r-2}\otimes H_{r-1}\otimes H_r.
\]
The node $v_{r-2}$ then holds no input. Its fixed local isometry, including any initialized workspace, followed by retaining only its message to $v_{r-1}$, defines an input-independent quantum channel. Tensoring this channel with the identity on $H_{r-1}\otimes H_r$ gives a channel $\Lambda$ satisfying $\sigma_x=\Lambda(\tau_x)$ for every $x$. Equivalently, \eqref{eq:window} can be written as
\[
\tr[E_y\sigma_x]=\tr[\Lambda^*(E_y)\tau_x],
\]
absorbing the local operation into the accepting POVM element on $\widetilde W$.
By contractivity of trace distance and Lemma~\ref{lem:dist}, the states $\tau_x$ have pairwise trace distance at least $\delta$. Applying the same packing argument in dimension
\[
\dim\widetilde W=2^{c(v_{r-2})+c(v_{r-1})+c(v_r)}
\]
proves the second assertion. In particular, at least one of these three nodes receives $\Omega(\log N)$ certificate qubits, proving \Cref{thm:equality-lower} even with unrestricted messages.
The restriction $r\ge3$ is essential here: for $r=2$, the input endpoints can send their strings directly to the middle node in one round, with no certificates.
\end{proof}

%=======================================

\section{Exponential quantum advantage}\label{sec:tree}
Our first separation is given by a graph property $\treeq_\mu$ whose
members are \emph{trees}. To define $\treeq_\mu$, we start with an
instance of $\eqtwo$ on a path, with one input string at each
endpoint. We remove the two input strings and encode each of them
in the structure of a tree attached to the corresponding endpoint.
The encoding has the following properties:
\begin{itemize}
\item All encoded graphs are trees (by definition).
    \item The shape of the trees can be verified with a deterministic $\dMA$ protocol.
    \item In each tree, the pair of encoded strings can be recovered
    uniquely from the isomorphism type (Lemma~\ref{lem:isoT}).
    
\end{itemize}

\subsection{Encoding $\eqtwo$ into trees}

Fix a positive integer $\mu\ge1$, known to the nodes, and an even integer $N\ge8$. Set
\[
L:=\left\lceil\log_2(N/2+1)\right\rceil,
\qquad \rho:=2\mu+1.
\]
The central-path length depends only on $\mu$. We first treat fixed $\mu$ and state the uniform parameter dependence at the end of the section.

We encode strings $z\in\{0,1\}^N$ of Hamming weight $N/2$ by their zero-run lengths. If the ones occur at positions $p_1<\cdots<p_{N/2}\in\{0,\dots,N-1\}$, define $r(z)=(r_0,\dots,r_{N/2})$ by
\[
r_i:=
\begin{cases}
p_1,&i=0,\\
p_{i+1}-p_i-1,&1\le i<N/2,\\
N-1-p_{N/2},&i=N/2.
\end{cases}
\]
This is a bijection between balanced $N$-bit strings and vectors in $\mathbb Z_{\ge0}^{N/2+1}$ summing to $N/2$.

The vertex $a_i$ will carry $r_i$ pendant leaves. To retain the order of these numbers under isomorphism, we attach a tag whose rooted shape identifies $i$: below $a_i$ is the path $a_i-s_i-t_i$, and the distinct-length arms below $t_i$ record the bits of $i$. The rigidity lemma will also exclude accidental tag branches elsewhere in the tree. This gives the following formal definition.

\begin{figure}[h]
\centering
\input{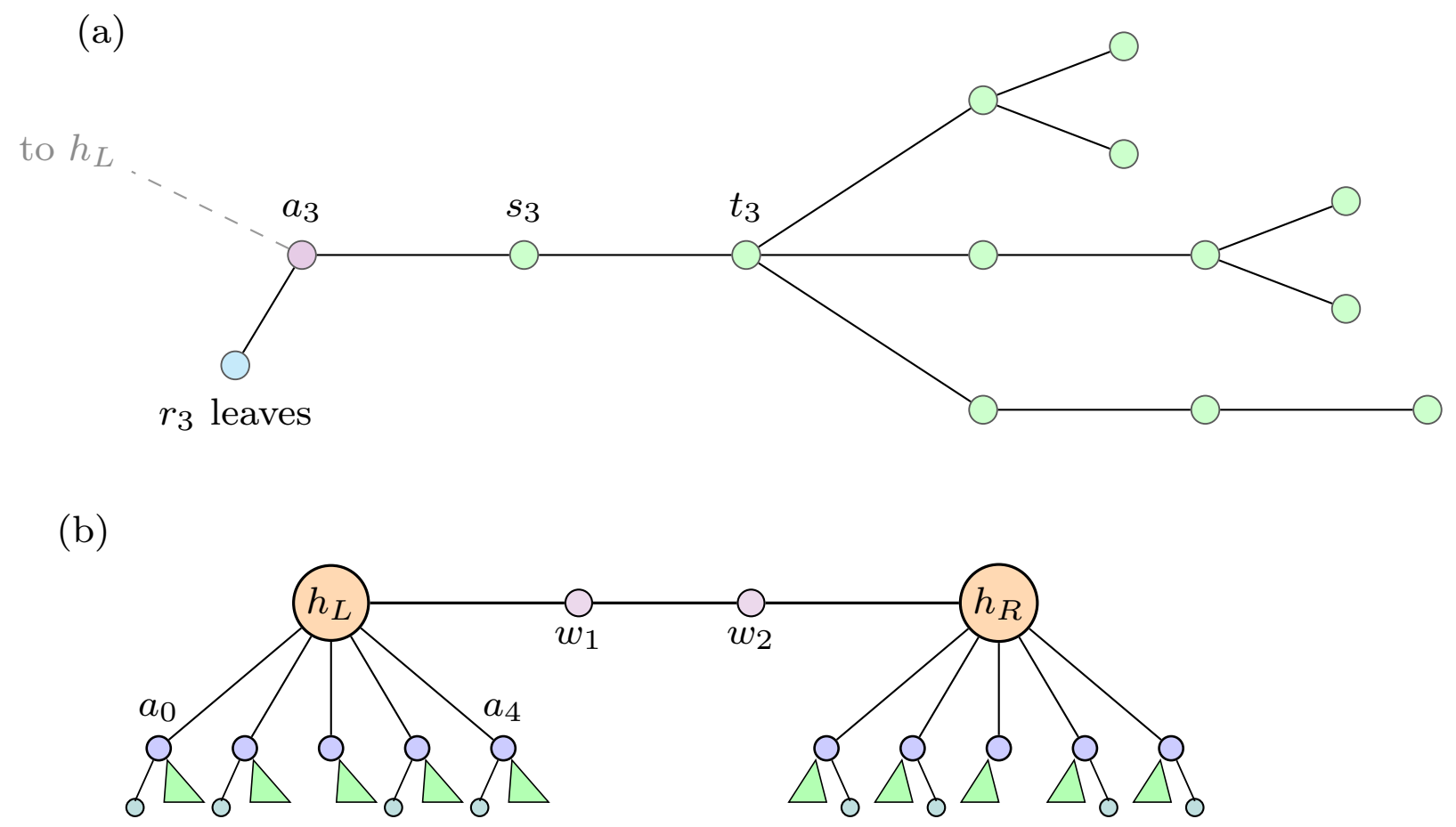}
\caption{
(a) The tag $\mathrm{Tag}(3)$ for $L=3$
($3=011_2$, with bits read from the least significant).
The paths starting at $t_3$ have lengths $1,2,3$;
the first two end in two extra leaves.
No other branch of an instance is isomorphic as a rooted tree to a tag
(Lemma~\ref{lem:tag}).
(b) The instance $T^{\mathrm{tr}}_{8,\mu}(z,z)$ for
$z=01011010$ and $r(z)=(1,1,0,1,1)$.
Here $\mu=1$, so the central path has $\rho=2\mu+1=3$ edges.
Each side has vertices $a_0,\dots,a_4$ carrying the prescribed
leaves and tags, drawn as triangles.
Each hub reconstructs all pairs $(i,\hat r_i)$ of its side
in one communication round.
}
\label{fig:treecomb}
\end{figure}

\begin{definition}\label{def:tag}
For $i\in\{0,\dots,N/2\}$, the rooted tree $\mathrm{Tag}(i)$
has a root $s$ with a single child $t$. For each $b\in[L]$,
a path of $b$ edges starts at $t$. These paths share only $t$.
If the $b$-th least significant bit of $i$ is $1$, we attach
two additional leaves to the last vertex of the corresponding
path. See Figure~\ref{fig:treecomb}(a).
\end{definition}

And using \Cref{def:tag} we can finally define the encoding of two input strings $z,z'$ into a tree.

\begin{definition}\label{def:tree}
For strings $z,z'\in\{0,1\}^N$ of Hamming weight $N/2$,
the tree $T^{\mathrm{tr}}_{N,\mu}(z,z')$ is constructed as follows.
Start with two vertices $h_L,h_R$ joined by the central path
$P=h_L,w_1,\dots,w_{\rho-1}, h_R$
of $\rho$ edges. Attach $N/2+1$ new vertices
$a_0,\dots,a_{N/2}$ to $h_L$. For each $i$, attach $r_i(z)$
leaves to $a_i$, and join $a_i$ by an edge to the root of
a disjoint copy of $\mathrm{Tag}(i)$.
Attach the same structure to $h_R$, using $r(z')$ in place
of $r(z)$. See Figure~\ref{fig:treecomb}(b).
\end{definition}
\begin{comment}

\end{comment}
 We use the following useful notation for the rest of the Section.
\begin{remark}\label{remark:notation}
    We call $a_i$ together with its leaves and its tag the $i$-th
\emph{gadget}. We call $h_L$ together with its gadgets the
\emph{left side}, and define the right side symmetrically.

Also, we denote as $T_L(z)$ the subtree consisting of $h_L$ and all
the gadgets attached to it, rooted at $h_L$; the subtree
$T_R(z')$ is defined symmetrically, with root $h_R$.

\end{remark}

Thus, $T_{N,\mu}^{\mathrm{tr}}(z,z')$ consists of the two rooted
trees $T_L(z)$ and $T_R(z')$, whose roots are joined by
the central path of $\rho$ edges.

\begin{fact}\label{fact:treecount}
For every positive integer $\mu$, every even $N\ge8$, and $z,z'\in\{0,1\}^N$ of Hamming weight $N/2$, the graph $T^{\mathrm{tr}}_{N,\mu}(z,z')$ is a tree. Its order is $n_\mu(N)=\Theta(N\log^2N)+2\mu$, with constants independent of $\mu$. For fixed $\mu$, this order depends only on $N$, not on $(z,z')$, and is strictly increasing in $N$.
\end{fact}

\begin{proof}
Define the values $m:=N/2$, $B_L:=L(L+1)/2$, and let $\mathrm{wt}(i)$
denote the Hamming weight of the binary expansion of $i$.
A path of $b$ edges starting at $t$ contributes exactly $b$
new vertices. Hence
\[
|\mathrm{Tag}(i)|=2+B_L+2\mathrm{wt}(i).
\]
One side contains its hub, $m+1$ vertices $a_i$, exactly $m$
pendant leaves at those vertices, and one copy of every tag.
Consequently, the exact order is
\[
n_\mu(N)=
2\left[
1+m+(m+1)(3+B_L)
+2\sum_{i=0}^{m}\mathrm{wt}(i)
\right]+\rho-1.
\]
For fixed $\mu$, this depends only on $N$. The bracketed expression is $\Theta(N\log^2N)$ with universal constants, and $\rho-1=2\mu$. This gives the stated additive dependence on $\mu$, and in particular $n_\mu(N)=\Theta(N\log^2N)$ for fixed $\mu$.

As even $N$ increases, $m$, $L$, $B_L$, and the sum of Hamming
weights are nondecreasing, while the $m$ term strictly increases.
When $L$ increases, adding leading zero bits does not change the
weights of the old indices. The central-path term is constant for fixed $\mu$. Thus $n_\mu(N)$ is strictly
increasing.
\end{proof}

\subsection*{Properties of $T_{N,\mu}^{\mathrm{tr}}(z,z')$ and the class $\treeq_\mu$}

Given that the trees $T_{N,\mu}^{\mathrm{tr}}(z,z')$ consist of a path of length $\rho$ with many $\mathrm{Tag}$ graphs adjacent to the endpoints, the verification of such structures and its connection with the original $\eqtwo$ problem relies on understanding how the $\mathrm{Tag}$ graphs encode the problem.
For a vertex $v$ of a tree $T$ and a neighbor $u$ of $v$, we refer to the
\emph{branch of $T$ at $v$ containing $u$} as the connected
component of $T-v$ containing $u$, rooted at $u$. Rooted isomorphism uses the notation $\cong_{\mathrm{rt}}$ from \Cref{sec:pre}.

\begin{lemma}\label{lem:tag}
Let $\mu\ge1$ and even $N\ge8$, let $z,z'\in\{0,1\}^N$ have Hamming weight $N/2$, and denote
$T:=T^{\mathrm{tr}}_{N,\mu}(z,z')$. The following properties hold.
\begin{enumerate}
    \item For any $i,j\in\{0,\dots, N/2\}$, $\mathrm{Tag}(i)\cong_{\mathrm{rt}}\mathrm{Tag}(j)$ if and only
    if $i=j$.
    \item For every vertex $v$ of $T$, every neighbor $u$
    of $v$, and every $i\in\{0,\dots,N/2\}$, the branch
    of $T$ at $v$ containing $u$ is isomorphic as a rooted tree to
    $\mathrm{Tag}(i)$ if and only if $v$ is an index-$i$
    vertex $a_i$, on either side, and $u$ is the root
    of the tag attached to $v$.
\end{enumerate}
\end{lemma}

\begin{proof}
For the first claim, consider the subtrees hanging below
node $t$ of any graph $\mathrm{Tag}(i)$. The subtree corresponding to arm $b\in[L]$ consists of
$b$ vertices in a path, possibly with two extra leaves at its last
vertex. Two such rooted subtrees are isomorphic exactly when
they have the same length (i.e. the same arm index $b$) and the same structure at the end.

Suppose first that there exists a rooted isomorphism between $\mathrm{Tag}(i)$ and
$\mathrm{Tag}(j)$. Such an isomorphism maps the node $t$ to its counterpart and matches the
subtrees below these vertices. It therefore preserves every length path and its number of leaves. The bits of $i$ and $j$ agree, so $i=j$.
The converse follows directly from the construction.

For the second claim, note that the branch at $a_i$ containing its tag
root is $\mathrm{Tag}(i)$ by construction. We show that no other
branch is a tag.

Define $m:=N/2$ and $B_L:=L(L+1)/2$. Since $N\ge8$, we have
$m\ge4$ and $L\ge3$. Every tag has at most $2+B_L+2L$
vertices. Moreover, every branch containing a hub or a central-path vertex
contains at least $m$ entire gadgets, and hence at least
$m(3+B_L)$ vertices. This is larger than every tag, since
\[
m(3+B_L)-(2+B_L+2L)
\ge 10+3B_L-2L>0.
\]
Thus such branches cannot be tags.

The remaining branches are contained in a single gadget.
Apart from the tag itself, they are exhausted by the following
cases.
\begin{itemize}
    \item A branch at a hub containing $a_i$.
    If $r_i\ge1$, its root $a_i$ has at least two children,
    whereas a tag root has one.
    If $r_i=0$, its root has the single child $s_i$, but
    $s_i$ has one child, whereas the child of a tag root has
    $L\ge3$ children.

    \item A branch at $a_i$ containing one of its pendant leaves,
    or a branch at the last vertex of a tag arm containing one
    of its extra leaves.
    Such a branch is a singleton, whereas a tag root has a child.

    \item A branch at the tag root containing $t_i$.
    Its root has $L\ge3$ children, whereas a tag root has one.

    \item A branch at $t_i$ containing the first vertex of an
    arm, or a branch at an arm vertex containing the next
    vertex.
    Such a branch is a path, possibly with two extra leaves
    at its end. Every vertex has at most two children, whereas
    a tag contains a vertex with $L\ge3$ children.

    \item An upward branch from a vertex inside the gadget.
    This contains $a_i$ and its hub, and is therefore excluded
    by the size argument above.
\end{itemize}
This exhausts all branches of $T$.
\end{proof}

Now we prove the following characterization that links isomorphism between trees in $\treeq_\mu$ with our equality problem.

\begin{lemma}\label{lem:isoT}
For every $\mu\ge1$, even $N,\hat N\ge8$, strings $z,z'\in\{0,1\}^N$ of Hamming weight $N/2$, and strings $\hat z,\hat z'\in\{0,1\}^{\hat N}$ of Hamming weight $\hat N/2$, it holds that
\(
T^{\mathrm{tr}}_{N,\mu}(z,z')
\cong
T^{\mathrm{tr}}_{\hat N,\mu}(\hat z,\hat z')
\)
if and only if $N=\hat N$ and
\[
(z,z')=(\hat z,\hat z')
\qquad\text{or}\qquad
(z,z')=(\hat z',\hat z).
\]
\end{lemma}

\begin{proof}
The parameters $N$ and $\hat N$ agree because the vertex count
is strictly increasing in $N$, by Fact~\ref{fact:treecount}.
Let $\varphi$ be an isomorphism.

By Lemma~\ref{lem:tag}, having a branch isomorphic as a rooted tree to
$\mathrm{Tag}(i)$ characterizes the two index-$i$ vertices,
one on each side. Hence $\varphi$ maps each $a_i$ to an
index-$i$ vertex of the target.

Distinct vertices $a_i,a_j$ of the same side are at distance
$2$, whereas vertices on opposite sides are at distance
$\rho+2\ge5$. After interchanging the left and right
designations in the target, if necessary, assume that
$\varphi$ maps the left $a_0$ to the left $\hat a_0$.
Every other left $a_i$ is at distance $2$ from $a_0$, rather than $\rho+2\ge5$. Its image, already restricted to the target's two index-$i$ vertices, therefore also lies on the left.

The vertex $h_L$ maps to the common neighbor of
$\varphi(a_0)$ and $\varphi(a_1)$, namely the target's $h_L$.
For each $i$, the tag root adjacent to $a_i$ maps to the tag
root adjacent to $\varphi(a_i)$, by Lemma~\ref{lem:tag}: the graph isomorphism induces a rooted isomorphism on each corresponding branch.
The remaining $r_i(z)$ neighbors of $a_i$ are its pendant
leaves. Consequently,
\[
r_i(z)=r_i(\hat z)
\qquad\text{for every }i.
\]
The run-length bijection gives $z=\hat z$.
The same argument on the right gives $z'=\hat z'$.
If the target's side designations were interchanged, the
conclusion is instead $(z,z')=(\hat z',\hat z)$.

Conversely, either equality of ordered pairs gives an
isomorphism directly, permuting interchangeable leaves and
exchanging the two sides when necessary.
\end{proof}

Now, we finally define the property $\treeq_\mu$. For any fixed positive integer $\mu\ge1$, we define the graph property $\treeq_\mu$ as follows:
\[
\treeq_\mu
:=
\left\{
T:
T\cong T^{\mathrm{tr}}_{N,\mu}(z,z)
\text{ for some even }N\ge8
\text{ and }z\text{ of weight }N/2
\right\}.
\]

\subsection*{Certifying the graphs $T_{N,\mu}^{\mathrm{tr}}(z,z')$}

We first give a deterministic one-round certification of the
graphs $T_{N,\mu}^{\mathrm{tr}}(z,z')$, without checking equality of the encoded strings. We formally analyze the more general property

\begin{equation*}
\begin{split}
\mathrm{Tree}_\mu:=\{T:\;&T\cong T_{N,\mu}^{\mathrm{tr}}(z,z')\text{ for some even }N\ge8,\\
&z,z'\in\{0,1\}^N\text{ with }\mathrm{wt}(z)=\mathrm{wt}(z')=N/2\}.
\end{split}
\end{equation*}

And we prove the following result.

\begin{lemma}\label{lem:shapeT}
The property $\mathrm{Tree}_\mu$ admits a deterministic one-round
$\dMA$ protocol on connected $n$-vertex graphs, using
$O(\log n)$ certificate bits per node and bits per message.
\end{lemma}

If all nodes accept in \Cref{lem:shapeT}, the certificates identify the left
and right trees and their roots $h_L,h_R$, together
with a valid parameter $\hat N$. During the same
communication round, $h_L$ reconstructs $\hat z$ from
the run lengths reported by its neighbors, and $h_R$
reconstructs $\hat z'$ in the same way. These strings
have Hamming weight $\hat N/2$ and satisfy
\[
G\cong T_{\hat N,\mu}^{\mathrm{tr}}(\hat z,\hat z').
\]

The local structure of each graph $T_{N,\mu}^{\mathrm{tr}}(z,z')$ allows us to give a deterministic classical certification for $\mathrm{Tree}_\mu$, although a bit tedious given the different sub-parts that we need to verify. Nevertheless, the following certification follows a basic idea: We can give to each node its role in the tree, i.e., to which part it belongs, and certify its correctness through local rules.

\begin{proof}[Proof of \Cref{lem:shapeT}]
Fix the constant $\mu\ge1$ and let $\rho:=2\mu+1$. By \Cref{def:tree}, in any yes instance of $\mathrm{Tree}_\mu$
the central path connecting the left and right trees has constant length $\rho$ and therefore it is fixed throughout the protocol, and does not depend on the input size.

\paragraph{Certificates and local rules.}
Each node receives a claimed even integer $8\le\hat N\le n$, encoded in a fixed $O(\log n)$-bit field, and a name from Table~\ref{tab:tree-rules}. It computes $\hat L:=\lceil\log_2(\hat N/2+1)\rceil$. Every gadget name contains its side $\sigma\in\{L,R\}$ and index $i\in\{0,\dots,\hat N/2\}$. Arm vertices and extra leaves additionally contain $b\in[\hat L]$, and an arm vertex also contains its position $p\in[b]$. Leaves sharing a parent and role have the same name. The only auxiliary field beyond these name coordinates is the run length $\hat r_i^\sigma\in\{0,\dots,\hat N/2\}$ at $a_i$.

\begin{table}[t]
\centering
\small
\renewcommand{\arraystretch}{1.25}
\begin{tabularx}{\textwidth}{@{}p{.20\textwidth}p{.16\textwidth}Y@{}}
\toprule
Name & Auxiliary field & Required multiset of neighboring names \\
\midrule
$h_L$ & none & One side-$L$ vertex $a_i$ for each $0\le i\le\hat N/2$, and $w_1$ \\
$h_R$ & none & One side-$R$ vertex $a_i$ for each $0\le i\le\hat N/2$, and $w_{\rho-1}$ \\
$w_j$, $1\le j<\rho$ & none & $w_{j-1},w_{j+1}$, interpreting $w_0=h_L$, $w_\rho=h_R$ \\
$a_i$ on side $\sigma$ & $\hat r_i^\sigma$ & $h_\sigma$, $s_i$, and $\hat r_i^\sigma$ pendant leaves of $a_i$ \\
Pendant leaf of $a_i$ & none & Only $a_i$ \\
Tag root $s_i$ & none & $a_i,t_i$ \\
$t_i$ & none & $s_i$ and the first vertex of each arm $b\in[\hat L]$ \\
Position $p$ of arm $b$ & none & Its predecessor ($t_i$ if $p=1$); its successor if $p<b$; at $p=b$, two extra leaves if the $b$-th least significant bit of $i$ is $1$, and none otherwise \\
Extra leaf of arm $b$ & none & Only position $b$ of arm $b$; this role is valid only if the $b$-th least significant bit of $i$ is $1$ \\
\bottomrule
\end{tabularx}
\caption{Classical fields and local rules for $\mathrm{Tree}_\mu$. Every gadget reference preserves the same side and index; arm predecessors, successors, and extra leaves also preserve the same arm index.}
\label{tab:tree-rules}
\end{table}

\paragraph{Verification.}
Before communication, each node checks the local conditions in (V1), including $n_\mu(\hat N)=n$, using the exact count in Fact~\ref{fact:treecount}. If no even $\hat N\ge8$ satisfies this equality, it rejects. Otherwise it broadcasts its certificate in one round and performs the remaining checks after reception:
\begin{itemize}
    \item[(V1)] 
    $\hat N$ is even, $8\le\hat N\le n$, $n_\mu(\hat N)=n$, and its name is valid,
    all fields required by that name are present,
    and all indices and run lengths are in their
    specified ranges. Every neighbor reports the
    same $\hat N$.

    \item[(V2)] The multiset of its neighbors' names
    equals the multiset required by its own certificate.

    \item[(V3)] If its name is $h_\sigma$, the run lengths
    obtained from its neighbors named $a_0,\dots,a_{\hat N/2}$
    satisfy
    \[
    \sum_{i=0}^{\hat N/2}\hat r_i^\sigma=\hat N/2.
    \]
\end{itemize}
A node accepts if and only if all applicable verifications pass.

\paragraph{Completeness.}
Suppose $G\in\mathrm{Tree}_\mu$.
Choose an isomorphism from $G$ to
$T_{N,\mu}^{\mathrm{tr}}(z,z')$ witnessing this membership.
The prover assigns every node the parameter $N$ and
the name of its image under this isomorphism.
Each vertex named $a_i$ additionally receives the
corresponding run length of $z$ or $z'$.
The neighboring names agree with the construction,
and the run lengths on each side sum to $N/2$.
Thus every node passes (V1)--(V3).

\paragraph{Soundness.}
The model tree has a unique name for every non-leaf vertex, and any repeated leaf name has a single parent. We use these names to build an injective homomorphism into $G$; exact degree matching and connectedness then force its image to be the whole graph, without needing global distance labels.

Suppose all nodes accept on a connected graph $G$.
By (V1) and the connectedness of $G$, all nodes report the same
even integer $\hat N\ge8$.

First, some node has name $h_L$.
Indeed, starting from any gadget vertex, the neighboring
names required by (V2) give a finite sequence leading
to its $a_i$ and then to its corresponding root.
From a vertex named $h_R$ or $w_j$, the required
central-path neighbors give a sequence ending at a
vertex named $h_L$.
Let $u_0$ be such a vertex.

Starting at $u_0$, follow the required neighboring names
\(w_1,\dots,w_{\rho-1},h_R,\)
and denote the last vertex by $u_\rho$.
These vertices are distinct because their names are
distinct.

By (V2), $u_0$ has exactly one neighbor named $a_i$
of side $L$ for each $i\in\{0,\dots,\hat N/2\}$.
Likewise, $u_\rho$ has exactly one neighbor named $a_i$
of side $R$ for each such $i$.
Let $\hat r_i^L$ and $\hat r_i^R$ be the run lengths
reported by these neighbors.
By (V3), they define strings
\[
\begin{aligned}
\hat z
&=
0^{\hat r_0^L}\,1\,0^{\hat r_1^L}\,1\cdots
1\,0^{\hat r_{\hat N/2}^L},\\
\hat z'
&=
0^{\hat r_0^R}\,1\,0^{\hat r_1^R}\,1\cdots
1\,0^{\hat r_{\hat N/2}^R}.
\end{aligned}
\]
Each string has exactly $\hat N/2$ zeros and $\hat N/2$
ones, so both belong to $\{0,1\}^{\hat N}$ and have
Hamming weight $\hat N/2$. Define
\(
T:=T_{\hat N,\mu}^{\mathrm{tr}}(\hat z,\hat z'),
\) rooted at $h_L$. We construct an injective map $\psi:V(T)\to V(G)$
that preserves names and edges.
Set $\psi(h_L):=u_0$, and proceed by increasing distance
from $h_L$ in $T$.

Suppose $\psi(x)$ has been defined.
For every child $y$ of $x$ whose name occurs only once
in $T$, define $\psi(y)$ to be the neighbor of $\psi(x)$
with that name.
For a group of children sharing a leaf name, match
these children bijectively with the neighbors of
$\psi(x)$ carrying that name.
The required neighbors exist, with exactly the required
multiplicities, by (V2).
In particular, the central path maps to the path chosen
above, and the vertices named $a_i$ map to the neighbors
of $u_0$ and $u_\rho$ from which the run lengths were read.
Their claimed run lengths therefore agree with those
used to construct $T$.

Every non-leaf name occurs only once in $T$. At $u_\rho=\psi(h_R)$, the neighbor named $w_{\rho-1}$ required by (V2) is necessarily the already visited vertex $\psi(w_{\rho-1})$: it is adjacent to $u_\rho$, and (V2) requires exactly one neighbor with that name. Thus expanding $h_R$ introduces only its gadget children and does not create a second copy of the central path.

This construction is injective.
Vertices with different names have different images.
The only names shared by several vertices of $T$
are leaf names, and all leaves with one such name have
a common parent; their images are distinct by the
chosen bijection.
Since every nonroot vertex of $T$ has a unique parent,
the construction defines $\psi$ on all of $V(T)$.

Every edge of $T$ maps to an edge of $G$.
Moreover, by (V2), for every $x\in V(T)$,
\(\deg_G(\psi(x))=\deg_T(x).\) By injectivity, the images of the neighbors of $x$
already account for all neighbors of $\psi(x)$.
Consequently, no additional edge is incident to any
vertex in the image of $\psi$.
In particular, no vertex outside the image is adjacent
to a vertex inside it.

The image is nonempty and $G$ is connected, so the image
is all of $G$.
Thus $\psi$ is an isomorphism, and
\(
G\cong T_{\hat N,\mu}^{\mathrm{tr}}(\hat z,\hat z').
\)
Therefore $G\in\mathrm{Tree}_\mu$, proving soundness.

The reconstruction above uses only the certificates
received by the two roots during the communication
round. Hence the roots reconstruct $\hat z$ and $\hat z'$
by the end of that same round.
By \Cref{lem:isoT}, their unordered pair is determined
by the isomorphism type of $G$.

\paragraph{Complexity.}
Each certificate contains a constant number of fields
encoding the parameter, the name, and, when applicable,
a run length.
The fixed field bounds $\hat N\le n$, all name ranges in Table~\ref{tab:tree-rules}, and $\rho<n$ on any size passing (V1) give $O(\log n)$ bits per node and per message on every instance, including no-instances. Larger or malformed encodings are rejected before communication.
\end{proof}

\subsection{Quantum advantage for $\treeq_\mu$}

We prove \Cref{thm:treeq-separation} by combining the quantum
upper bound of \Cref{thm:treeQ} with the classical lower bound
of \Cref{thm:treeC}.

\begin{lemma}\label{thm:treeQ}
For every fixed integer $\mu\ge1$, the property $\treeq_\mu$
admits a one-round $\dQMA$ protocol on connected $n$-vertex
graphs with perfect completeness, soundness error at most
$1/3$, and $O(\mu\log n)$ certificate qubits per node and qubits per individual message. The implicit constant is independent of $\mu$; for fixed $\mu$ this is $O(\log n)$.
\end{lemma}

\begin{proof}
Fix $\mu\ge1$ and let $\rho:=2\mu+1$.
We combine the structural certification of
\Cref{lem:shapeT} with the path equality protocol of
\Cref{thm:equality-uniform}, applied to the central path of
$\rho$ edges.
The key point that allows this combination in one round is
that the equality protocol needs the endpoint strings
only for its final local tests.
All operations performed before receiving the messages
are independent of those strings.

\paragraph*{Certificates}
The prover supplies each node with the classical certificate
of \Cref{lem:shapeT}.
The nodes named $h_L,w_1,\dots,w_{\rho-1},h_R$ in this protocol also receive
the quantum registers prescribed by the equality protocol of \Cref{thm:equality-uniform},
using the claimed string length $\hat N$ and path length
$\rho$ fixed in \Cref{lem:shapeT}.
Use $m_\rho:=\lceil70(\rho+1)\ln3\rceil$ repetitions of Protocol~\ref{protocol-equality}. Because local (V1) checks uniquely determine $\hat N$ from known $n$ and $\mu$, register dimensions are bounded before any quantum operation. Pad fingerprints to a common width $\ell(n)=O(\log n)$, large enough for every $N\le n$, and reserve at most $2m_\rho\ell(n)$ quantum certificate qubits at each node. Its measured role selects the prescribed blocks within this fixed budget (one block per repetition at $h_R$, two at each other central-path node); unused padding is ignored. These budgets do not depend on the validity of the instance.

Each node named $h_L$ or $w_j$ additionally receives the
port leading to its claimed successor on the path \(h_L,w_1,\dots,w_{\rho-1},h_R.\)
 This port number uses $O(\log n)$ bits and allows the node
to send its quantum registers before learning its
neighbors' names.

\paragraph*{Verification}
Before communication, every node measures the classical
part of its certificate in the computational basis.
It first checks the local part of (V1), its successor port when applicable, and that the active register format has exactly the number and widths of blocks prescribed for its measured role. If a check fails, it records rejection and applies no dimension-dependent quantum operation. Well-formed nodes proceed with their locally prescribed blocks; rejecting nodes may send dummy quantum messages.
During the communication round, every node sends its
measured classical certificate, including its successor
port when applicable, to each neighbor.
At the same time, the nodes named $h_L$ or $w_j$ perform the
random swaps prescribed by the equality protocol (\Cref{protocol-equality}) and
send their quantum registers through their
claimed successor ports.
These operations require neither $\hat z$ nor $\hat z'$.

After receiving the messages, every node performs
checks (V1)--(V3) from the structural certification from \Cref{lem:shapeT}.
Every node records the local port on which each incoming classical and quantum message arrives. Each sender checks that the classical certificate received on its designated successor port carries the required successor name. Each receiver in $\{w_1,\dots,w_{\rho-1},h_R\}$ identifies the unique port carrying its predecessor's classical certificate, checks that the expected quantum registers arrived on that same port, and uses those registers in its quantum test.

Before applying a SWAP or endpoint test, each receiver also checks that the incoming quantum block count and widths agree with $m_\rho$ and $\ell(n)$. A failed classical, routing, or dimension check causes rejection without applying that test. If all these verifications pass, the structural certification
identifies a single central path, and the quantum
registers have been transmitted along that path as
required by the equality protocol.

The roots $h_L$ and $h_R$ can now reconstruct their
respective strings.
Indeed, $h_L$ receives exactly one pair
$(i,\hat r_i^L)$ for each $i\in\{0,\dots,\hat N/2\}$,
and reconstructs
\[
\hat z
=
0^{\hat r_0^L}\,1\,0^{\hat r_1^L}\,1\cdots
1\,0^{\hat r_{\hat N/2}^L}.
\]
The root $h_R$ reconstructs $\hat z'$ in the same way.
The structural verification ensures that these run lengths
agree with the actual gadgets.
Although each root receives many classical messages,
every individual message has only $O(\log n)$ bits,
and local computation and storage are unrestricted.

The path vertices then complete the local tests of the
equality protocol, using $|h_{\hat z}\rangle$ and $|h_{\hat z'}\rangle$
for the projection of the endpoints.
Therefore, the roots learn their strings before computing
the only tests that depend on them.
Every node accepts if and only if all its structural,
routing, and quantum checks accept.

\paragraph{Completeness.}
Suppose $G\in\treeq_\mu$, and choose an isomorphism
$G\cong T_{N,\mu}^{\mathrm{tr}}(z,z)$.
The prover supplies the corresponding structural
certificates and successor ports, together with the
honest quantum certificates for equality of $z$ with
itself.
All structural and routing checks pass.
The roots reconstruct $z$, and perfect completeness
of the equality protocol ensures that every quantum
test accepts with certainty.

\paragraph{Soundness.}
Suppose $G\notin\treeq_\mu$, and consider any joint
quantum certificate supplied by the prover.
Condition on the outcomes of all initial measurements
of the classical certificate parts.

For these fixed outcomes, the structural and routing
checks are deterministic.
If any of them fails, some node rejects with certainty.
Otherwise, \Cref{lem:shapeT} gives
\[
G\cong T_{\hat N,\mu}^{\mathrm{tr}}(\hat z,\hat z'),
\]
where $\hat z$ and $\hat z'$ are the strings reconstructed
by the two roots.
These strings must be distinct: if they were equal,
this isomorphism would imply $G\in\treeq_\mu$.

Conditioned on the classical outcomes, the quantum
registers on the central path may still be in an
arbitrary joint state, including a state entangled
across nodes and repetitions.
The soundness guarantee of \Cref{thm:equality-uniform} applies
to every such state.
Since the routing checks ensure that the registers
follow the prescribed path, the probability that all
quantum tests accept is at most $1/3$.
Averaging over the classical measurement outcomes
proves the same bound for the original certificate.

\paragraph{Complexity.}
The structural certificates and successor ports use
$O(\log n)$ classical bits per node.
The equality protocol uses $O(\rho)$ repetitions,
with padded $O(\log n)$-qubit registers, giving $O(\rho\log n)$ certificate qubits in the fixed budget and qubits per path-edge message. Only the central-path quantum slots are used; the remaining active fields and messages are classical. Since $\rho=2\mu+1$, the maximum certificate size, including padding, and individual message size are
\[
O(\log n+\rho\log n)=O(\mu\log n),
\]
with a universal implicit constant.
\end{proof}

Now we prove the classical lower bound of \Cref{thm:treeq-separation}.

\begin{theorem}\label{thm:treeC}
For every fixed integer $\mu\ge1$, any $\mu$-round $\dMA$
protocol for $\treeq_\mu$ with completeness $c$ and soundness error $s$ such that $2c-1>s$ has worst-case
certificate size $\Omega(n/(\mu\log^2 n))$ bits. The implicit constant is independent of $\mu$.
\end{theorem}

\begin{proof}
Fix $\mu\ge1$ and a $\mu$-round $\dMA$ protocol $\Pi$
for $\treeq_\mu$, with completeness $c$ and soundness
error $s$ satisfying $2c-1>s$.
Recall that the central path has length $\rho=2\mu+1$.
We follow the fooling-set argument of Fraigniaud
et al.~\cite[Theorem~9]{fraigniaud2021distributed}.

Fix an even integer $N\ge8$.
By \Cref{fact:treecount}, all trees
$T_{N,\mu}^{\mathrm{tr}}(z,z')$ encoding $N$-bit strings $z$ and $z'$ with Hamming weight $N/2$
have the same number $n$ of vertices, regardless of the strings $z$ and $z'$.
Let $b:=b(n)$ be the worst-case certificate size of $\Pi$
on $n$-vertex instances.

In this argument, the main idea is to ``combine'' two instances $T_{N,\mu}^{\mathrm{tr}}(z,z)$ and $T_{N,\mu}^{\mathrm{tr}}(t,t)$ into a new instance $T_{N,\mu}^{\mathrm{tr}}(z,t)$, taking the left tree from the first instance and the right tree from the second. In order to apply the argument, we need to choose identifiers and ports consistently so that, when the certificates are combined as described below, the nodes in each tree retain their $\mu$-round views from the corresponding original instance. Agreement of the certificates on the central path will ensure that the appropriate views are also preserved at the internal path vertices.

We therefore assign the identifiers and ports as follows. Given an instance $T_{N,\mu}^{\mathrm{tr}}(z,z')$,
both the left and right trees have the same number of vertices for every
string (with Hamming weight $N/2$): its tags are fixed, and the total number
of pendant leaves encoding the string is $N/2$.
We can therefore reserve disjoint identifier intervals
for the two sides and the internal path vertices, using
$O(n)$ identifiers in total as follows: First, we can keep the identifiers of $h_L,h_R$ and all $w_j$ fixed. Second, within each side, enumerate the vertices other than its root in a fixed order determined by the encoded string,\footnote{Order the gadgets by increasing $i$. Within each gadget, list $a_i$, its pendant leaves, $s_i$, $t_i$, and then the tag arms by increasing $b$. List the vertices of each arm by increasing $p$, followed by its extra leaves, if any. Leaves sharing a parent may be listed in any order, since they have identical neighborhoods.} and assign them the remaining identifiers from the interval reserved for that side in increasing order.
Third, choose the internal port numbering of each side in the
same way, and keep the ports along the central path,
including its attachments to $h_L$ and $h_R$, fixed.
Consequently, whenever the same string occurs on the
same side of two instances, that side has identical
identifiers and ports.
All common parameters, including $n$, are also unchanged.

For each balanced string $z\in\{0,1\}^N$, denote by $\Gamma_z$ a certificate assignment under which all nodes of $T_{N,\mu}^{\mathrm{tr}}(z,z)$ accept with probability at least $c$ in $\Pi$. Such an assignment exists by completeness.

The following claim states that different balanced strings cannot produce the same certificates in the nodes of the central path. Recall that by construction, the central path has the same length (for a fixed $\mu$) and the identifiers and ports are the same.

\begin{claim}\label{claim:accept-cert}
    Let $z,t\in\{0,1\}^N$ be two different balanced strings. Then 
    \[
\bigl(
\Gamma_z(w_1),\dots,\Gamma_z(w_{2\mu})
\bigr)\neq \bigl(
\Gamma_{t}(w_1),\dots,\Gamma_{t}(w_{2\mu})
\bigr),
\]
where $w_1,\dots, w_{2\mu}$ are the nodes of the central path of both $T_{N,\mu}^{\mathrm{tr}}(z,z)$ and $T_{N,\mu}^{\mathrm{tr}}(t,t)$.
\end{claim}

\begin{proof}[Proof of Claim~\ref{claim:accept-cert}]

Suppose that two distinct strings $z,t$ give the same
tuple in the central path.
Consider $T_{N,\mu}^{\mathrm{tr}}(z,t)$, which is a no-instance
by \Cref{lem:isoT}.
Partition its vertices into a left group consisting of
the entire left side and $w_1,\dots,w_\mu$, and a right
group consisting of all remaining vertices.
Give the left group its certificates from $\Gamma_z$
and the right group its certificates from $\Gamma_t$.
On $w_1,\dots,w_{2\mu}$, the two assignments agree
by assumption.

Every vertex of the left group has the same
radius-$\mu$ neighborhood, including identifiers, ports,
and certificates, as in
$T_{N,\mu}^{\mathrm{tr}}(z,z)$ with $\Gamma_z$.
Indeed, the right root is at distance at least
$\rho-\mu=\mu+1$ from this group.
Any internal path vertex visible across the cut has
index at most $2\mu$, so its certificate belongs to
the common tuple.

Similarly, every vertex of the right group has the
same radius-$\mu$ neighborhood as in
$T_{N,\mu}^{\mathrm{tr}}(t,t)$ with $\Gamma_t$.
The left root is at distance at least $\mu+1$ from
this group, and every certificate visible across the
cut belongs to the common block.
The side structures and their identifiers and ports
agree by construction.

Couple the mixed execution and the two diagonal
executions using the same public random tape and the
same private tape at each corresponding identifier.
Each individual execution still has the prescribed
distribution of public and private randomness.
Since information crosses at most one edge per round,
the matching neighborhoods imply matching decisions
after $\mu$ rounds.

The probability that all vertices of the left group
accept in the mixed execution is therefore at least $c$:
whenever all vertices accept in the first diagonal
execution, this group accepts as well.
The same reasoning applies to the right group using
the second diagonal execution.
Without assuming independence between these events,
\[
\Pr\!\left[
\text{all vertices accept on }
T_{N,\mu}^{\mathrm{tr}}(z,t)
\right]
\ge 1-(1-c)-(1-c)
=2c-1>s.
\]
This contradicts soundness and proves that the tuples
are distinct.

\end{proof}

There are $\binom{N}{N/2}$ balanced strings, and a binary certificate of length at most $b$ has
\[
\sum_{i=0}^{b}2^i=2^{b+1}-1
\]
possible values, including the empty string.
Thus the injectivity proved in Claim~\ref{claim:accept-cert} gives
\[
\binom{N}{N/2}
\le (2^{b+1}-1)^{2\mu}
<2^{2\mu(b+1)},
\]
and hence
\[
b>
\frac{1}{2\mu}\log_2\binom{N}{N/2}-1.
\]

The central binomial coefficient is the largest term
in the binomial expansion of $2^N$, so
\[
2^N
=\sum_{i=0}^{N}\binom{N}{i}
\le (N+1)\binom{N}{N/2}.
\]
Consequently,
\[
b>
\frac{N-\log_2(N+1)}{2\mu}-1.
\]
Since $\mu$ is fixed, this gives $b=\Omega(N/\mu)$
as $N$ grows.

Finally, \Cref{fact:treecount} gives
$n=\Theta(N\log^2N)$, and therefore
$\log n=\Theta(\log N)$ and
$N=\Theta(n/\log^2n)$.
It follows that
\[
b:=b(n)=\Omega\!\left(\frac{n}{\mu\log^2n}\right).
\]
The implicit constants are universal: the only requirement for absorbing the additive term in the counting bound is $N/\mu\to\infty$.
\end{proof}

\paragraph{Growing round budgets.}
The same construction and estimates apply uniformly to growing known $\mu$ along the constructed instance sizes. Indeed, $n=n_\mu(N)=\Theta(N\log^2N)+2\mu$. If $\mu=o(n/\log^2n)$, then $2\mu=o(n)$, $N=\Theta(n/\log^2n)$, and $\mu=o(N)$. The counting bound still gives $\Omega(n/(\mu\log^2n))$ bits, while the quantum budget remains $O(\mu\log n)$ qubits. Thus these cost bounds extend to this range of parameterized instances.

\section{Graph Symmetry}\label{sec:sym}

Our second separation concerns graph symmetry, the existence of a nontrivial automorphism. Let $\sym$ be the class of connected graphs having such an automorphism. For a known integer $D\ge1$, we also consider the restricted symmetry class
\[
\symk:=\{G\in\sym:\operatorname{rad}(G)\le D\}.
\]
As usual in local certification, we assume that the input graph is connected, but the radius condition is part of the property being certified for $\symk$. We use in our argument two implications of the condition $\operatorname{rad}(G)\le D$: Every instance of $\symk$ has diameter at most $2D$ and a vertex of eccentricity at most $D$.

The quantum protocols certify a proposed automorphism by comparing strings that describe the neighborhoods of vertices and their proposed images. We use the distance-independent equality protocol for $\sym$, and the fingerprint protocol for $\symk$, where the comparisons can follow short paths. For the classical lower bounds, we construct a family in which a nontrivial automorphism exists exactly when two attached graphs are isomorphic.

The reduction runs one comparison of $N_*=\Theta(n\log n)$ bits for each vertex, along a simple path of length at most $n-1$. Up to $n$ such paths can cross one vertex or edge. Applying the $O(r\log N_*)$ fingerprint cost to paths with $r=\Theta(n)$ therefore gives only a $\widetilde O(n^2)$ worst-case bound under this congestion. The relay-based approach of~\cite{hasegawa2024power} likewise retains a linear-in-$N_*$ certificate bound at relay vertices and does not give a subquadratic guarantee from this composition.

Protocol~\ref{new-protocol-equality} instead uses $\widetilde O(N_*^{2/3})$ qubits at every path vertex, independently of the path length. Multiplying by the congestion gives $\widetilde O(n^{5/3})$. For radius at most $D$, comparison paths can have length at most $2D$, and Protocol~\ref{protocol-equality} gives $O(D\log n)$ per path and $O(nD\log n)$ in total.

Both compositions fit in one round because all pre-communication quantum operations and outgoing registers are independent of the endpoint strings. Classical headers and quantum registers can be sent together; the endpoints construct their neighborhood strings upon reception and only then perform the input-dependent tests. We use this property explicitly below.

\subsection*{Classical certificates and local verifications}

We first describe the classical part of our protocols.
Its purpose is to certify the paths along which the equality
protocols will run and to construct the strings compared at
their endpoints. The difference between the protocol for $\sym$ and $\symk$ is minimal, since for the latter we only need to give a few extra certificates, and the radius condition will imply the improved complexity; therefore, we present both protocols, proving \Cref{thm:quan-upper-sym} and \Cref{cor:quan-upper-symk} together.

Fix a connected $n$-vertex graph $G$, with $n\ge2$, and denote by
$U$ the known identifier range, where $U=n^{O(1)}$.
We use the known $n$ and $U$ convention from \Cref{sec:definition}, so all string lengths and quantum register budgets below are determined before communication. If $n$ is not provided, the classical scaffold can certify it by adding a common claimed order and subtree-size fields to (L2): neighbors agree on the claim, each vertex checks one plus the sizes of children pointing to it, and the root checks that its size equals the common claim. These are $O(\log n)$-bit fields on an accepting assignment. We analyze the quantum upper bounds with $n$ and $U$ known, as stated in the model.

Define also for every $n$-node graph $G$ the function
\[
h(G):=
\begin{cases}
n-1,&\text{for }\sym,\\
\min\{2D,n-1\},&\text{for }\symk.
\end{cases}
\]

The following lemma states the guarantees of the classical verification.
Once these classical verifications pass, verifying the remaining string
equalities using our quantum protocol from \Cref{thm:equality-uniform} will certify a nontrivial automorphism.

\begin{lemma}\label{lem:sym-classical}
There is a deterministic one-round $\dMA$ protocol using
$O(n\log n)$ certificate bits per node
with the following guarantees.

If all nodes accept, the certificates define
functions $\pi,\sigma:V(G)\to[U]$, a spanning tree rooted at
a vertex $v^*$ satisfying
$\pi(v^*)\ne\mathrm{id}(v^*)$, and exactly one simple path
$P_v$ starting at each vertex $v$, of length at most $h$, and at most
$n$ paths pass through any vertex.
For $\symk$, acceptance also guarantees
$\operatorname{rad}(G)\le D$.

By the end of the verification round, every vertex $v$ constructs strings
$X_v,Y_v$ of length $O(n\log n)$ with the following properties:
\begin{enumerate}
    \item If all nodes accept and
    \(
    X_v=Y_{\tau(v)}\)
    for every $v$,
    where $\tau(v)$ is the endpoint of $P_v$, then $\tau$
    is a nontrivial automorphism of $G$.

    \item If $G\in\sym$, or $G\in\symk$ for the
    bounded-radius version, there is a certificate assignment
    under which all nodes accept and all these equalities hold.
\end{enumerate}
\end{lemma}

\begin{proof}
We describe the certificates and their verification first.

\paragraph*{Certificates}
Each node $v$ receives the following classical data.
\begin{description}
    \item[(L1)]
    For $\symk$ only, the certificate contains a root
    identifier $\mathrm{rt}_\ell$ and a label
    \[
    \ell(v)\in\{0,\dots,\min\{D,n-1\}\}.
    \]

    \item[(L2)]
    The certificate contains a root identifier $\mathrm{rt}$,
    a depth $d(v)\in\{0,\dots,n-1\}$, and a parent port,
    or $\bot$ at depth zero.\footnote{This is a standard spanning-tree certification scheme (see, e.g.,~\cite{korman2010proof}).
We include its short description and verification
for completeness.}

    \item[(L3) ]
    The certificate contains two identifier values
    $\pi(v),\sigma(v)\in[U]$, intended to identify the images
    of $v$ under an automorphism and its inverse.

    \item[(L4) ]
    The certificate contains at most $n$ path records, with
    at most one record for each path identifier $c\in[U]$.
    A record has the form
    \[
    (c,t,j,\mathrm{pred},\mathrm{succ}),
    \qquad 0\le j\le t\le h.
    \]
    Here $t$ is the claimed path length and $j$ is the
    node's position.
    Each predecessor or successor field is either $\bot$
    or a pair consisting of a local port and a claimed
    neighbor identifier.

    A record has $\mathrm{pred}=\bot$ if and only if $j=0$,
    and $\mathrm{succ}=\bot$ if and only if $j=t$.
    A position-zero record must have $c=\mathrm{id}(v)$,
    and every node must have such a record for its own
    identifier.
\end{description}
All fields use $O(\log n)$ bits.
Malformed certificates are rejected.

\paragraph*{Verification}
Every node sends its certificate and identifier to
each neighbor.
Each message also specifies the local port through which
it is sent, allowing reciprocal port pointers to be checked
after reception.
The nodes perform the following verifications.
\begin{description}
    \item[(L1)]
    For $\symk$, neighboring nodes must report the same
    $\mathrm{rt}_\ell$.
    Each node checks that $\ell(v)=0$ if and only if
    $\mathrm{id}(v)=\mathrm{rt}_\ell$, and that a node
    with positive label has a neighbor whose label is
    one smaller.

    \item[(L2)]
    Neighboring nodes must report the same $\mathrm{rt}$.
    Each node checks that $d(v)=0$ if and only if
    $\mathrm{id}(v)=\mathrm{rt}$.
    At depth zero its parent must be $\bot$; otherwise,
    its parent port must lead to a neighbor of depth
    $d(v)-1$.
    The depth-zero node additionally checks
    $\pi(v)\ne\mathrm{id}(v)$.

    \item[(L3)]
    Each node checks that $\pi(v)$ and $\sigma(v)$
    lie in $[U]$.

    \item[(L4)]
    Every predecessor or successor port must lead to
    the neighbor with the claimed true identifier.
    A successor must hold a record with the same $c,t$,
    position $j+1$, and a predecessor pointer back to
    the current node.
    A predecessor must similarly hold the record at
    position $j-1$ with a successor pointer back.
    The exchanged port numbers allow these pointers
    to be checked against the actual edge used.
\end{description}
Each node accepts if and only if all the above verifications hold.

After the reception of the certificates, the nodes also construct the strings
that will be compared by the quantum protocol.
Set
\[
\ell_*:=\lceil\log_2(U+n+2)\rceil,
\qquad
N_*:=(n+2)\ell_*\in O(n\log n).
\]
Use $\ell_*$ bits for each integer field.
To encode two identifiers, a degree, and a neighbor list,
sort the list and pad it with zeros to $n-1$ entries.
Including the degree makes this encoding unambiguous.
Define
\[
\begin{aligned}
X_v&:=\operatorname{enc}\bigl(
\mathrm{id}(v),\pi(v),\deg(v),
(\pi(u))_{u\in N(v)}
\bigr),\\
Y_v&:=\operatorname{enc}\bigl(
\sigma(v),\mathrm{id}(v),\deg(v),
(\mathrm{id}(u))_{u\in N(v)}
\bigr).
\end{aligned}
\]
The lists are treated as multisets and sorted as above.
Every string has length $N_*$.
Each node knows its own certificate and degree, and receives
its neighbors' claimed images and true identifiers during
the round.
It can therefore construct both strings by the end of
that round.

\paragraph{Completeness.}
Suppose $G$ belongs to the property under consideration,
and choose a nontrivial automorphism $\alpha$.
Choose a vertex $v^*$ moved by $\alpha$, supply correct
spanning-tree data rooted there, and set
\[
\pi(v):=\mathrm{id}(\alpha(v)),
\qquad
\sigma(v):=\mathrm{id}(\alpha^{-1}(v)).
\]
For each $v$, choose a shortest path from $v$ to $\alpha(v)$
and supply its records and ports.
These paths are simple and have length at most $n-1$.

For $\symk$, additionally choose a vertex of eccentricity
at most $D$ and provide its true distance labels for (L1).
The diameter is then at most $2D$, so every chosen shortest
path has length at most $\min\{2D,n-1\}=h$.

All classical checks pass.
Moreover, $\tau(v)=\alpha(v)$.
The inverse identities and preservation of degrees and
neighborhoods give
\[
X_v=Y_{\alpha(v)}=Y_{\tau(v)}
\qquad\text{for every }v.
\]

\paragraph{Soundness.}
Suppose all nodes accept.

For $\symk$, connectedness and (L1) make
$\mathrm{rt}_\ell$ common to all nodes.
A node of minimum label must have label zero, since otherwise
it would need a neighbor with smaller label.
Distinct identifiers make this node unique.
Following decreasing labels reaches it from every node
within $D$ steps.
Thus $\operatorname{rad}(G)\le D$.

The same argument for (L2) gives a unique depth-zero
vertex $v^*$.
The spanning-tree construction is standard and known to correctly construct a spanning tree if all the verification holds \cite{korman2010proof}.
The pointers, therefore, form a spanning tree.
The root check gives
$\pi(v^*)\ne\mathrm{id}(v^*)$.

We next show that the certificates describe simple paths.
Start at a vertex $v$ with position zero in its certificate
and follow the successor pointers.
The checks ensure that successive vertices report the same
path length $t$, while their positions increase by one.
At position $t$, the successor is $\bot$, so the path ends.
A vertex cannot appear twice, since its certificate would
then have to specify two different positions on the same
path, which is not allowed.
Thus, this is a simple path of length $t\le h$.

Conversely, consider any vertex with a certificate for a
path with identifier $c$, and follow the predecessor pointers.
Positions decrease by one until reaching position zero.
The vertex at this position must have identifier $c$,
so the starting vertex is unique.
The predecessor and successor pointers agree at every step,
so following successors from this starting vertex leads
back to the original vertex.
Thus, every certificate for $c$ describes a position on
this same path.
Since each certificate specifies at most one successor,
the path cannot branch.
There is exactly one path for each starting vertex,
so at most $n$ paths pass through any vertex or edge.

Finally, suppose that $X_v=Y_{\tau(v)}$ for every $v$.
The first fields give
\[
\sigma(\tau(v))=\mathrm{id}(v).
\]
If $\tau(v)=\tau(u)$, these equalities imply
$\mathrm{id}(v)=\mathrm{id}(u)$, and hence $v=u$.
Thus $\tau$ is injective and, since $V(G)$ is finite,
bijective.

The second fields give
$\pi(v)=\mathrm{id}(\tau(v))$.
Equality of the degrees and neighbor lists gives
\[
\{\pi(u):u\in N(v)\}
=
\{\mathrm{id}(z):z\in N(\tau(v))\}
\]
as multisets.
Substituting the second-field equalities and using distinct
identifiers yields
\[
\tau(N(v))=N(\tau(v))
\qquad\text{for every }v.
\]
By injectivity, $u\in N(v)$ if and only if
$\tau(u)\in N(\tau(v))$.
Therefore $\tau$ is an automorphism.
Furthermore,
\[
\mathrm{id}(\tau(v^*))=\pi(v^*)\ne\mathrm{id}(v^*),
\]
so it is nontrivial.

\paragraph{Complexity.}
Each path record contains a constant number of
$O(\log n)$-bit fields, and each node holds at most $n$
records.
The remaining certificate fields use $O(\log n)$ bits.
Thus the certificate size is $O(n\log n)$ bits per node.

Every node sends these data, its true identifier, and
the sending port to each neighbor.
Each individual message therefore has $O(n\log n)$ bits.
The strings are constructed locally after reception and
require no additional communication.
\end{proof}

\subsection{Quantum upper bounds for symmetry}

We prove \Cref{thm:quan-upper-sym}, which we re-state below.
\quantumsymmetry*
\begin{proof}
We combine the classical certification above with one equality
protocol along each certified path. As in the protocol for
$\treeq_\mu$, the endpoint strings are learned during the
communication round, before the final tests that depend on them.

\paragraph*{Certificates}
The prover supplies each node with the classical certificate fields
(L2)--(L4) from \Cref{lem:sym-classical}.
For each positive-length path through a node, it also supplies
separate quantum registers for comparing $X_v$ at the starting
vertex $v$ with $Y_{\tau(v)}$ at the endpoint.

These are the registers of the distance-independent equality
protocol of \Cref{new-protocol-equality}.
Each execution uses the $K$ label-register pairs of \Cref{new-protocol-equality}, with $K$ and $L$ determined solely by the known input length $N_*$. Each node reserves $n$ path slots, each of $2K\lceil\log L\rceil$ qubits, and its at most $n$ measured records select distinct active slots. Unused slots and the slots of length-zero paths are ignored. This gives a fixed $O(nN_*^{2/3}(\log(N_*+2))^{5/3})$-qubit budget even on no-instances. A local record $(c,t,j,\mathrm{pred},\mathrm{succ})$ determines whether the node acts as the starting vertex ($j=0$), an internal vertex ($0<j<t$), or the endpoint ($j=t$), and determines the outgoing port. It does not determine a repetition count in this protocol.

\paragraph*{Verification}
Before communication, every node measures the classical part
of its certificate in the computational basis.
It checks every record's local ranges and ports and its prescribed quantum register format before any pre-communication quantum operation. A malformed record or an incompatible block count or width causes rejection, and no dimension-dependent operation is applied to that slot. Valid slots execute strictly according to their own measured records, before neighboring records are known.
In the single communication round, it sends its measured
classical data and true identifier to every neighbor,
including the sending port as described above.

At the same time, it performs the operations prescribed before
communication by each equality execution and sends the
designated quantum registers through that path's successor port.
These operations are independent of $X_v$ and $Y_{\tau(v)}$.
Registers and random coins for different paths are separate,
and no quantum register is copied.
A node that records a quantum rejection still carries out the
remaining prescribed operations and communication, and rejects
at the end.

After reception, every node performs verification checks (L2)--(L4). In particular, (L4) deterministically detects any discrepancy between adjacent records in the path identifier, length, positions, or predecessor/successor data.
It also checks that the quantum registers prescribed for each
path arrived through its predecessor port, with the required
sizes and path identifiers.
A failed structural, routing, or register-dimension check causes rejection without applying the affected final quantum test.
If all such checks pass, \Cref{lem:sym-classical} identifies
the paths, and the quantum registers have followed them
as required.

The nodes now construct their strings from the received
classical data.
On each positive-length path, the endpoints perform the
remaining input-dependent tests with $X_v$ and $Y_{\tau(v)}$,
and the internal vertices complete their local tests.
On a length-zero path, its sole vertex checks $X_v=Y_v$
directly.
Each node accepts if and only if all its classical, routing,
and quantum checks accept.
All operations fit in one communication round.

\paragraph{Completeness.}
Suppose $G\in\sym$, and choose a nontrivial automorphism $\alpha$.
Choose a vertex $v^*$ moved by $\alpha$, supply correct
spanning-tree data rooted there, and set
\[
\pi(v):=\mathrm{id}(\alpha(v)),
\qquad
\sigma(v):=\mathrm{id}(\alpha^{-1}(v)).
\]
For each $v$, choose a shortest path from $v$ to $\alpha(v)$
and supply its correct certificates and ports.
Its length is at most $n-1$.
All classical checks pass, and the automorphism identities
give $X_v=Y_{\alpha(v)}$ for every $v$.
Honest quantum certificates make every positive-length
comparison accept with certainty by perfect completeness
of the equality protocol.
Length-zero comparisons also accept.

\paragraph{Soundness.}
Consider a no-instance of $\sym$ and any joint quantum
certificate.
Condition on all outcomes of the initial measurements of
the classical parts.
For these fixed outcomes, the classical and routing checks
are deterministic.
If any fails, some node rejects with certainty.

Otherwise, the certified paths and $\tau$ are well defined
by \Cref{lem:sym-classical}.
Since $G$ is asymmetric, the same lemma implies that some
vertex $v$ satisfies $X_v\ne Y_{\tau(v)}$.

If this path has length zero, its local comparison rejects
with certainty.
Otherwise, consider the registers and tests assigned to
this path.
Conditioned on the classical outcomes, their initial reduced
state is arbitrary: it may be mixed and entangled across
nodes and repetitions, and with the registers of other paths.
The equality protocol's soundness guarantee applies to every
such state, so the probability that all tests on this path
accept is at most $1/3$.

The prescribed operations on other paths act on separate
registers and do not control the operations on this path.
Ignoring their outcomes therefore leaves exactly the
marginal execution just analyzed.
Since acceptance of the network requires this path to accept,
the network accepts with probability at most $1/3$.
Averaging over the initial classical outcomes gives the same
bound for the original certificate.

\paragraph{Complexity.}
Each node receives $O(\log n)$ certificate bits for each
of the at most $n$ paths passing through it.
The remaining classical data use $O(\log n)$ bits per node.
Thus, each classical certificate and each individual
classical message have size $O(n\log n)$ bits.

The equality protocol uses
\[
O\!\left(N_*^{2/3}(\log(N_*+2))^{5/3}\right)
=
O\!\left(n^{2/3}(\log n)^{7/3}\right)
\]
certificate qubits at each vertex of a path and qubits
per message on each of its edges.
By \Cref{lem:sym-classical}, at most $n$ paths pass through
any vertex or edge.
Including the classical certificates gives
$O(n^{5/3}(\log n)^{7/3})=\widetilde O(n^{5/3})$
certificate qubits per node and qubits per message.
\end{proof}

We next consider $\symk$, where the graph is also required
to have radius at most $D$.
The classical certification of \Cref{lem:sym-classical}
verifies this bound and limits the comparison paths
to length at most $2D$.
We can therefore use the $\bigo{r\log N}$-qubit equality protocol
of \Cref{thm:equality-uniform} along these paths,
obtaining the following improved upper bound.

\quantumsymmetrybounded*

\begin{proof}
We use the construction of \Cref{thm:quan-upper-sym},
including the radius certificate fields and verification checks (L1) from
\Cref{lem:sym-classical}, and requiring all certified paths
to have length at most $h=\min\{2D,n-1\}$.
On each positive-length path, we replace the distance-independent protocol by the $\bigo{r\log N}$-qubit equality protocol of \Cref{thm:equality-uniform}. All positive-length paths use the common repetition count
\[
m_h:=\lceil70(h+1)\ln3\rceil,
\]
which is at least the count required for any $1\le t\le h$. Fix a common fingerprint width $\ell_{\mathrm{fp}}=O(\log(N_*+2))=O(\log n)$. A node reserves $n$ slots of at most $2m_h\ell_{\mathrm{fp}}$ qubits each, with fingerprints padded to this width. Its local measured record fixes its role and successor port; terminal roles use only one block per repetition and ignore padding. Before communication it validates the local record and block format, then performs only the operations prescribed by that record. After communication, verification check (L4) confirms agreement of the records, and incoming widths and block counts are checked before any SWAP test. A failure causes rejection without an incompatible quantum operation. Thus inconsistent claims cannot invalidate the pre-communication schedule or increase its register budget.
Its operations before receiving messages do not depend on
the endpoint strings, so the same scheduling argument applies:
the endpoints construct their strings during the communication
round and then perform the tests that depend on them.
Length-zero paths are checked directly as before.

\paragraph{Completeness.}
Suppose $G\in\symk$.
Choose a nontrivial automorphism and the corresponding
classical certificates as in \Cref{lem:sym-classical},
using true distance labels from a center for (L1).
Since every vertex is within distance $D$ of this center,
any two vertices are at distance at most $2D$.
Thus the shortest paths to the automorphism images have
length at most $h$.
All classical verifications pass, and all endpoint strings agree.
Perfect completeness of the fingerprint equality protocol
therefore makes every node accept with certainty.

\paragraph{Soundness.}
Condition on the initial measurements of the classical
certificates, as in the proof of \Cref{thm:quan-upper-sym}.
If a classical or routing check fails, some node rejects
with certainty.
Otherwise, \Cref{lem:sym-classical} guarantees
$\operatorname{rad}(G)\le D$.
A no-instance must therefore be asymmetric, and the same
lemma gives a path with unequal endpoint strings.
A length-zero comparison rejects directly.
For a positive-length path, \Cref{thm:equality-uniform}
bounds its acceptance probability by $1/3$.
The argument of \Cref{thm:quan-upper-sym} applies also here,
including when certificates are entangled across paths.
Averaging over the classical outcomes gives soundness
error at most $1/3$.

\paragraph{Complexity.}
The classical certificates and individual classical messages
use $O(n\log n)$ bits by \Cref{lem:sym-classical}.
The common count $m_h=O(h)$ gives $O(h\log(N_*+2))=O(D\log n)$ certificate qubits per reserved path slot and qubits per path-edge message, and
at most $n$ paths pass through any vertex.
Including the classical certificates therefore gives
$O(nD\log n)$ certificate qubits per node and qubits per individual message, including the unused-slot padding.
\end{proof}

\subsection{A hard family for symmetry}\label{sec:symLB}

We use one family to prove both classical lower bounds of \Cref{thm:BC}. The instances will be graphs denoted as $B_\rho(H, H')$, consisting of two asymmetric graphs $H$ and $H'$ joined by a central path of length $\rho$.

Technically, the two graphs will have a minimum degree at least $3$, so an automorphism must preserve the degree-$2$ internal vertices of the central path. This will ensure that a nontrivial automorphism exists exactly when the two attached graphs agree, and we will apply a fooling-set argument. Before describing the lower bound argument, we define these graphs formally and state some useful properties. 

Given a positive integer $k$, consider the isomorphism classes
of connected asymmetric graphs on $k$ vertices with diameter
at most $2$ and minimum degree at least $3$. Let $\mathcal F_k$ denote a family containing exactly one
representative from each of these classes.
For each $H\in\mathcal F_k$, fix an ordering of $V(H)$
and let $a_H$ denote its first vertex. The following Fact holds.

\begin{fact}\label{fact:count}
As $k\to\infty$, we have $\log_2|\mathcal F_k|\ge(1-o(1))k^2/2$.
\end{fact}

\begin{proof}
Let $G\sim G(k,1/2)$ be an Erd\H{o}s--R\'enyi random graph,
i.e., its vertex set is $\{1,\dots,k\}$ and each possible
edge is present independently with probability $1/2$.
We first show that, with probability $1-o(1)$ as
$k\to\infty$, $G$ is isomorphic to a graph in
$\mathcal F_k$, and then use this estimate to count
the corresponding isomorphism classes.

We begin with connectivity and diameter.
For two distinct vertices $u,v$, each other vertex $\omega\neq u,v$ is
adjacent to both with probability $1/4$, independently
of the others.
The probability that $u,v$ have no common neighbor is
therefore $(3/4)^{k-2}$.
By a union bound, the probability that there exists a pair of vertices with no
common neighbor is at most $\binom{k}{2}(3/4)^{k-2}=o(1)$.
Otherwise, every pair is joined by a path of length at
most $2$, so $G$ is connected and has diameter at most $2$.

For the minimum-degree condition, a fixed vertex has degree
less than $3$ exactly when $0$, $1$, or $2$ of its $k-1$
possible incident edges are present.
Counting these choices and applying a union bound over
the vertices gives
\[
\Pr[\text{some vertex has degree less than }3]
\le
k\,2^{-(k-1)}
\left(1+(k-1)+\binom{k-1}{2}\right)
=o(1).
\]
Finally, $G$ is asymmetric with probability $1-o(1)$
by~\cite{erdos1963asymmetric}. Combining these three conditions, another union bound shows that
all the required properties hold simultaneously with
probability $1-o(1)$.
Thus, with this probability, $G$ is isomorphic to a
member of $\mathcal F_k$.

To turn this probability estimate into a count, observe
that all $2^{\binom{k}{2}}$ labeled graphs on
$\{1,\dots,k\}$ are equally likely.
Hence at least $(1-o(1))2^{\binom{k}{2}}$ of them satisfy
the requirements.
Each isomorphism class contains at most $k!$ such graphs,
since there are only $k!$ ways to assign the vertex labels.
Dividing by this number gives a lower bound on
$|\mathcal F_k|$.
Taking logarithms and using
$\log_2(k!)\le k\log_2 k=o(k^2)$, we obtain
\[
\log_2|\mathcal F_k|
\ge
\binom{k}{2}-\log_2(k!)+\log_2(1-o(1))
=
(1-o(1))\,\frac{k^2}{2}.
\]
\end{proof}
\begin{definition}\label{def:sym-family}
Given $H,H'\in\mathcal F_k$ and $\rho\ge3$, the graph $B_\rho(H,H')$ consists of disjoint copies of $H$ and $H'$ whose distinguished vertices are joined by the central path \(a_H,p_1,\cdots,p_{\rho-1},a_{H'}\) of $\rho$ edges. Its number of vertices is $n=2k+\rho-1$.
\end{definition}

We can characterize when each graph $B_\rho(H,H')$ belongs to $\sym$.

\begin{lemma}\label{lem:pin}
For $H,H'\in\mathcal F_k$ and $\rho\ge3$, the graph $B_\rho(H,H')$ has a nontrivial automorphism if and only if there is an isomorphism $H\to H'$ mapping $a_H$ to $a_{H'}$. In particular, $B_\rho(H,H')\in\sym$ if and only if $H=H'$.
\end{lemma}

\begin{proof}
A rooted isomorphism extends to a nontrivial automorphism by applying the isomorphism and its inverse on the two copies and reversing the central path.

Conversely, let $\varphi$ be an automorphism of $B_\rho(H,H')$. Its degree-$2$ vertices are exactly $p_1,\dots,p_{\rho-1}$, since every vertex of either copy has degree at least $3$. Thus $\varphi$ preserves this set. Its induced graph is a path, so $\varphi$ either fixes its vertices pointwise or reverses their order.

If it fixes them pointwise, it fixes both attachment vertices: $a_H$ is the neighbor of $p_1$ outside the path interior, and similarly for $a_{H'}$. Removing the path interior leaves the two connected copies, each of which must then be preserved. Their asymmetry forces the restriction of $\varphi$ to each copy to be the identity, so $\varphi$ is the identity on the entire graph.

If it reverses the path, it exchanges the attachment vertices and restricts to an isomorphism $H\to H'$ mapping $a_H$ to $a_{H'}$. Distinct representatives in $\mathcal F_k$ are not isomorphic, while for $H=H'$ the identity between the two copies preserves their distinguished vertices. This proves the characterization.
\end{proof}

Finally, the following direct observation holds, which implies that $B_\rho(H,H')$ belongs to $\symk$ when it belongs to $\sym$ and the conditions of the below fact hold.
\begin{fact}\label{fact:ecc}
Every graph $B_\rho(H,H')$ has radius at most $\lceil\rho/2\rceil+2$. In particular, if $\rho\in\{2\mu+2,2\mu+3\}$ and $\mu\le D-4$, every such graph has radius at most $D$.
\end{fact}

\begin{proof}
A vertex in the middle of the central path between the nodes $a_H$ and $a_{H'}$ is at distance at most $\lceil\rho/2\rceil$ from both $a_H$ and $a_{H'}$. Every vertex of either graph $H$ or $H'$ is within distance $2$ of its respective vertex $a_H$ or $a_{H'}$, because the copy has diameter at most $2$. Hence this vertex in the middle has eccentricity at most $\lceil\rho/2\rceil+2$. For either value of $\rho$, this is at most $\mu+4\le D$.
\end{proof}

\subsection{Classical lower bounds for symmetry}

Fraigniaud et al.~\cite[Corollary~3.2]{fraigniaud2019merlin}
established an $\Omega(n^2)$ certificate lower bound for
one-round $\dMA$ protocols for unrestricted symmetry.
The following theorem extends this bound to multiple
verification rounds and shows that the hardness persists
under a radius bound.

\classicalsymmetry*

\begin{proof}
Fix such a protocol $\Pi$, for either property. For each sufficiently large $n$, define $k:=\frac{n-\rho+1}{2}$ and
\[
\rho:=
\begin{cases}
2\mu+2,&n\text{ odd},\\
2\mu+3,&n\text{ even}.
\end{cases}
\]
Then $k$ is an integer, and $\mu=o(n)$ gives $k=\Theta(n)$. Consider the $n$-vertex graphs $B_\rho(H,H')$ with $H,H'\in\mathcal F_k$. By \Cref{lem:pin}, these belong to $\sym$ exactly when $H=H'$. For $\symk$, \Cref{fact:ecc} ensures that every graph in the family satisfies the radius condition, so membership in $\symk$ is also equivalent to $H=H'$.

We apply the same fooling-set argument as for $\treeq_\mu$, following~\cite[Theorem~9]{fraigniaud2021distributed}. As there, identifiers and ports must be chosen consistently so that combining the left copy of one instance with the right copy of another preserves the relevant local views.

Assign identifiers $1,\dots,k$ to the left copy in its fixed vertex order, and $k+1,\dots,2k$ to the right copy. In particular, the two attachment vertices always have identifiers $1$ and $k+1$. Give $p_j$ identifier $2k+j$. Fix the central-path ports, including port $1$ at each attachment vertex. At $a_H$ and $a_{H'}$, assign ports $2,\dots,\deg_H(a_H)+1$ and $2,\dots,\deg_{H'}(a_{H'})+1$, respectively, to the neighbors inside their copy in increasing identifier order. Every other vertex $u$ in a copy uses ports $1,\dots,\deg(u)$ in increasing neighbor identifier order. Thus the identifiers and ports within a copy depend only on that copy and its side, and the path identifiers and ports are the same throughout the family. All instances also have the same known parameters, including $n$.

Let $b=b(n)$ be the worst-case certificate size of $\Pi$. For each $H\in\mathcal F_k$, choose a certificate assignment $\Gamma_H$ under which all nodes of $B_\rho(H,H)$ accept with probability at least $c$. Such an assignment exists by completeness, and is fixed before the random coins are sampled.

\begin{claim}\label{claim:sym-certificates}
For distinct $H,H'\in\mathcal F_k$, it holds 
\[
\bigl(\Gamma_H(p_1),\dots,\Gamma_H(p_{2\mu})\bigr)
\neq
\bigl(\Gamma_{H'}(p_1),\dots,\Gamma_{H'}(p_{2\mu})\bigr).
\]
\end{claim}

\begin{proof}[Proof of \Cref{claim:sym-certificates}]
Suppose the tuples agree. Consider the no-instance $B_\rho(H,H')$. Partition its vertices into a \textit{left group} $L$ consisting of the vertices in $H$ plus $p_1,\dots,p_\mu$, and a \textit{right group} $R$ containing the remaining vertices. Give to the nodes in $L$ their certificates from $\Gamma_H$ and to the nodes in $R$ their certificates from $\Gamma_{H'}$. Notice that on $p_1,\dots,p_{2\mu}$, the assignments agree by assumption. The remaining internal vertices $p_{2\mu+1}$ and, when present, $p_{2\mu+2}$ are at distance at least $\mu+1$ from $L$ and lie in $R$ with certificates from $\Gamma_{H'}$, so they need not appear in the common tuple.

Every node in $L$ has the same radius-$\mu$ initial view as in $B_\rho(H,H)$ with $\Gamma_H$. Indeed, the vertex $a_{H'}$ is at distance at least $\rho-\mu\ge\mu+2$ from the nodes in this set, and every certificate on the path visible across the cut has index at most $2\mu$, so belongs to the common tuple. The nodes in $H$ and their identifiers and ports agree by construction. Similarly, by an analogous argument, every node in $R$ has the same radius-$\mu$ initial view as in $B_\rho(H',H')$ with $\Gamma_{H'}$. 

We compare the executions on $B_\rho(H,H')$, $B_\rho(H,H)$,
and $B_\rho(H',H')$ using the same public random bits
in all three and the same private random bits at nodes
with the same identifier.
Each execution still has the prescribed distribution
of randomness.
By locality, every node in $L$ makes the same decision
on $B_\rho(H,H')$ as its counterpart on $B_\rho(H,H)$.
Likewise, every node in $R$ makes the same decision
on $B_\rho(H,H')$ as its counterpart on $B_\rho(H',H')$.

By completeness of $\Pi$, and the fact that $B_\rho(H,H)$ is a yes-instance, the probability that all nodes in $L$ accept in the execution of $B_\rho(H,H')$ is therefore at least $c$, and the same holds for the nodes in $R$. Therefore,
\[
\Pr\!\left[
\text{all nodes accept on }B_\rho(H,H')
\right]
\ge
1-(1-c)-(1-c)
=
2c-1>s.
\]
This contradicts the soundness error $s$ of $\Pi$: $B_\rho(H,H')$ is a no-instance and therefore all the nodes should accept with probability at most $s$.
\end{proof}

A binary certificate of length at most $b$ has $2^{b+1}-1$ possible values, including its length. By \Cref{claim:sym-certificates}, the $|\mathcal F_k|$ instances of the form $B_\rho(H,H)$ require distinct tuples of $2\mu$ such certificates. Hence
\[
|\mathcal F_k|
\le
(2^{b+1}-1)^{2\mu}
<
2^{2\mu(b+1)},
\]
and therefore
\[
b>
\frac{\log_2|\mathcal F_k|}{2\mu}-1.
\]
By \Cref{fact:count}, $\log_2|\mathcal F_k|=\Omega(k^2)$. Since $k=\Theta(n)$ and $\mu=o(n)$, the term $k^2/\mu$ tends to infinity, so the additive constant is negligible. It follows that
\[
b(n)=\Omega(k^2/\mu)=\Omega(n^2/\mu).
\]
The proof applies to both properties $\sym$ and $\symk$.
\end{proof}

Combining \Cref{thm:quan-upper-sym}, \Cref{cor:quan-upper-symk}, and \Cref{thm:BC}, for every fixed classical round budget $\mu\ge1$, graph symmetry has a quantum protocol using $O(n^{5/3}(\log n)^{7/3})$ qubits per node, whereas every classical protocol with the stated completeness and soundness requires $\Omega(n^2)$ certificate bits. For fixed $D\ge5$ and $1\le\mu\le D-4$, the corresponding bounds for $\symk$ are $O(n\log n)$ qubits and $\Omega(n^2)$ bits. The quantum upper bounds also apply to individual message sizes. The bounded-radius comparison also applies to growing $D=D(n)$ and $\mu=\mu(n)=o(n)$ with $D\ge5$ and $\mu\le D-4$. It gives a strict quantum advantage whenever $nD\log n=o(n^2/\mu)$, equivalently $\mu D\log n=o(n)$.

\paragraph*{Acknowledgements} Masayuki Miyamoto was supported by JSPS KAKENHI Grant Number JP25K24383 and MEXT Supporting Pioneering Research through AI for 1,000 Discovery challenges Program (SPReAD) Japan Grant Number JPMXP1726277558.

\paragraph*{AI disclosure}
We used GPT-5.6 Sol to assist with constructing the protocol described in Section~\ref{sec:new-equality-protocol}. ChatGPT with GPT-6 Astra and GPT-6.1 Sol were used to assist with the writing.
The authors verified the correctness and originality of all content, including references, and take full responsibility for them.

\bibliographystyle{alpha}
\bibliography{refs.bib}

\appendix

\section{Proof of Claim~A}
\setcounter{claim}{0}

The following lemma gives an inequality for two adjacent
terms of $H_t$ that we will use to prove Claim~\ref{cl:A}.
 
\begin{lemma}\label{lem:three}
$H_3=(I-\Pi^{1,2})+(I-\Pi^{2,3})$ has no eigenvalue in the open interval $(0,1/2)$. Equivalently,
\[
(I-\Pi^{1,2})(I-\Pi^{2,3})+(I-\Pi^{2,3})(I-\Pi^{1,2})\;\succeq\;-\tfrac12H_3 .
\]
\end{lemma}
 
\begin{proof}
Recall from the proof of Lemma~\ref{lem:ground-space} that $\ket\Omega=L^{-1/2}\sum_a\ket{aa}$ and $\ket{s_{ab}}=(\ket{ab}+\ket{ba})/\sqrt2$, $a<b$, form an orthonormal basis of $W_L$, so that $\Pi_L=\ket\Omega\bra\Omega+\sum_{a<b}\ket{s_{ab}}\bra{s_{ab}}$. Hence
\begin{equation}\label{eq:Pi}
\Pi_L\ket{aa}=L^{-1/2}\ket\Omega=\frac1L\sum_b\ket{bb},\qquad
\Pi_L\ket{ab}=\ket{s_{ab}}/\sqrt2=\tfrac12(\ket{ab}+\ket{ba})\quad(a\neq b).
\end{equation}

To study the eigenvalues of $H_3$, we decompose $(\mathbb C^L)^{\otimes3}$ into
orthogonal subspaces that $H_3$ maps into themselves. For $1\le a<b<c\le L$, define the subspace $E_{abc}$ of
$(\mathbb C^L)^{\otimes3}$ by
\[
E_{abc}=\mathrm{span}\{\ket{abc}, \ket{acb}, \ket{bac}, \ket{bca}, \ket{cab}, \ket{cba}\},
\]
and for an index $k$ define the subspace $E_{k}$ of $(\mathbb C^L)^{\otimes 3}$ by
\[
E_{k}=\mathrm{span}\{\ket{kkk}, \ket{aak}, \ket{aka}, \ket{kaa} \text{ with $a\neq k$} \}.
\]
Every basis vector $\ket{xyz}$ lies in exactly one of
these subspaces: in $E_{abc}$ for some $a<b<c$ if $x,y,z$ are distinct, and
otherwise in $E_k$ for the index $k$ that occurs once
or three times among $x,y,z$. Thus we have the orthogonal decomposition
\[
(\mathbb C^L)^{\otimes3}
=
\left(\bigoplus_{1\le a<b<c\le L} E_{abc}\right)
\oplus
\left(\bigoplus_{k=1}^{L} E_k\right).
\] By \eqref{eq:Pi}, $\Pi^{1,2}$ maps a basis vector $\ket{xyz}$ to $\frac12(\ket{xyz}+\ket{yxz})$ if $x\neq y$ and to $\frac1L\sum_b\ket{bbz}$ if $x=y$. Hence $\Pi^{1,2}$ maps each of the subspaces $E_{abc}$ and $E_k$ into itself, and the same argument applied to registers $2$ and $3$ shows that $\Pi^{2,3}$ does too. Therefore $H_3=2I-\Pi^{1,2}-\Pi^{2,3}$ maps each of the subspaces $E_{abc}$, $E_k$ into itself, and since the subspaces form an orthogonal decomposition of $(\mathbb C^L)^{\otimes3}$, the eigenvalues of $H_3$ are the eigenvalues of the restrictions of $H_3$ to the subspaces. It therefore suffices to show that the restriction of $H_3$ to each of the subspaces $E_{abc}$ and $E_k$ has no eigenvalue in $(0,\frac12)$.

\paragraph*{Eigenvalues on $E_{abc}$.}

On $E_{abc}$ the three registers carry distinct symbols, so \eqref{eq:Pi} gives $\Pi^{1,2}\ket{xyz}=\frac12(\ket{xyz}+\ket{yxz})$ for every basis vector, that is, $\Pi^{1,2}=\frac12(I+F_{1,2})$ on $E_{abc}$, where $F_{1,2}$ exchanges the first two registers, and likewise $\Pi^{2,3}=\frac12(I+F_{2,3})$. Hence $H_3=\frac12(2I-F_{1,2}-F_{2,3})$ on $E_{abc}$. Write $\ket{w_1},\dots,\ket{w_6}$ for $\ket{abc},\ket{bac},\ket{bca},\ket{cba},\ket{cab},\ket{acb}$ in this order, and consider the index modulo $6$. Then $F_{1,2}$ exchanges $\ket{w_1}$ with $\ket{w_2}$, $\ket{w_3}$ with $\ket{w_4}$, and $\ket{w_5}$ with $\ket{w_6}$, while $F_{2,3}$ exchanges $\ket{w_2}$ with $\ket{w_3}$, $\ket{w_4}$ with $\ket{w_5}$, and $\ket{w_6}$ with $\ket{w_1}$. Consequently
\[
(F_{1,2}+F_{2,3})\ket{w_\ell}=\ket{w_{\ell-1}}+\ket{w_{\ell+1}}\qquad(1\le\ell\le6),
\]
and for $0\le j\le5$ and $\omega=e^{2\pi i/6}$ the vector $\sum_{\ell=1}^6\omega^{j\ell}\ket{w_\ell}$ is an eigenvector of $F_{1,2}+F_{2,3}$ with eigenvalue $\omega^{-j}+\omega^{j}=2\cos(\pi j/3)$, because 
\begin{align*}
(F_{1,2}+F_{2,3})\sum_{\ell}\omega^{j\ell}\ket{w_\ell}&=
\sum_\ell\omega^{j\ell}(\ket{w_{\ell-1}}+\ket{w_{\ell+1}})\\&=\sum_\ell(\omega^{j(\ell+1)}+\omega^{j(\ell-1)})\ket{w_\ell}\\
&=(\omega^{j}+\omega^{-j})\sum_\ell\omega^{j\ell}\ket{w_\ell}.
\end{align*}
 These six vectors $\{\sum_\ell\omega^{j\ell}\ket{w_\ell}\}_{0 \le j \le5}$ are pairwise orthogonal, since $\sum_{\ell=1}^6\omega^{(j'-j)\ell}=0$ for $j\neq j'$, so they form an eigenbasis of $E_{abc}$. The eigenvalues of $F_{1,2}+F_{2,3}$ on $E_{abc}$ are therefore $2\cos(\pi j/3)$ for $0\le j\le5$, namely $2,1,-1,-2,-1,1$, and those of $H_3=\frac12(2I-F_{1,2}-F_{2,3})$ are $0,\frac12,\frac32,2,\frac32,\frac12$. Thus, none of them lies in $(0,\frac12)$.

\paragraph*{Eigenvalues on $E_k$.}
We first compute the action of $H_3$ on the basis vectors
of $E_k$. By \eqref{eq:Pi}, for every $a\neq k$,
\begin{equation}\label{eq:Ek}
\begin{aligned}
H_3\ket{aak}
&=\frac32\ket{aak}-\frac12\ket{aka}
  -\frac1L\sum_{b=1}^L\ket{bbk},\\
H_3\ket{aka}
&=\ket{aka}-\frac12\ket{aak}-\frac12\ket{kaa},\\
H_3\ket{kaa}
&=\frac32\ket{kaa}-\frac12\ket{aka}
  -\frac1L\sum_{b=1}^L\ket{kbb},\\
H_3\ket{kkk}
&=2\ket{kkk}
  -\frac1L\sum_{b=1}^L\ket{bbk}
  -\frac1L\sum_{b=1}^L\ket{kbb}.
\end{aligned}
\end{equation}

The sums $\sum_b\ket{bbk}$ and $\sum_b\ket{kbb}$
are independent of $a$.
Let $\ket{\Sigma}$ and $\ket{\Delta}$ be their sum
and their difference:
\[
\begin{aligned}
\ket{\Sigma}
&=\sum_{b=1}^L\bigl(\ket{bbk}+\ket{kbb}\bigr),\\
\ket{\Delta}
&=\sum_{b=1}^L\bigl(\ket{bbk}-\ket{kbb}\bigr)
 =\sum_{a\neq k}\bigl(\ket{aak}-\ket{kaa}\bigr).
\end{aligned}
\]
Thus,
\[
\sum_{b=1}^L\ket{bbk}
=\frac12\bigl(\ket{\Sigma}+\ket{\Delta}\bigr),
\qquad
\sum_{b=1}^L\ket{kbb}
=\frac12\bigl(\ket{\Sigma}-\ket{\Delta}\bigr).
\]
To cancel the terms proportional to $\ket{\Sigma}$
in the calculation in \eqref{eq:eig} below, define
\[
H'=H_3+\frac1{2L}\ket{\Sigma}\bra{\Sigma}.
\] 
We show below that
\begin{equation}\label{eq:Hpos}
H'\succeq\tfrac12I\quad\text{on }E_k.
\end{equation}
We first derive from \eqref{eq:Hpos} that $H_3$ has no eigenvalue in $(0,\frac12)$ on $E_k$. By the definition of $H'$, every $\ket\phi\in E_k$ with $\braket{\Sigma|\phi}=0$ satisfies
\begin{equation}\label{eq:form}
\bra\phi H_3\ket\phi=\bra\phi H'\ket\phi \ge \tfrac12\|\ket{\phi}\|^2 .
\end{equation}
Suppose that $\ket{\psi_1},\ket{\psi_2}\in E_k$
are two orthonormal eigenvectors of $H_3$, both with
eigenvalues less than $\frac12$.
If either eigenvector is orthogonal to $\ket\Sigma$,
take that eigenvector as $\ket\phi$.
Otherwise, set
\[
\ket\phi
=
\frac{\ket{\psi_1}}{\braket{\Sigma|\psi_1}}
-
\frac{\ket{\psi_2}}{\braket{\Sigma|\psi_2}}.
\]
In this case,
\[
\braket{\Sigma|\phi}
=
\frac{\braket{\Sigma|\psi_1}}
     {\braket{\Sigma|\psi_1}}
-
\frac{\braket{\Sigma|\psi_2}}
     {\braket{\Sigma|\psi_2}}
=1-1=0,
\]
and $\ket\phi\neq0$ because $\ket{\psi_1}$ and
$\ket{\psi_2}$ are linearly independent.
Thus, in either case, $\ket\phi$ is a nonzero vector
in $E_k$ orthogonal to $\ket\Sigma$.
Since it is a linear combination of eigenvectors
of $H_3$ with eigenvalues less than $\frac12$,
\[
\bra\phi H_3\ket\phi<\tfrac12\|\ket{\phi}\|^2,
\]
contradicting \eqref{eq:form}.
Therefore, at most one eigenvalue of $H_3$ on $E_k$ is less than $\frac12$.

On the other hand, $0$ is an eigenvalue of $H_3$ on $E_k$:
adding the first three lines of \eqref{eq:Ek} gives
$H_3(\ket{aak}+\ket{aka}+\ket{kaa})
=\ket{aak}+\ket{kaa}-\frac1L\ket\Sigma$
for every $a\neq k$, and together with the last line
of \eqref{eq:Ek},
\[
H_3\Bigl(
\ket{kkk}+\sum_{a\neq k}
\bigl(\ket{aak}+\ket{aka}+\ket{kaa}\bigr)
\Bigr)
=
2\ket{kkk}
+\sum_{a\neq k}\bigl(\ket{aak}+\ket{kaa}\bigr)
-\tfrac{1+(L-1)}{L}\ket\Sigma
=
\ket\Sigma-\ket\Sigma=0.
\]
Hence $H_3$ has no eigenvalue in $(0,\frac12)$ on $E_k$. This completes the proof.

\paragraph*{Proving \eqref{eq:Hpos}.}

It remains to prove \eqref{eq:Hpos}. By \eqref{eq:Ek} and the definitions of $H'$ and
$\ket\Sigma$, for every $a\neq k$ we can find the following five eigenvectors of $H'$ whose eigenvalues are all at least $1/2$. In the $\ket{kkk}$ calculation, $\langle\Sigma|kkk\rangle=2$, since the two terms $\ket{bbk}$ and $\ket{kbb}$ both equal $\ket{kkk}$ when $b=k$; this explains the coefficient $2/(2L)$ below.
\begin{equation}\label{eq:eig}
\begin{aligned}
H'\bigl(\ket{aak}+2\ket{aka}+\ket{kaa}\bigr)
&=\tfrac12\bigl(\ket{aak}+2\ket{aka}+\ket{kaa}\bigr)
  -\tfrac1L\ket\Sigma+\tfrac{2}{2L}\ket\Sigma\\
&=\tfrac12\bigl(\ket{aak}+2\ket{aka}+\ket{kaa}\bigr),\\
H'\bigl(\ket{aak}-\ket{aka}+\ket{kaa}\bigr)
&=2\bigl(\ket{aak}-\ket{aka}+\ket{kaa}\bigr)
  -\tfrac1L\ket\Sigma+\tfrac{2}{2L}\ket\Sigma\\
&=2\bigl(\ket{aak}-\ket{aka}+\ket{kaa}\bigr),\\
H'\ket{kkk}
&=2\ket{kkk}-\tfrac1L\ket\Sigma
  +\tfrac{2}{2L}\ket\Sigma
 =2\ket{kkk},\\
H'\ket\Delta
&=\left(\tfrac32-\tfrac{L-1}{L}\right)\ket\Delta
 =\left(\tfrac12+\tfrac1L\right)\ket\Delta,\\
H'\left(\ket{aak}-\ket{kaa}-\frac{\ket\Delta}{L-1}\right)
&=\tfrac32\bigl(\ket{aak}-\ket{kaa}\bigr)
  -\tfrac1L\ket\Delta
  -\frac1{L-1}\left(\tfrac12+\tfrac1L\right)\ket\Delta\\
&=\tfrac32\left(\ket{aak}-\ket{kaa}
               -\frac{\ket\Delta}{L-1}\right).
\end{aligned}
\end{equation}
We now check that the eigenvectors in \eqref{eq:eig}
span $E_k$. For every $a\neq k$, a linear combination of the fourth
and fifth eigenvectors in \eqref{eq:eig} gives
\[
\ket{aak}-\ket{kaa}
=
\left(\ket{aak}-\ket{kaa}-\frac{\ket\Delta}{L-1}\right)
+\frac{\ket\Delta}{L-1}.
\]
Then, all of $\ket{aak},\ket{aka},\ket{kaa}$ can be reconstructed by linear combinations of
\[
\ket{aak}+2\ket{aka}+\ket{kaa},\qquad
\ket{aak}-\ket{aka}+\ket{kaa},\qquad
\ket{aak}-\ket{kaa}.
\]
Since $\ket{kkk}$ is also included in~\eqref{eq:eig}, these eigenvectors
span $E_k$.
Therefore, \eqref{eq:Hpos} follows.
\end{proof}
 
\begin{claim}[Restated]
For every integer $t\ge2$, $H_t\succeq\dfrac{4}{t^2}\,(I-G_t)$.
\end{claim}

\begin{proof}[Proof of Claim~\ref{cl:A}]
Let $h_i=I-\Pi^{i,i+1}$ for $1\le i<t$, so that $H_t=\sum_{i=1}^{t-1}h_i$. 
Set $\kappa_i=i(t-i)$ for $1\le i<t$, with $\kappa_0=\kappa_t=0$, and define
\[
D=\sum_{i=1}^{t-1}\kappa_ih_i.
\]
Let $P$ be the projector onto $\ker D$.
We first prove $D\succeq I-P$.
Expanding the square,
\begin{equation}\label{eq:square}
\begin{aligned}
D^2
&=\sum_{i=1}^{t-1}\kappa_i^2h_i
+\sum_{i=1}^{t-2}\kappa_i\kappa_{i+1}(h_ih_{i+1}+h_{i+1}h_i)
+\sum_{\substack{1\le i,j<t\\|i-j|\ge2}}\kappa_i\kappa_jh_ih_j\\
&\succeq\sum_{i=1}^{t-1}\kappa_i^2h_i
-\frac12\sum_{i=1}^{t-2}\kappa_i\kappa_{i+1}(h_i+h_{i+1})\\
&=\sum_{i=1}^{t-1}\kappa_i
\left(\kappa_i-\frac{\kappa_{i-1}+\kappa_{i+1}}2\right)h_i\\
&=\sum_{i=1}^{t-1}\kappa_ih_i=D.
\end{aligned}
\end{equation}
Here, the inequality uses Lemma~\ref{lem:three}, applied to
registers $i,i+1,i+2$ for each $1\le i\le t-2$,
and the fact that $h_ih_j\succeq0$ when $|i-j|\ge2$,
since $h_i$ and $h_j$ act on disjoint pairs of registers. Since $D$ is positive semidefinite, it has an orthonormal basis $\ket{z_1},\ket{z_2},\dots$ of eigenvectors, say $D\ket{z_j}=\lambda_j\ket{z_j}$ with $\lambda_j \ge 0$, and \eqref{eq:square} gives $\lambda_j^2=\bra{z_j}D^2\ket{z_j}\ge\bra{z_j}D\ket{z_j}=\lambda_j$, so $\lambda_j=0$ or $\lambda_j\ge1$. Expanding an arbitrary vector as $\ket x=\sum_jc_j\ket{z_j}$,
\[
\bra xD\ket x=\sum_j\lambda_j|c_j|^2\ \ge\ \sum_{j:\lambda_j\neq0}|c_j|^2=\bra x(I-P)\ket x ,
\]
that is, $D\succeq I-P$. 

Since $\kappa_i>0$ for $1\le i<t$,
\[
\ker D = \ker \sum_i \kappa_i h_i = \ker \sum_i h_i =\ker H_t.
\]
Hence, by Lemma~\ref{lem:ground-space}, $\ker H_t=S_t$, meaning $I-P=I-G_t$.
Finally, $i(t-i)\le t^2/4$ for $1\le i<t$, so
\[
\frac{t^2}{4}H_t = \frac{t^2}{4}\sum_i h_i
\succeq D
\succeq I-P
=I-G_t.
\]
This proves Claim~\ref{cl:A}.
\end{proof}

\section{Proof of Claim B}
\begin{claim}[Restated]
For $1\le j\le s-2$, the operator $\gamma_j+\gamma_{j+1}$ has no eigenvalue in the open interval $(0,3/4)$.
\end{claim}

Fix $1\le j\le s-2$.
The operator $\gamma_j+\gamma_{j+1}$ acts only on
the three consecutive blocks $j,j+1,j+2$.
Thus, we relabel these blocks as $1,2,3$, and write this operator as
$A=(I-G_{2m}^{1,2})+(I-G_{2m}^{2,3}).$
To prove Claim~\ref{cl:B}, it suffices to show that, for every vector
$\ket{\psi}$ on blocks $1,2,3$,
\begin{equation}\label{eq:claim-b-bound}
\bra{\psi}A\ket{\psi}
\ge \frac9{11}\|(I-G_{3m})\ket{\psi}\|^2.
\end{equation}
Indeed, $A$ vanishes on $S_{3m}$, so any eigenvector of
the Hermitian operator $A$ with a nonzero eigenvalue
is orthogonal to $S_{3m}$.
Applying \eqref{eq:claim-b-bound} to such an eigenvector
shows that its eigenvalue is at least $9/11>3/4$,
as required in Claim B.

The following lemma will be used to prove
\eqref{eq:claim-b-bound}.

\begin{lemma}\label{lem:gram}
Let $Z=\{u\in\{0,1\}^L:u_1=0\}$.
Fix a finite-dimensional Hilbert space of arbitrary dimension.
For each $u\in Z$, let $\ket{a_u}$ be a vector in that space.
Then
\[
\frac9{10}\sum_{u\in Z}\|\ket{a_u}\|^2
\le
\left\|\sum_{u\in Z}
\ket{\phi_u}^{\otimes m}\otimes\ket{a_u}\right\|^2
\le
\frac{11}{10}\sum_{u\in Z}\|\ket{a_u}\|^2.
\]
In particular, the vectors $\{\ket{\phi_u}^{\otimes m}\}_{u \in Z}$ are linearly independent.
\end{lemma}

\begin{proof}[Proof of Lemma~\ref{lem:gram}]
For a fixed $u\in Z$, there are $\binom{L-1}{\ell}$
strings $v\in Z$ that differ from $u$ in exactly $\ell$ bits,
and $\langle\phi_u|\phi_v\rangle=1-2\ell/L$.
Combining the terms indexed by $\ell$ and $L-\ell$ uses $\binom{L-1}{\ell}+\binom{L-1}{L-\ell}=\binom L\ell$. When $L$ is even, the middle term $\ell=L/2$ vanishes because $1-2\ell/L=0$. Thus
\begin{equation}\label{eq:label-overlap-bound}
\begin{aligned}
\sum_{\substack{v\in Z\\v\ne u}}
|\langle\phi_u|\phi_v\rangle|^m
&=\sum_{\ell=1}^{L-1}\binom{L-1}{\ell}
  \left|1-\frac{2\ell}{L}\right|^m\\
&=\sum_{1\le\ell<L/2}\binom{L}{\ell}
  \left(1-\frac{2\ell}{L}\right)^m
\le\sum_{\ell\ge1}\bigl(Le^{-2m/L}\bigr)^\ell
\le\frac1{10}.
\end{aligned}
\end{equation}
Here, the inequalities use $\binom{L}{\ell}\le L^\ell$,
$1-x\le e^{-x}$, and
$Le^{-2m/L}\le e^{-12/5}<1/11$.
We now have
\begin{align*}
&\left|
\left\|\sum_{u\in Z}\ket{\phi_u}^{\otimes m}
                       \otimes\ket{a_u}\right\|^2
-\sum_{u\in Z}\|\ket{a_u}\|^2
\right|\\
&=\left|
\sum_{\substack{u,v\in Z\\u\ne v}}
\langle\phi_u|\phi_v\rangle^m\langle a_u|a_v\rangle
\right|\\
&\le
\sum_{\substack{u,v\in Z\\u\ne v}}
|\langle\phi_u|\phi_v\rangle|^m
\|\ket{a_u}\|\,\|\ket{a_v}\|
\qquad\substack{\text{(triangle inequality and Cauchy--Schwarz)}}\\
&\le
\frac12\sum_{\substack{u,v\in Z\\u\ne v}}
|\langle\phi_u|\phi_v\rangle|^m
\bigl(\|\ket{a_u}\|^2+\|\ket{a_v}\|^2\bigr)
\qquad(2ab\le a^2+b^2)\\
&=
\sum_{u\in Z}\|\ket{a_u}\|^2
\sum_{\substack{v\in Z\\v\ne u}}
|\langle\phi_u|\phi_v\rangle|^m
\\
&\le\frac1{10}\sum_{u\in Z}\|\ket{a_u}\|^2
\qquad\text{(by }\eqref{eq:label-overlap-bound}).
\end{align*}
The two inequalities in Lemma~\ref{lem:gram} follow.
Taking $\ket{a_u}\in\mathbb C$ in the lower bound proves that
$\{\ket{\phi_u}^{\otimes m}:u\in Z\}$ is linearly independent.
\end{proof}

We are now ready to prove Claim B.

\begin{proof}[Proof of \eqref{eq:claim-b-bound}]
It suffices to
construct $\ket{\chi}\in S_{3m}$ with
$\|\ket{\psi}-\ket{\chi}\|^2\le(11/9)\bra{\psi}A\ket{\psi}$ since $\|\ket{\psi}-\ket{\chi}\|^2 \ge \|(I-G_{3m})\ket{\psi}\|^2$.

We write
\[
\bra{\psi}A\ket{\psi}
=
\|\ket{\psi}-G_{2m}^{1,2}\ket{\psi}\|^2
+
\|\ket{\psi}-G_{2m}^{2,3}\ket{\psi}\|^2.
\]
To bound these two squared norms from below, we first
decompose $\ket{\psi}$ according to $S_m$ and $S_m^\perp$
on their common block $2$.
Complementing all bits of $u$ negates $\ket{\phi_u}$,
and $Z$ contains one string from each complementary pair, so
$S_m=\operatorname{span}\{\ket{\phi_u}^{\otimes m}:u\in Z\}$.
We can therefore write
\[
\begin{gathered}
\ket{\psi}=\ket{\psi_\perp}
+\sum_{u\in Z}\ket{\phi_u}^{\otimes m}_{2}\otimes\ket{a_u}_{1,3},
\qquad \ket{a_u}\in(\mathbb C^L)^{\otimes2m},\\
\ket{\psi_\perp}\in
(\mathbb C^L)^{\otimes m}\otimes S_m^\perp\otimes(\mathbb C^L)^{\otimes m}.
\end{gathered}
\]
The vector $\ket{\psi_\perp}$ is orthogonal to every sum of vectors
with a factor from $S_m$ on block $2$.

To obtain $\ket{\chi}\in S_{3m}$, we project blocks $1$ and $3$
of each $\ket{a_u}$ onto the span of $\ket{\phi_u}^{\otimes m}$.
Let $Q_u=(\ket{\phi_u}\bra{\phi_u})^{\otimes m}$ be the
orthogonal projector onto that span, and set
\[
\ket{\chi}=\sum_{u\in Z}\ket{\phi_u}^{\otimes m}_{2}
\otimes\bigl((Q_u\otimes Q_u)\ket{a_u}\bigr)_{1,3}.
\]
Each summand is a scalar multiple of $\ket{\phi_u}^{\otimes3m}$,
so $\ket{\chi}\in S_{3m}$. Subtracting $\ket{\chi}$ from $\ket{\psi}$
and applying Lemma~\ref{lem:gram} gives the following upper bound:
\begin{equation}\label{eq:claim-b-distance-upper}
\begin{aligned}
\|(I-G_{3m})\ket{\psi}\|^2
&=\|(I-G_{3m})(\ket{\psi}-\ket{\chi})\|^2
&&\bigl(G_{3m}\ket{\chi}=\ket{\chi}\bigr)\\
&\le\|\ket{\psi}-\ket{\chi}\|^2
&&\bigl(\|I-G_{3m}\|\le1\bigr)\\
&=\|\ket{\psi_\perp}\|^2+
\left\|\sum_{u\in Z}\ket{\phi_u}^{\otimes m}_{2}
\otimes\bigl((I-Q_u\otimes Q_u)\ket{a_u}\bigr)_{1,3}\right\|^2
&&\text{(orthogonality)}\\
&\le\|\ket{\psi_\perp}\|^2+
\frac{11}{10}\sum_{u\in Z}\|(I-Q_u\otimes Q_u)\ket{a_u}\|^2
&&\text{(Lemma~\ref{lem:gram}).}
\end{aligned}
\end{equation}
For each $u\in Z$, using $I-Q_u\otimes Q_u=(I-Q_u)\otimes I+Q_u\otimes(I-Q_u)$, 
\begin{equation}\label{eq:claim-b-two-projections}
\begin{aligned}
\|(I-Q_u\otimes Q_u)\ket{a_u}\|^2
&=\|((I-Q_u)\otimes I)\ket{a_u}\|^2
 +\|(Q_u\otimes(I-Q_u))\ket{a_u}\|^2\\
&\le\|((I-Q_u)\otimes I)\ket{a_u}\|^2
 +\|(I\otimes(I-Q_u))\ket{a_u}\|^2
.
\end{aligned}
\end{equation}
The projectors $G_{2m}^{1,2}$ and $G_{2m}^{2,3}$ project blocks
$1,2$ and $2,3$, respectively, onto $S_{2m}$ and act as the identity
on the remaining block. By the spanning description of $S_{2m}$ above,
there are $\ket{b_u},\ket{c_u}\in(\mathbb C^L)^{\otimes m}$ with
\[
\begin{aligned}
G_{2m}^{1,2}\ket{\psi}
&=\sum_{u\in Z}\ket{\phi_u}^{\otimes m}_{1}
 \otimes\ket{\phi_u}^{\otimes m}_{2}\otimes\ket{b_u}_{3},\\
G_{2m}^{2,3}\ket{\psi}
&=\sum_{u\in Z}\ket{c_u}_{1}
 \otimes\ket{\phi_u}^{\otimes m}_{2}\otimes\ket{\phi_u}^{\otimes m}_{3}.
\end{aligned}
\]
Subtract the two displayed expansions from that of $\ket{\psi}$:
\begin{align*}
\bra{\psi}A\ket{\psi}
&=\|(I-G_{2m}^{1,2})\ket{\psi}\|^2
 +\|(I-G_{2m}^{2,3})\ket{\psi}\|^2
\qquad\substack{I-G_{2m}^{1,2},\ I-G_{2m}^{2,3}\\
                 \text{are orthogonal projectors}}\\
&=2\|\ket{\psi_\perp}\|^2
 +\left\|\sum_{u\in Z}\ket{\phi_u}^{\otimes m}_{2}\otimes
 \bigl(\ket{a_u}-\ket{\phi_u}^{\otimes m}\otimes\ket{b_u}\bigr)_{1,3}
 \right\|^2\\
&\quad+\left\|\sum_{u\in Z}\ket{\phi_u}^{\otimes m}_{2}\otimes
 \bigl(\ket{a_u}-\ket{c_u}\otimes\ket{\phi_u}^{\otimes m}\bigr)_{1,3}
 \right\|^2
\qquad\text{(orthogonality)}\\
&\ge2\|\ket{\psi_\perp}\|^2+
 \frac9{10}\sum_{u\in Z}\Bigl(
 \|\ket{a_u}-\ket{\phi_u}^{\otimes m}\otimes\ket{b_u}\|^2
 +\|\ket{a_u}-\ket{c_u}\otimes\ket{\phi_u}^{\otimes m}\|^2\Bigr)
\qquad\text{(Lemma~\ref{lem:gram})}\\
&\ge2\|\ket{\psi_\perp}\|^2+
 \frac9{10}\sum_{u\in Z}\Bigl(
 \|((I-Q_u)\otimes I)\ket{a_u}\|^2
 +\|(I\otimes(I-Q_u))\ket{a_u}\|^2\Bigr)
\qquad\substack{(I-Q_u)\ket{\phi_u}^{\otimes m}=0\\
             \|(I-Q_u)\otimes I\|\le1\\
             \|I\otimes(I-Q_u)\|\le1}\\
&\ge2\|\ket{\psi_\perp}\|^2+
 \frac9{10}\sum_{u\in Z}\|(I-Q_u\otimes Q_u)\ket{a_u}\|^2
\qquad\text{by }\eqref{eq:claim-b-two-projections}\\
&\ge\frac9{11}\left(\|\ket{\psi_\perp}\|^2+
 \frac{11}{10}\sum_{u\in Z}\|(I-Q_u\otimes Q_u)\ket{a_u}\|^2\right)
\qquad\bigl(2\ge9/11\bigr)\\
&\ge\frac9{11}\|\ket{\psi}-\ket{\chi}\|^2
 \ge\frac9{11}\|(I-G_{3m})\ket{\psi}\|^2
\qquad\text{by }\eqref{eq:claim-b-distance-upper}.
\end{align*}
\end{proof}

\end{document}